\documentclass[sn-mathphys-num]{sn-jnl}% Math and Physical Sciences Numbered Reference Style 
\usepackage{graphicx}%
\usepackage{multirow}%
\usepackage{amsmath,amssymb,amsfonts}%
\usepackage{amsthm}%
\usepackage{mathrsfs}%
\usepackage[title]{appendix}%
\usepackage{xcolor}%
\usepackage{textcomp}%
\usepackage{manyfoot}%
\usepackage{booktabs}%
\usepackage{algorithm}%
\usepackage{algorithmicx}%
\usepackage{algpseudocode}%
\usepackage{listings}%
\usepackage{tabularx}
\theoremstyle{thmstyleone}%
\theoremstyle{thmstyletwo}%

\theoremstyle{thmstylethree}%

\begin{document}

\title[Sequential Bayesian Registration for Functional Data]{Sequential Bayesian Registration for Functional Data}


\author[1]{\fnm{Yoonji} \sur{Kim}}\email{yoonj2kim@gmail.com}

\author[2]{\fnm{Oksana A.} \sur{Chkrebtii}}\email{oksana@stat.osu.edu}

\author[1]{\fnm{Sebastian A.} \sur{Kurtek}}\email{kurtek.1@stat.osu.edu}

\affil*[1]{\orgdiv{Department of Statistics}, \orgname{The Ohio State University}, \orgaddress{\street{1958 Neil Avenue}, \city{Columbus}, \postcode{43210}, \state{Ohio}, \country{USA}}}


\abstract{In many modern applications, discretely-observed data may be naturally understood as a set of functions. Functional data often exhibit two confounded sources of variability: amplitude ($y$-axis) and phase ($x$-axis). The extraction of amplitude and phase, a process known as registration, is essential in exploring the underlying structure of functional data in a variety of areas, from environmental monitoring to medical imaging. Critically, such data are often gathered sequentially with new functional observations arriving over time. 
Despite this, existing registration procedures do not sequentially update inference based on the new data, requiring model refitting. To address these challenges, we introduce a Bayesian framework for sequential registration of functional data, which updates statistical inference as new sets of functions are assimilated. This Bayesian model-based sequential learning approach utilizes sequential Monte Carlo sampling to recursively update the alignment of observed functions while accounting for associated uncertainty. 
 Distributed computing significantly reduces computational cost relative to refitting the model using an iterative method such as Markov chain Monte Carlo on the full data. Simulation studies and comparisons reveal that the proposed approach performs well even when the target posterior distribution has a challenging structure. We apply the proposed method to three real datasets: (1) functions of annual drought intensity near Kaweah River in California, (2) annual sea surface salinity functions near Null Island, and (3) a sequence of repeated patterns in electrocardiogram signals.}




\maketitle

\section{Introduction}

In various real world problems, the goal of statistical analysis is to discover and explore patterns in the trajectories formed by the temporal evolution of a variable of interest. This type of data is commonly referred to as functional data, and is often recorded on a very fine temporal grid. The use of multivariate statistical methods to analyze functional data is inappropriate for two main reasons: (1) failure to account for the underlying infinite-dimensional structure of the data, and (2) inability to appropriately model strong temporal dependence within each functional observation \citep{ferraty2006,horvath2012}. 
This has given rise to the field of functional data analysis (FDA), which provides a comprehensive framework for statistical modeling, summarization, analysis and visualization of data that comes in the form of functions \citep{ramsay1991,ramsay2005}.

\begin{figure}[!t]
  \centering\includegraphics[width=10.5cm]{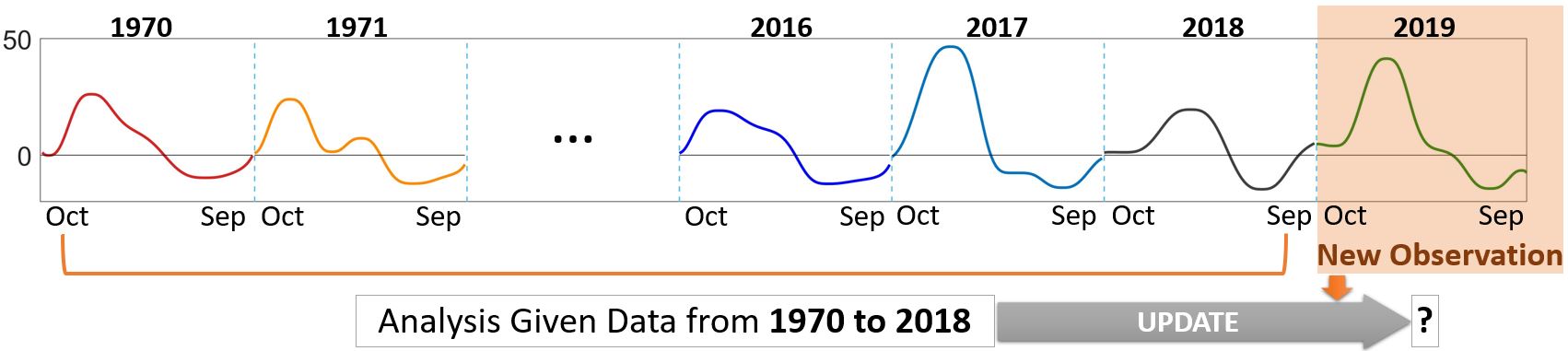}
  \caption{Visualization of sequential learning for trajectories of annual drought intensity near Kaweah River in California during years 1970 to 2019.}
  \label{fig:motiv_ex}
\end{figure}

An important and common feature of functional data is that sampling is often automated or conducted over long periods of time, so that new observed functions arrive sequentially. In general, there are two different approaches to perform statistical analysis for such an expanding collection of data in the finite or infinite-dimensional settings. The first is to implement the full spectrum of statistical analysis every time a new observation arrives, referred to as \textit{batch learning}. The second is to update the existing analysis by accounting for the new data, referred to as  \textit{sequential learning} or online learning \citep{hoi2018}. Most existing FDA techniques are designed for batch learning, meaning that they are performed once a given number of functional data is collected, and the analysis must be repeated on the entire sample as more data arrives. 
This fails to account for the sequential way in which functional data is often gathered, with the sample size increasing over time in many application domains such as environmental monitoring or biomedical imaging. For example, trajectories of annual temperature, or other measures related to the environment, are formulated through sets of repeated measurements on an annual basis; in medicine, biosignals such as electrocardiogram (ECG) signals or gait measurements contain repetitions of a particular pattern wherein each repetition can be interpreted as an observation. We provide a visualization of sequential learning for trajectories of annual drought intensity near Kaweah River in California in Figure \ref{fig:motiv_ex} (see Section \ref{sec:drought} for details).
In such scenarios, new functional observations are added to existing data sequentially, so the statistical analysis pipeline must be modified to allow for updating and monitoring of inferential results as the collection of data expands.

\subsection{Sequential learning}
A sequential learning method seeks to update current inferential results to include new data.
The Bayesian approach is well-suited to this problem because it (1) provides a systematic way to assimilate new data by updating the posterior distribution as new data arrives, and (2) allows the user to keep track of structured uncertainty. 
In most scenarios, the posterior distribution does not have a closed form, so inference is based on estimates of posterior features obtained from posterior samples. Perhaps the most widely used sampling-based method for Bayesian inference is Markov chain Monte Carlo (MCMC), which is a batch learning algorithm; as such, every time new data arrives, MCMC sampling must be repeated using the full data, resulting in inefficient computation. On the other hand, sequential Bayesian learning via, e.g., sequential Monte Carlo (SMC) can assimilate new data as it arrives. Unlike MCMC, which targets a fixed posterior density, SMC defines a sequence of intermediate target densities, each represented with a set of weighted samples, or \textit{particles}, that are perturbed and reweighted to represent the next density in the sequence \citep{andrieu1999,storvik2002,lopes2011}. When these intermediate target densities correspond to posteriors under different data availability scenarios, SMC becomes a sequential learning algorithm. 

In addition to the efficiency of sequentially updating inference as new data arrives, SMC can exploit distributed computing to speed up implementation relative to MCMC, because each SMC particle is updated independently. SMC also performs well for sampling from challenging target posteriors, e.g., in the presence of posterior multimodality, because the intermediate sequence of target densities can act as a bridge between the prior and a challenging posterior distribution \citep{schweizer2012, paulin2019, dai2020}. 
The potential for particle degeneracy, i.e., when particle weights become very small in certain situations and lead to large sampling variance in the SMC estimator, can be ameliorated through resampling or via techniques such as block sampling \citep{gilks2001,chopin2002,delmoral2006,kantas2013,beskos2015}. While SMC methods for sequential Bayesian learning have been used for assimilating multivariate data \citep{lopes2011, liu2001}, our focus in this work is on a natural inferential problem arising in FDA.

\subsection{Functional data registration}
A common challenge in FDA is the presence of two confounded sources of variability: amplitude and phase \citep{ramsay1998}. 
Examining Figure \ref{fig:motiv_ex}, we note that the functions contain similar shape features, e.g., number of local extrema, but the timing of the features is not the same along the temporal $x$-axis across all observations. For example, drought intensity tends to increase sharply early in each hydrological year. Then, around the month of December, drought intensity decreases following a small peak. Thus, the variation in the data can be attributed to two sources: (1) the magnitude of function values (drought intensity) termed amplitude ($y$-axis) variability, and (2) the timing of amplitude features (e.g., local extrema) termed phase ($x$-axis) variability. Phase variation may be an inherent feature of data or the result of measurement error, and can be regarded as nuisance or a quantity of interest depending on the application. Importantly, phase variation cannot be ignored when performing statistical analysis of functional data as this may lead to misleading results \citep{marron2015}. Instead, amplitude and phase components of functions should first be estimated through a process called registration. A common registration approach is to consider the observed functions as deformed versions of an unknown template function, and to extract their phase components via horizontal synchronization to an estimate of this template.

There are many methods for functional data registration \citep{ramsay1998, tang2008, srivastava2011b}. Here, we briefly review a small subset. Landmark-based registration focuses on synchronization of (a small number of) points, which represent important features of the observed functions, e.g., local extrema \citep{ramsay2005, kneip2000}. While conceptually simple, these approaches rely on a faithful specification of landmarks, which is often difficult and time consuming. Metric-based registration uses a distance on the function space of interest to achieve horizontal synchronization of entire functions, i.e., it does not require landmark specification. However, the distance must satisfy a key invariance property (see Section \ref{subsec:regist} for details), which is not the case for the standard $\mathbb{L}^2$ distance commonly used in FDA.  As an alternative, \cite{srivastava2011b} proposed the extended Fisher-Rao (eFR) metric, which is equivalent to the $\mathbb{L}^2$ distance under a simple transformation of the original functions, called the square-root velocity transformation. The resulting metric-based registration method is commonly referred to as elastic. Bayesian model-based registration of functional data has been explored relatively recently. In this setting, the main challenge lies in specifying an appropriate prior distribution over the phase component of functional data. \cite{telesca2008} were the first to approach registration from this perspective and modeled phase via (constrained) B-splines. \cite{lu2017} explicitly considered the geometry of the representation space of the phase component and specified a Gaussian process prior on this space. \cite{horton2021} also modeled phase via a Gaussian process, but additionally allowed the incorporation of landmark information in the registration process. Finally, \cite{cheng2016} and \cite{bharath2020} used the Dirichlet distribution as a prior model on consecutive increments of discretized phase functions. An extension of Bayesian registration to sparse/fragmented functional data was developed by \cite{matuk2021}. Importantly, all of the aforementioned model-based approaches rely on batch learning, and in particular MCMC, for inference.

\subsection{Summary of proposed approach and paper organization}
Motivated by data collection scenarios such as the one presented in Figure \ref{fig:motiv_ex}, we propose a novel sequential Bayesian learning approach for registration of functional data. The proposed approach updates the posterior over all unknown model components, including the template function, all of the phase components, and the observation error variance, when a new function arrives. 
This framework leverages SMC to efficiently update the joint posterior distribution over the template function and the phase components associated with each observation as new data arrives. To the best of our knowledge, this is the first sequential inference strategy for this statistical problem.

The proposed computational approach is applicable to general registration models for functional data, though the specific model we consider is built on the state-of-the-art framework of \cite{lu2017}. In this model, a template function captures amplitude variation across observations, and individual phase functions account for the phase of each observed function. For each phase, we employ a low-dimensional piecewise linear prior model \citep{bharath2020}.
We further address the challenge posed by the increasing dimension of the state space corresponding to the addition of an unknown phase for the incoming data.  
In addition to using a low-dimensional prior on phase, we propose an efficient stochastic initialization strategy for the new components of each SMC particle. 

The rest of this paper is organized as follows. Section \ref{sec:model} provides a brief review of elastic registration and introduces the Bayesian registration model.
Sections \ref{sec:smc} and \ref{sec:algo} describe SMC methods for state spaces of fixed and increasing dimension and the proposed SMC Bayesian registration algorithm enabling sequential inference as new functional data is observed, respectively. Section \ref{sec:numerical_applications} presents simulations and real data examples. We close with a brief discussion in Section \ref{sec:summary}. Appendix A in the supplement contains the detailed sequential Bayesian registration algorithm (Algorithm 1). Appendices B and C present additional registration results for annual sea surface salinity functions considered in Section \ref{sec:SSS} and segmented PQRST complexes considered (and defined) in Section \ref{sec:ecg}, respectively. Appendix D compares effective sample size for MCMC-based batch learning and the proposed sequential approach. Finally, Appendix E considers comprehensive sensitivity analyses. The code to reproduce results in this manuscript can be found at \url{https://github.com/yoonj2kim/Bayesian-sequential-registration}.

%\section{Background material}
\section{Functional registration model}
\label{sec:model}
%\section{Bayesian model for registration of functional data}

We first provide an overview of the elastic registration framework. For brevity, we only present concepts relevant to the Bayesian hierarchical model for elastic registration of functions, which is based on the square-root velocity function representation; we refer the interest reader to \citet{srivastava2016} for further details. The choice of this representation is motivated by the desirable properties of the extended Fisher-Rao metric as described in Section \ref{subsec:regist}. The presented model is built on the general structure of the observation model in \cite{lu2017}, while utilizing different priors including a low-dimensional prior distribution over phase.

\subsection{Elastic registration}
\label{subsec:regist}

We restrict our attention to absolutely continuous functions with domain $[0,1]$, resulting in the representation space $\mathcal{F}=\{f:[0,1]\rightarrow\mathbb{R}\mid f\textnormal{ is absolutely continuous}\}$. The domain can be further extended to any $[a,b] \subset \mathbb{R},\ a<b.$
The phase component of a function $f\in\mathcal{F}$ is denoted by $\gamma$ and is an element of  $\Gamma=\{\gamma:[0,1]\rightarrow[0,1]\mid\gamma(0)=0,\ \gamma(1)=1,\ 0<\dot{\gamma}<\infty\}$, where $\dot{\gamma}$ is the derivative of $\gamma$. The main goal of registration is to estimate the phase components $\gamma_1,\dots,\gamma_n$ of a set of functions $f_1,\ldots,f_n$, such that $f_i\circ\gamma_i,\ i=1,\dots,n$ are horizontally synchronized, i.e., their features are well-aligned. The composition of $f$ and $\gamma$ is usually referred to as domain warping. Metric-based registration utilizes a distance on $\mathcal{F}$ to quantify the quality of alignment between two functions, and estimation of phase is carried out by minimizing this distance over elements of $\Gamma$. The chosen distance must satisfy $d(f_1,f_2)=d(f_1\circ\gamma,f_2\circ\gamma)$ for $f_1,\ f_2\in\mathcal{F}$ and any $\gamma\in\Gamma$, i.e., it must be invariant to simultaneous domain warping. Crucially, the commonly used $\mathbb{L}^2$ distance does not satisfy this property.

\cite{srivastava2011b} proposed a formulation of metric-based registration using the extended Fisher-Rao (eFR) Riemannian metric. The eFR distance is preserved under simultaneous domain warping, i.e., $d_{eFR}(f_1,f_2)=d_{eFR}(f_1\circ\gamma,f_2\circ\gamma)$ for $f_1,\ f_2\in\mathcal{F}$ and any $\gamma\in\Gamma$. Since $d_{eFR}$, and thus the resulting registration problem, are not computationally tractable, Srivastava et al. introduced a transformation that allows the distance to be computed in closed form. The \textit{square-root velocity function (SRVF)} representation given by the mapping $Q:\mathcal{F}\rightarrow\mathcal{Q}$ is defined as $Q(f) = \text{sign}(\dot{f})\sqrt{|\dot{f}|}=: q$ for $f\in\mathcal{F}$; the derivative of $f$, $\dot{f}$, is approximated using finite differencing. Given the starting point $f(0)$, the mapping $Q$ is bijective and the original function can be reconstructed using $f(t) = Q^{-1}(f(0),q)(t)=f(0)+\int_0^tq(s)|q(s)|ds$. Further, under this mapping, the eFR metric simplifies to the $\mathbb{L}^2$ metric, i.e., $d_{eFR}(f_1,f_2)=d_{\mathbb{L}^2}(Q(f_1),Q(f_2))$ for $f_1,\ f_2,\in\mathcal{F}$, and the resulting space of SRVFs $\mathcal{Q}$ is a subset of $\mathbb{L}^2([0,1],\mathbb{R})$. 
The domain warping $f\circ\gamma$ of a function $f\in\mathcal{F}$ by $\gamma\in\Gamma$ can be mapped to the SRVF space via $(q,\gamma):=(q\circ\gamma)\sqrt{\dot\gamma}$ \ ($q=Q(f)$).
Finally, the amplitude of a function $f$ can be formally defined through its SRVF $q$ as the equivalence class $[q]=\{(q,\gamma)\mid\gamma\in\Gamma\}$; the set of all amplitudes is the quotient space $\mathcal{Q}/\Gamma$.

\begin{figure}[!t]
\graphicspath{{figures/real1_metric_registration}}
  \centering
  \begin{tabularx}{12cm} { 
  |>{\centering\arraybackslash}X 
   >{\centering\arraybackslash}X 
   >{\centering\arraybackslash}X 
   >{\centering\arraybackslash}X| }
    \hline
    (a)&(b)&(c)&(d)
    \\
    \begin{minipage}{2.5cm}
    \centering\includegraphics[width=2.5cm]{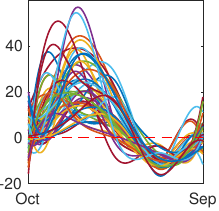}
    \end{minipage}
    &\begin{minipage}{2.5cm}
    \centering\includegraphics[width=2.5cm]{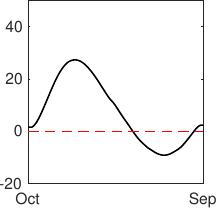}
    \end{minipage}
    &\begin{minipage}{2.5cm}
    \centering\includegraphics[width=2.5cm]{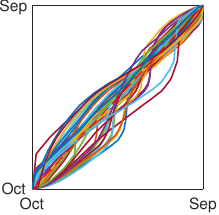}
    \end{minipage}
    &\begin{minipage}{2.5cm}
    \centering\includegraphics[width=2.5cm]{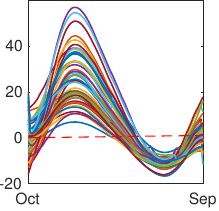}
    \end{minipage}
    \\\hline
  \end{tabularx}
  \caption{Metric-based registration of trajectories of annual drought intensity near Kaweah River in California during years 1970 to 2019. (a) Observed functional data. (b) Estimated template function. (c) Estimated phase components. (d) Registered functions using the phase components in (c).}
    \label{fig:metric_registration}
\end{figure}

The SRVF representation is then used to define the metric-based registration problem as follows.
For registration of two functions, $f_1,\ f_2\in\cal{F}$, we set $f_1$ as the reference function and find the optimal phase component of $f_2$ that minimizes the $\mathbb{L}^2$ distance between their SRVFs $q_1,\ q_2\in\cal{Q}$, i.e., $\gamma^*=\arg\min_{\gamma\in\Gamma}d_{\mathbb{L}^2}(q_1,(q_2,\gamma))$.
When multiple functions $f_1,\ldots,f_n$ are given, we register them to (an estimate of) a representative of the mean equivalence class, referred to as a template, rather than an arbitrarily chosen reference function.
The SRVF of the template function, denoted by $q_\mu$, is estimated using $[q_\mu]=\arg\min_{[q_\mu]\in\mathcal{Q}/\Gamma}\sum_{i=1}^n\min_{\gamma\in\Gamma}d_{\mathbb{L}^2}(q_\mu,(q_i,\gamma))^2$; the solution is an entire equivalence class. For identifiability, one generally selects $q_\mu\in[q_\mu]$ such that the average of the phase components, estimated via $\gamma_i^*=\arg\min_{\gamma_i\in\Gamma}d_{\mathbb{L}^2}(q_\mu,(q_i,\gamma_i))$,  $i=1,\ldots,n$, is the identity $\gamma_{id}(t)=t$. For visualization, the SRVF of the template, $q_\mu$, is mapped to $\mathcal{F}$ using $Q^{-1}$. Figure \ref{fig:metric_registration} presents an illustration of metric-based registration under the elastic framework.

\subsection{Observation model}
Denote the functional data by $f_1,\ldots,f_n\in\mathcal{F}$ and the corresponding SRVFs by $q_1,\ldots,q_n$ $\in\mathcal{Q}$. 
We assume that each datum $f_i$ is a noisy deformation of a latent template function $f_\mu\in\mathcal{F}$ whose SRVF is $q_\mu\in\mathcal{Q}$. The deformation of the $i$th function is expressed through the unknown phase $\gamma_i\in\Gamma$.   
The observation error is assumed to be additive in the SRVF space, so that the $i$th function's SRVF can be written as  $q_i=(q_\mu,\gamma_i^{-1})+\epsilon_i,\ i=1,\dots,n$,
where $\epsilon_i$ is an error process,  and $(q,\gamma)=(q\circ\gamma)\sqrt{\dot{\gamma}}$ denotes the SRVF of the domain warping by $\gamma$ of a function $f\in\cal{F}$ with SRVF $q \in \mathcal{Q}$. 
At the implementation stage, the functional data is discretized across the time domain over a fine grid $[t] = \left(t_1=0,\ldots,t_M=1\right)^\top$, 
where $t_1<t_2<\cdots<t_M$.
Assuming that the error follows a Gaussian process with white noise covariance structure, the observation model is
\begin{equation}
    q_i([t])|q_\mu,\gamma_i,\sigma^2 \overset{ind}{\sim} N((q_\mu,\gamma_i^{-1})([t]),\sigma^2I_M),\ \ \  i=1,\ldots,n,
\end{equation}
where $I_M$ is the $M\times M$ identity matrix. While we assume that the variance is the same across all functions, this assumption may be relaxed; other covariance structures can also be accommodated by the model.

\subsection{Signal components - template function and phase}
We next present the process model involving the unknown quantities $q_\mu$ and $\gamma_i,\ i=1,\dots,n$ (and $\sigma^2$). 
We assume that the SRVF template $q_\mu$ is a linear combination of $B$ basis functions, 
\begin{equation}
    q_\mu(t) = \sum_{b=1}^B c_b\phi_b(t).
\end{equation}
The number and type of basis functions $\phi_1,\ldots,\phi_B:[0,1]\rightarrow\mathbb{R}$ depend on the desired number of features, e.g., the number of local extrema, and the smoothness of the SRVF template. We wish to choose $B$ that is large enough to reproduce in the template prominent features present in the functional data, but not so large as to capture variation due to noise.  Specific choices of basis are discussed in Section \ref{sec:numerical_applications} where we apply this modeling framework to simulated and real data; in all experiments, we use cubic B-splines, which resulted in satisfactory registration results. Other popular choices of bases, e.g., Fourier, which are not employed in our analyses, are discussed in Appendix E.3 in the supplement in the context of sensitivity analyses.
Each phase component, $\gamma_i$, is a strictly increasing function with $\gamma_i(0)=0$ and $\gamma_i(1)=1$. We adopt a piecewise linear model introduced by \cite{bharath2020}, which, while potentially low-dimensional, is still flexible enough to capture phase variation among functional data. The $M_\gamma$-dimensional partition of the domain $[0,1]$ is prespecified as $0=s_1<s_2<\cdots<s_{M_\gamma-1}<s_{M_\gamma}=1$, and the phase model
$\gamma_i = d^{-1}(\textbf{d}_i),\ i = 1, \ldots, n$, is the linear interpolation of phase with increments
$\textbf{d}_i=d(\gamma_i) = (\gamma_i(s_2),\ldots,\gamma_i(s_m)-\gamma_i(s_{m-1}),\ldots,1-\gamma_i(s_{M_\gamma-1}) )^\top$.

\subsection{Prior model}

Prior choice for the template coefficients $\textbf{c}=\left(c_1,\ldots,c_B\right)^\top$ is application-specific, and adaptive models may be considered (see, e.g., \citet{LangBrezger2004}). For generality, we assume a multivariate normal model with mean $0_B$ and diagonal covariance $\Sigma_c$. 
This is equivalent to a $B$-dimensional Gaussian process prior on the SRVF template function \citep{lu2017}. The prior hyperparameters for $\textbf{c}$ are taken to be fixed. Sensitivity analysis to different choices of $\Sigma_c$ is presented in Appendix E.1 in the supplement.

The vector of phase increments $\textbf{d}_i$ must be restricted to the $(M_\gamma-1)$-dimensional simplex. Thus, we assign a Dirichlet distribution as a prior model:
\begin{align}
    \textbf{d}_i|\textbf{u}_i &\overset{ind}{\sim} Dir(\kappa \textbf{u}_i),\ \ \ i=1,\dots,n,
\end{align}
where $\textbf{u}_i =(u_{i(2)},\ldots,u_{i(m-1)}-u_{i(m)},\ldots,1-u_{i(M_\gamma-1)})^\top$ and $u_{i(2)},\ldots,u_{i(M_\gamma-1)}$ are order statistics of a random sample drawn from the uniform distribution on $[0,1]$. 
The constant $\kappa$ serves as a concentration parameter and is fixed. Note that this prior is centered around identity warping $\gamma_{id}$, which corresponds to the case where the template is not warped. Finally, phase increment vectors for the $n$ functions are assumed to be \textit{a priori} independent. In Appendix E.2 in the supplement, we study sensitivity of the posterior to different choices of the partition size $M_\gamma$. 

For the auxiliary parameter $\sigma^2$, the error variance in the SRVF space, we assume an inverse-gamma prior distribution, which is conjugate for the multivariate normal likelihood. The shape and scale hyperparameters $\alpha_\sigma$ and $\beta_\sigma$ are fixed.

\section{Sequential Monte Carlo}
\label{sec:smc}

%This section reviews SMC methods for state spaces of fixed and increasing dimension \citep{liu1998,delmoral2006}, which are adopted to perform sequential inference on the proposed model.

Sequential Monte Carlo (SMC) refers to a class of sampling algorithms targeting a sequence of prespecified distributions \citep{gordon1993}. In the Bayesian inferential setting, these consist of either a sequence of posterior distributions or some transformation thereof. 
Suppose the target distributions have densities $\eta_\tau$, $\tau\in\mathbb{N}_+$ over the state variables $\theta_\tau$, and are defined on a measurable space $(E_\tau,\mathcal{E}_\tau)$.
SMC is a sequential version of importance sampling that generates a set of weighted samples, which are used to approximate features of each intermediate target distribution \citep{gordon1993}. This allows the user to track uncertainty while updating inference recursively, or sequentially annealing challenging posterior distributions.
For an intermediate probability density $\eta_\tau$, samples are first randomly drawn from a different distribution, termed the importance distribution with density denoted by $g_\tau$, which is easy to sample from and is available in closed form.
These samples, also known as \textit{particles}, are then reweighted to reflect the shape of $\eta_\tau$.
To make the description more precise, let $\theta_\tau^{(j)},\ j=1,\ldots,J$ represent $J$ samples drawn from the importance distribution. Then, their corresponding weights are computed as $w_\tau^{(j)}\propto\eta_\tau(\theta_\tau^{(j)})/g_\tau(\theta_\tau^{(j)})$ and subsequently normalized, so that
the pairs $\{(\theta_\tau^{(j)},w_\tau^{(j)}),\ j=1,\ldots,J\}$ form a collection of weighted samples from  $\eta_\tau(\cdot)$.
Crucially, the next density $\eta_{\tau+1}$ in the sequence can then be sampled recursively starting from the weighted pairs from the importance density $g_{\tau+1} = \eta_{\tau}$, and so on.
We present a brief overview of SMC for two different Bayesian inference scenarios that are utilized in the proposed registration approach: (1) a state space with increasing dimension and target distributions having fixed marginal densities across the sequence, and (2) a fixed-dimensional state space and subsequent target distributions that are similar. 

\subsection{SMC for state space with increasing dimension}
\label{sec:smc_liu}
SMC on a state space $E_\tau$ of increasing dimension, i.e., $\dim\left(E_{\tau}\right) < \dim\left(E_{\tau+1}\right)$, is often of interest, such as when the target sequence of distributions consists of posteriors over an increasing number of unknown model components. 
Assume that we are given a set of weighted samples, $\{(\theta_\tau^{(j)},w_\tau^{(j)}),\ j=1,\ldots,J\}$, drawn from the distribution with density $\eta_\tau$, and we aim to modify the weights and particles such that they approximate the next target distribution in the sequence with density $\eta_{\tau+1}$.
Suppose that the state variable at time $\tau+1$ is obtained by appending a new variable $\tilde\theta$ to the previous state, $\theta_{\tau+1}=(\theta_{\tau},\tilde\theta)$, and the marginal density of $\theta_\tau$ at time $\tau+1$ is equal to the density at time $\tau$: $\eta_\tau(\theta_\tau)=\eta_{\tau+1}(\theta_\tau)$.
Interest lies in approximating $\eta_{\tau+1}$ using $\{(\theta_\tau^{(j)},w_\tau^{(j)}),\ j=1,\ldots,J\}$ as well as random samples from the conditional distribution of $\tilde\theta$ given $\theta_\tau$. \cite{liu1998} present an SMC sampler for such a scenario, which is described below.

%Since the marginal distribution of $\theta_\tau$ does not change in this scenario from time $\tau$ to $\tau+1$, we may keep the existing sample and append new particles corresponding to $\tilde\theta$ generated through a Markov transition kernel ${K}_{\tau+1}(\tilde\theta\mid\theta_{\tau})$. An efficient choice for the transition kernel is one that targets the conditional distribution of $\tilde\theta$ given $\theta_\tau$. Then, the resulting augmented particles are $\theta_{\tau+1}^{(j)}=(\theta_{\tau}^{(j)},\tilde{\theta}^{(j)}),\ j=1,\ldots,J$. The unnormalized weights of $\theta_{\tau+1}^{(j)},\ j=1,\ldots,J$ are updated using
Since the marginal distribution of $\theta_\tau$ does not change from time $\tau$ to $\tau+1$, we update the existing samples to $\theta_{\tau+1}^{(j)}=(\theta_{\tau}^{(j)},\tilde{\theta}^{(j)}),\ j=1,\ldots,J$ by appending $\tilde\theta^{(j)}$ generated via $\tilde{K}_{\tau+1}(\tilde{\theta}^{(j)}\mid\theta_\tau^{(j)})$. This defines a Markov transition kernel ${K}_{\tau+1}(\theta_{\tau+1}^{(j)}\mid\theta_{\tau}^{(j)}):=\eta_{\tau}(\theta_{\tau}^{(j)})\tilde{K}_{\tau+1}(\tilde{\theta}^{(j)}\mid\theta_\tau^{(j)})$ on  
$E_{\tau +1} \times \mathcal{E}_{\tau +1}$. An efficient choice for $\tilde{K}_{\tau+1}$ is one that targets the conditional density of $\tilde\theta$ given $\theta_\tau$. The unnormalized weights of $\theta_{\tau+1}^{(j)},\ j=1,\ldots,J$ are updated using
    \begin{equation}
    \label{equ:liuweight}
        w_{\tau+1}^{(j)} \propto\frac{\eta_{\tau+1}\left(\theta_{\tau+1}^{(j)}\right)}{{g}_{\tau+1}\left(\theta_{\tau+1}^{(j)}\right)}=w_{\tau}^{(j)}\frac{\eta_{\tau+1}\left(\theta_{\tau+1}^{(j)}\right)}{\eta_{\tau}\left(\theta_\tau^{(j)}\right)\tilde{K}_{\tau+1}\left(\tilde{\theta}^{(j)}\mid\theta_\tau^{(j)}\right)},    
    \end{equation}
where ${g}_{\tau+1}$ denotes the importance density for sampling $\theta_{\tau+1}^{(j)}$; once computed, the weights can be normalized. This framework is widely used for dynamical systems with state space models increasing in dimension \citep{liu1998}.

If we recursively update weighted samples targeting a long sequence of distributions with increasing state space dimension, at some point the weighted samples may poorly estimate the target distribution due to the curse of dimensionality.
For example, when using $J$ weighted samples to approximate the target distribution at time $\tau$, fewer particles retain large weights as the dimension increases with $\tau$ for subsequent target distributions.
This is known as the degeneracy problem.
To measure potential degeneracy, we compute the effective sample size (ESS) using the magnitude of the normalized weights, $ESS=\left(\sum_{j=1}^J(w_\tau^{(j)})^2\right)^{-1}$. 
If ESS is small, a large portion of the weighted samples have small weights suggesting particle degeneracy. One solution is to remove particles with small weights and duplicate ones with large weights through resampling \citep{gilks2001,chopin2002}.
A common approach draws random samples from the multinomial distribution with weights serving as parameters; the resampled particles are then assigned equal weights \citep{gordon1993}.
The overall procedure is as follows: (1) generate samples $\theta_{\tau+1}$ via Markov transition kernel $K_{\tau+1}$, 
%draw samples of $\tilde{\theta}$ through a Markov transition kernel $K_{\tau+1}$,
 (2) update the weights using Equation \ref{equ:liuweight}, and (3) resample the weighted samples if ESS $<J/2$.

\subsection{SMC for state space with fixed dimension}
\label{sec:smc_delmoral}
Another scenario of interest is when the sequence of distributions $\eta_\tau$, $\tau\in\mathbb{N}_+$ are defined on the same space $E$, and thus have fixed dimension. This occurs, for example, when we wish to sample from a challenging target posterior distribution by building a sequence, or bridge, of intermediate distributions that are similar to one another, such as annealed versions of the posterior density.
Assume that we have a set of weighted samples from a distribution at time $\tau$ and we aim to perturb them toward the target distribution at time $\tau+1$.
We assume that the dimensions of the state variables at times $\tau$ and $\tau+1$ are the same, i.e., $\text{dim}(\theta_\tau)=\text{dim}(\theta_{\tau+1})$, and that the adjacent distributions are similar to each other, i.e., $\eta_\tau(\theta_\tau)\approx\eta_{\tau+1}(\theta_{\tau+1})$.
Given a set of weighted samples at time $\tau$, $\{(\theta_\tau^{(j)},w_\tau^{(j)}),\ j=1,\ldots,J\}$, we perturb them toward the target distribution $\eta_{\tau+1}$ via a transition kernel $K_{\tau+1}(\theta_{\tau+1}^{(j)}\mid\theta_\tau^{(j)})$ defined on $E\times \mathcal{E}$.
A natural choice for $K_{\tau+1}$ is an MCMC kernel. The weight update is given by
\begin{equation}
w_{\tau+1}^{(j)}\propto\frac{\eta_{\tau+1}\left(\theta_{\tau+1}^{(j)}\right)}{g_{\tau+1}\left(\theta_{\tau+1}^{(j)}\right)}=\frac{\eta_{\tau+1}\left(\theta_{\tau+1}^{(j)}\right)}{\int g_{\tau}\left(\theta_{\tau}\right)K_{\tau+1}\left(\theta_{\tau+1}^{(j)}\mid\theta_\tau\right)d\theta_\tau},\ \ \  j=1,\dots,J.
\end{equation}
This requires updating the importance density which is, however, often intractable as it involves integration that does not have a closed form solution. 
To circumvent this issue, \cite{delmoral2006} calculate the weights through the use of a backward kernel
%First, existing particles are perturbed toward the target distribution $\eta_{\tau+1}$ through the forward transition kernel $K_{\tau+1}\left(\theta_{\tau+1}^{(j)}\mid\theta_\tau^{(j)}\right)$.
%The updated weight for each particle may then be computed through the use of a backward kernel 
$L_\tau(\theta_\tau^{(j)}\mid\theta_{\tau+1}^{(j)})$, defined on $E \times \mathcal{E}$, as
    \begin{equation}
    \label{eq:intractable}
        w_{\tau+1}^{(j)} \propto w_\tau^{(j)}\frac{\eta_{\tau+1}\left(\theta_{\tau+1}^{(j)}\right)L_\tau\left(\theta_\tau^{(j)}\mid\theta_{\tau+1}^{(j)}\right)}{\eta_{\tau}\left(\theta_\tau^{(j)}\right)K_{\tau+1}\left(\theta_{\tau+1}^{(j)}\mid\theta_\tau^{(j)}\right)},\ \ \ j=1,\dots,J,
    \end{equation}
if the transition kernel has a closed form. If $K_{\tau+1}$ is an MCMC kernel that is invariant to $\eta_{\tau+1}$, and the adjacent target distributions are similar, \cite{delmoral2006} proposed using the following approximate backward kernel that avoids explicit evaluation of $K_{\tau+1}$ and $L_\tau$ in the weight update: $L_\tau(\theta_{\tau}^{(j)}\mid\theta_{\tau+1}^{(j)}) = \frac{\eta_{\tau+1}(\theta_{\tau}^{(j)})K_{\tau+1}(\theta_{\tau+1}^{(j)}\mid\theta_{\tau}^{(j)})}{\eta_{\tau+1}(\theta_{\tau+1}^{(j)})},\ j=1,\dots,J$. For this choice of backward kernel, the weight update in Equation \ref{eq:intractable} simplifies to
    \begin{equation}
    \label{equ:delmoralweight}
        w_{\tau+1}^{(j)} \propto w_{\tau}^{(j)}\frac{\eta_{\tau+1}\left(\theta_\tau^{(j)}\right)}{\eta_{\tau}\left(\theta_\tau^{(j)}\right)},\ \ \ j=1,\dots,J,
    \end{equation}
%which is easily computed. 
%This approximate weight update allows us to use transition kernels whose densities are intractable, such as the Metropolis-Hastings kernel. % requires integration which is not computable in closed form.
and the weights are then normalized.

\section{SMC algorithm for registration of functional data}
\label{sec:algo}

In this section, we present a sequential Bayesian registration approach for the model introduced in Section \ref{sec:model}, noting that it can be straightforwardly extended to a wider variety of models for functional data, for which the state space similarly increases in dimension as new data arrives; a detailed  algorithm is provided in Appendix A in the supplement. In the following, we will use the subscript $1:n$ to denote a set of objects indexed from $1$ through $n$, so that, for example, $f_{1:n}=\{f_1,\ldots,f_n\}$ represents the first $n$ observed functions and $\textbf{d}_{1:n}=\{\textbf{d}_1,\ldots,\textbf{d}_n\}$ represents the increments of the phase components for the first $n$ functions. 
We introduce a sequential Monte Carlo algorithm that updates the posterior distribution over the template function, all of the phase components, and the error variance when the state space of the phase components increases as new functions arrive.
Let $\ldots, \pi(\textbf{c},\textbf{d}_{1:n},\sigma^2\mid f_{1:n}),\ \pi(\textbf{c},\textbf{d}_{1:n+1},\sigma^2\mid f_{1:n+1}),\ \pi(\textbf{c},\textbf{d}_{1:n+2},\sigma^2\mid f_{1:n+2}),\ldots$ denote the sequence of target posterior densities over all of the parameters given an increasing number of functions, and suppose that we have a large number $J$ of weighted samples $\{(\textbf{c}^{(j)},\textbf{d}^{(j)}_{1:n},\sigma^{2(j)},w^{(j)}),\ j=1,\ldots,J\}$ approximating the posterior $\pi(\textbf{c},\textbf{d}_{1:n},\sigma^2\mid f_{1:n})$, where $\textbf{c}$ is the vector of basis coefficients defining the template function and $\sigma^2$ is the observation error variance. 
The goal is to update these particles and weights to generate a new weighted sample $\{(\textbf{c}^{(j)},\textbf{d}^{(j)}_{1:n+1},\sigma^{2(j)},w^{(j)}),\ j=1,\ldots,J\}$, approximating the posterior  $\pi(\textbf{c},\textbf{d}_{1:n+1}, \sigma^2\mid f_{1:n+1})$, as a new function $f_{n+1}$ becomes part of the data. 

\subsection{Stochastic initialization of new phase components}
Since additional parameters are required in the model when a new function, $f_{n+1}$, is observed, we consider the increasing state space dimension setting described in Section \ref{sec:smc_liu}. The first step (lines 4-10 in Algorithm 1) augments the previous particles with phase component $\textbf{d}_{n+1}^{(j)}$ for the new function $f_{n+1}$, and updates their weights. To enhance the efficiency of the sampling algorithm, we initialize this new component in a region of high posterior probability. The initialization kernel leaves the existing particle components unchanged and draws the phase increment $\textbf{d}_{n+1}^{(j)}$ from a distribution centered around the increment $d\left(\hat{\gamma}_{n+1}^{(j)}\right)$ obtained by optimally aligning $f_{n+1}$ to the template $q_\mu^{(j)}=\sum_{b=1}^Bc_b^{(j)}\phi_b$ as follows.
\begin{enumerate}
\item Compute the phase function $\hat{\gamma}_{n+1}^{(j)}$ that provides the optimal alignment to the template $q_\mu^{(j)}=\sum_{b=1}^Bc_b^{(j)}\phi_b$ by solving the optimization problem $\hat\gamma_{n+1}^{(j)}=\arg\min_{\gamma\in\Gamma}$ $\|q_\mu^{(j)}-(q_{n+1}\circ \gamma)\sqrt{\dot{\gamma}}\|^2$ via the Dynamic Programming algorithm \citep{bertsekas2000}. Each resulting $\hat\gamma_{n+1}^{(j)}$ is a piecewise linear function that is finely sampled on the domain $[0,1]$. For the next step, we further approximate $\hat\gamma_{n+1}^{(j)}$ over the prespecified partition $0=s_1<s_2<\ldots<s_{M_\gamma-1}<s_{M_\gamma}=1$ using a least squares procedure.
\item Draw a random sample $\textbf{d}^p$ from $Dir((\kappa_{ini}/(M_\gamma-1))1_{M_\gamma-1})$ and
initialize the new phase component by composition via $\gamma^{(j)*} = \hat{\gamma}_{n+1}^{(j)} \circ d^{-1}(\textbf{d}^p)$. \end{enumerate}
The sampling density \citep{zang2021} of these new phase increments $\textbf{d}_{n+1}^{(j)} = d(\gamma^{(j)*})$ is
\begin{align}
\tilde{K}(\textbf{d}^{(j)}_{n+1}\mid\textbf{c}^{(j)},f_{n+1}) = \left[\prod_{m=2}^{M_\gamma}\left(\left(\hat{\gamma}_{n+1}^{(j),-1}\right)'\circ \gamma^{(j)*}\right)(s_m)\right]Dir(\textbf{d}^p;\frac{\kappa_{ini}}{M_\gamma-1}1_{M_\gamma-1}), \label{transition_kernel}
\end{align}
where $(\hat{\gamma}_{n+1}^{(j),-1})'$ is the derivative of $\hat{\gamma}_{n+1}^{(j),-1}$, and the concentration parameter $\kappa_{ini}$ is user-selected. We provide a sensitivity analysis to the choice of $\kappa_{ini}$ in Appendix E.4 in the supplement. 
Next, we update the weight of each particle while accounting for the increased dimension of the parameter space, following the approach of \cite{liu1998}. Denote the likelihood by $h$, the prior density by $p$, the importance sampling density by $g$, and the initialization kernel density by $\tilde{K}$. The importance sampling density after augmentation is related to the previous importance sampling density via $g(\textbf{c}^{(j)},\textbf{d}_{1:n+1}^{(j)},\sigma^{2(j)}\mid f_{1:n+1}) = g(\textbf{c}^{(j)},\textbf{d}_{1:n}^{(j)},\sigma^{2(j)}\mid f_{1:n})\tilde{K}(\textbf{d}_{n+1}^{(j)}\mid\textbf{c}^{(j)},f_{n+1})$. The new weights are then given by
\begin{small}
\begin{align}
 \label{equ:weight}
    \tilde{w}^{(j)} &\propto \frac{\pi\left(\textbf{c}^{(j)},\textbf{d}_{1:n+1}^{(j)},\sigma^{2(j)}\mid f_{1:n+1}\right)}{g\left(\textbf{c}^{(j)},\textbf{d}_{1:n+1}^{(j)},\sigma^{2(j)}\mid f_{1:n+1}\right)}
    \propto \frac{h\left(f_{1:n+1}\mid\textbf{c}^{(j)},\textbf{d}_{1:n+1}^{(j)},\sigma^{2(j)}\right)p\left(\textbf{c}^{(j)},\textbf{d}_{1:n+1}^{(j)},\sigma^{2(j)}\right)}{g\left(\textbf{c}^{(j)},\textbf{d}_{1:n}^{(j)},\sigma^{2(j)}\mid f_{1:n}\right)\tilde{K}\left(\textbf{d}_{n+1}^{(j)}\mid\textbf{c}^{(j)},f_{n+1}\right)}\nonumber\\
    &\propto w^{(j)}\frac{h\left(f_{n+1}\mid\textbf{c}^{(j)},\textbf{d}_{n+1}^{(j)},\sigma^{2(j)}\right)p\left(\textbf{d}_{n+1}^{(j)}\right)}{\tilde{K}\left(\textbf{d}^{(j)}_{n+1}\mid\textbf{c}^{(j)},f_{n+1}\right)},
\end{align}
\end{small}for each $j=1,\dots,J$, where $w^{(j)}$ denotes the weight before augmentation. Thus, the resulting weighted sample, composed of the augmented particles and updated weights, now targets the posterior density given $f_{1:n+1}$.

\subsection{Particle perturbation and centering}

While the weighted samples $\{(\textbf{c}^{(j)},\textbf{d}^{(j)}_{1:n+1},\sigma^{2(j)},\tilde{w}^{(j)}),\ j=1,\ldots,J\}$ already approximate the target posterior $\pi(\textbf{c},\textbf{d}_{1:n+1},\sigma^{2}\mid f_{1:n+1})$, the presence of particles with small weights as well as low particle diversity, can result in SMC estimators with large variance. Thus, we  resample the particles if ESS falls below $J/2$ using a multinomial distribution with parameters $\tilde{w}^{(j)},\ j=1,\ldots,J$, assigning equal weights to the resampled particles. Furthermore, since resampling can result in multiple copies of particles, we use MCMC perturbations to further diversify the particle locations. Numerical experiments suggest that performing this step guards against particle degeneracy in subsequent updates. Since perturbation occurs on a state space of fixed dimension, we use the tools described in Section \ref{sec:smc_delmoral}. We begin by perturbing particle components $\textbf{c}^{(j)}$ and $\textbf{d}_{1:n+1}^{(j)}$, given $\sigma^{2(j)}$, $j=1,\ldots,J$ using a Metropolis-Hastings (MH) transition kernel (lines 17-35 in Algorithm 1). We adopt the approximate backward kernel approach of \cite{delmoral2006} to avoid computing the density of the MH kernel in the weight update calculation. Because the additional perturbation still targets the desired posterior $\pi(\textbf{c},\textbf{d}_{1:n+1},\sigma^{2}\mid f_{1:n+1})$, i.e.  $\eta_{\tau+1} = \eta_{\tau}$ in Equation \ref{equ:delmoralweight}, the weight update simplifies to $\tilde{\tilde{w}}^{(j)}=\tilde{w}^{(j)}$. 
This argument also enables the application of multiple subsequent MH perturbation steps to further diversify the sample, if desired. Our implementation performs 30 MH perturbation steps. The proposal distribution for the template coefficients (line 18 in Algorithm 1) is multivariate normal $N(\textbf{c}^{(j)},\widehat{\Sigma}_c)$ centered at the $j$th particle, $\textbf{c}^{(j)}$, with empirical covariance $\widehat{\Sigma}_c=\frac{1}{J-1}\sum_{j=1}^J\tilde{w}^{(j)}\textbf{c}^{(j)}\textbf{c}^{(j)\top}$. Phase increment proposals are generated by composing $\gamma^{(j)}$ with the linear interpolation of a $Dir(\frac{\kappa}{M_\gamma-1}1_{M_\gamma-1})$ realization, where $\kappa$ is fixed (lines 26-29 in Algorithm 1). An analogous update involving $\sigma^{2(j)}$ is performed using a Gibbs sampling kernel following the centering step described below.

After each full update, we perform a centering step to ensure identifiability of the template function (line 36 in Algorithm 1; see Section \ref{subsec:regist} for a brief discussion). This step constrains the mean of the phase components of the data with respect to the template function to be the identity $\gamma_{id}$. To do this, we first compute the sample average of the phase components $\gamma_{1:n+1}^{(j)}=d^{-1}(\textbf{d}_{1:n+1}^{(j)})$ \citep{srivastava2011b} and then apply its inverse to each template function $q_\mu^{(j)}=\sum_{b=1}^Bc_b^{(j)}\phi_b$ and each  phase $\gamma_{1:n+1}^{(j)}$. 
We then update the weights accordingly. 
As this step utilizes a deterministic kernel, and the importance density and likelihood are invariant to simultaneous warping, the weight update is based on the ratio of prior densities evaluated at the centered particles and their uncentered counterparts. 
To do this, we first use the centered template function and phase components to construct the centered particles $\{(\tilde{\textbf{c}}^{(j)},\tilde{\textbf{d}}_{n+1}^{(j)}),\ j=1,\ldots,J\}$. We then update the weights (line 37 in Algorithm 1) using
\begin{equation}
    \label{equ:weight_center}
    \tilde{w}^{(j)} \gets \tilde{w}^{(j)}\frac{p\left(\tilde{\textbf{c}}^{(j)},\tilde{\textbf{d}}_{1:n+1}^{(j)}\right)}{p\left(\textbf{c}^{(j)},\textbf{d}_{1:n+1}^{(j)}\right)},\ \ \ j=1,\dots,J,
\end{equation}
which are normalized subsequently.
Assuming that the initial samples are centered (this is the standard approach in MCMC-based batch learning for registration of functional data), the sample average of the phase components does not deviate very much from the identity after each new function is observed. This fact, coupled with our specification of diffuse priors, ensures that the weights do not change much due to this centering step. Thus, to improve computational efficiency, one may choose to not update the weights of the particles after the centering step.

After updating the template and phase components of the weighted particles, the components corresponding to the error variance $\sigma^{2}$ are perturbed via a Gibbs update (lines 38-39 in Algorithm 1). This is done by sampling $\tilde{\sigma}^{2(j)}$, for each $j=1,\dots,J$,  from the full-conditional distribution,
$\pi(\sigma^{2}\mid\tilde{\textbf{c}}^{(j)},\tilde{\textbf{d}}^{(j)}_{1:n+1},f_{1:n+1})$, which is inverse-gamma with parameters $\alpha_\sigma+\frac{(n+1)M}{2}$ and $\beta_\sigma+\sum_{i=1}^{n+1}\sum_{m=1}^{M}\left(q_i(t_m)-(\tilde{q}_\mu^{(j)},\tilde{\gamma}^{-1(j)}_i)(t_m)\right)^2$,
%\begin{small}
%\begin{equation*}
%    \sigma^2 \mid \tilde{\textbf{c}}^{(j)},\tilde{\textbf{d}}^{(j)}_{1:n+1},f_{1:n+1}\sim IG\left(\alpha_\sigma+\frac{(n+1)M}{2},\beta_\sigma+\sum_{i=1}^{n+1}\sum_{m=1}^{M}\left(q_i(t_m)-(\tilde{q}_\mu^{(j)},\tilde{\gamma}^{-1(j)}_i)(t_m)\right)^2\right),
%\end{equation*}
%\end{small}
where $\tilde{q}^{(j)}_\mu=\sum_{b=1}^B\tilde c^{(j)}_b\phi_b$ and $\tilde{\gamma}^{(j)}_i = d^{-1}(\tilde{\textbf{d}}^{(j)}_i),\ i=1,\ldots,n+1$.
Again adopting the approximate backward kernel approach of \cite{delmoral2006}, the weights remain approximately unchanged after the Gibbs update. 

Following the above resampling, perturbation, and centering steps with associated weight updates, we obtain a collection of $J$ weighted samples $\{(\tilde{\textbf{c}}^{(j)},\tilde{\textbf{d}}^{(j)}_{1:n+1},\tilde\sigma^{2(j)},\tilde{w}^{(j)}),\ j=1,\ldots,J\}$ approximating the target posterior distribution $\pi(\textbf{c},\textbf{d}_{1:n+1},\sigma^{2}\mid f_{1:n+1})$. 
%The algorithm for the full sampling procedure described in this section is provided in Appendix A of the supplement.

\section{Simulations and real data examples}
\label{sec:numerical_applications}
We begin in Section \ref{sec:viz} by discussing visualization and summarization of posterior uncertainty resulting from our analyses. The following sections describe in detail the application of the proposed sequential Bayesian elastic registration framework to two simulated examples and three real data studies. For these analyses we assume that the template function is a weighted sum of cubic B-splines with equally-spaced knots, as they are sufficiently flexible to capture the expected variation in the template.
% without over-fitting. 
Our model also accommodates the use of other basis sets. The number of bases is chosen to reflect the complexity of amplitude features in the data without over-fitting. For the simulated examples, we use the same number of basis functions to model the template as were used to simulate the data. In the following examples, we set $\kappa_{ini} = 100$ for the initialization kernel of each new phase component in the SMC algorithm. Our implementation employs parallel computing with 12 workers.

\subsection{Summarization and visualization of target posterior distribution}
\label{sec:viz}

An important consideration is how to visualize marginal posteriors over template and phase in their respective function spaces. To extract posterior features we first map the particle components $\tilde{\textbf{c}}^{(j)}$ and $\tilde{\textbf{d}}^{(j)}_{i}$ to the template SRVF $\tilde{q}_\mu^{(j)} = \sum_{b=1}^B \tilde{c}_b^{(j)}\phi_b \in \mathcal{Q}$ and the phase $\tilde{\gamma}_i^{(j)} = d^{-1}(\tilde{\textbf{d}}^{(j)}_{i}),\ i = 1,\ldots,n,$ $j=1,\ldots,J$, respectively.
The posterior mean of $q_\mu$ is estimated using the weighted sample average $\sum_{j=1}^{J} \tilde{w}^{(j)}\tilde{q}_\mu^{(j)} =: \bar q_\mu$, while its variance is estimated using the weighted sample average of the squared eFR distance of each particle from the posterior mean, i.e., $\sum_{j=1}^J \tilde{w}^{(j)} \int_0^1(\tilde{q}_\mu^{(j)}(t)-\bar{q}_\mu(t))^2 dt$. Pointwise estimates may be used to visualize summaries of uncertainty at specific times, and posterior summaries for phase are computed similarly.

%The marginal posteriors over $q_\mu$ and $\gamma_{1:n}$ may be visualized using spaghetti plots of the corresponding particle components with partial transparency proportional to each particle's weight.
We use smooth line plots of multiple superimposed trajectories across time, known as \emph{spaghetti plots}, to visualize marginal posteriors over $q_\mu$ and $\gamma_{1:n}$ by plotting each SMC particle across time with partial transparency proportional to the weights of the individual particles. This enables visualization of posterior features such as dispersion and clustering, which are difficult to summarize in the functional setting.
Visualizing the uncertainty in the \emph{shape} of the template function can be more intuitive in the original function space $\mathcal{F}$ rather than the SRVF space $\mathcal{Q}$. However, since SRVFs do not contain translation information, the spaghetti plot representation in the original space can appear misleading because starting points must be chosen arbitrarily. To circumvent this issue, we propose to visualize the leading sample principal components computed using the template particles to illustrate uncertainty in template shape. We first compute the $M\times M$ pointwise covariance of the marginal posterior of the template function in the SRVF space, $\Sigma_q = \sum_{j=1}^J \tilde{w}^{(j)} (\tilde{q}_\mu^{(j)}([t])-\bar{q}_\mu([t]))(\tilde{q}_\mu^{(j)}([t])-\bar{q}_\mu([t]))^\top$ where $\bar{q}_\mu$ is the marginal posterior mean, and perform singular value decomposition, $\Sigma_q = U\Delta V^\top$, where the diagonal entries of $\Delta$ are the singular values of $\Sigma_q$ and the columns of $U$ are orthonormal bases $u_1,\ldots,u_M$. We may then obtain positive/negative directions along the $k$th principal component (in units of standard deviation) from the marginal posterior mean by $\bar q_\mu +/- \alpha\sqrt{\Delta_{k}} u_k$, where $\alpha$ is a positive number and $\Delta_k$ is the $k$th diagonal entry of $\Delta$. We can visualize the uncertainty in template shape by mapping the functions represented by these vectors to the original function space $\mathcal{F}$ via the inverse mapping $Q^{-1}$, where the template function value at $t = 0$ is taken as the sample average $\frac{1}{n}\sum_{i=1}^n f_i(0)$. This visualization approach is used in Section \ref{sec:ecg}.

\subsection{Simulated example 1}

\begin{figure}[!t]
\graphicspath{{figures/sim1}}
  \centering
  \begin{tabularx}{13.3cm} { 
  |>{\centering\arraybackslash}X 
   >{\centering\arraybackslash}X 
   >{\centering\arraybackslash}X 
   >{\centering\arraybackslash}X| }
    \hline
    \multicolumn{4}{|c|}{(a)}
    \\
    \begin{minipage}{2.5cm}
    \centering\includegraphics[width=2.5cm]{forig1.pdf}
    \end{minipage}
    &\begin{minipage}{2.5cm}
    \centering\includegraphics[width=2.5cm]{forig2.pdf}
    \end{minipage}
    &\begin{minipage}{2.5cm}
    \centering\includegraphics[width=2.5cm]{forig3.pdf}
    \end{minipage}
    &\begin{minipage}{2.5cm}
    \centering\includegraphics[width=2.5cm]{forig4.pdf}
    \end{minipage}
    \\\hline
    \multicolumn{4}{|c|}{(b)}
    \\
    \begin{minipage}{2.5cm}
    \centering\includegraphics[width=2.5cm]{d1.pdf}
    \end{minipage}
    &\begin{minipage}{2.5cm}
    \centering\includegraphics[width=2.5cm]{d2.pdf}
    \end{minipage}
    &\begin{minipage}{2.5cm}
    \centering\includegraphics[width=2.5cm]{d3.pdf}
    \end{minipage}
    &\begin{minipage}{2.5cm}
    \centering\includegraphics[width=2.5cm]{d4.pdf}
    \end{minipage}
    \\\hline
    %\multicolumn{3}{|c|}{(c) }
    %\\\hline
    %\multicolumn{1}{|c}{}
    (c)&\multicolumn{1}{|c}{(d)}&\multicolumn{1}{|c}{(e)}&\multicolumn{1}{|c|}{(f)}\\
    \begin{minipage}{2.5cm}
    \centering\includegraphics[width=2.5cm]{qmean.pdf}
    \end{minipage}
    &
    \multicolumn{1}{|c}{\begin{minipage}{2.5cm}
    \centering\includegraphics[width=2.5cm]{fmean.pdf}
    \end{minipage}}
    &\multicolumn{1}{|c}{\begin{minipage}{2.5cm}
    \centering\includegraphics[width=2.5cm]{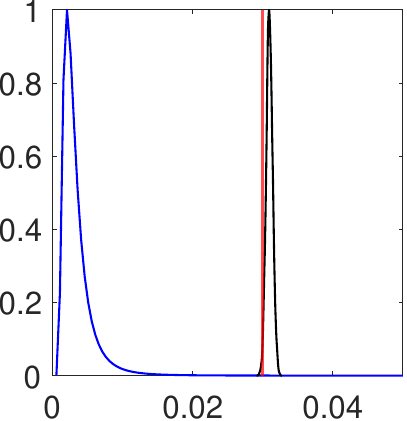}
    \end{minipage}}
    &\multicolumn{1}{|c|}{\begin{minipage}{2.5cm}
    \centering\includegraphics[width=2.5cm]{ESS.pdf}
    \end{minipage}}
    \\\hline
  \end{tabularx}
  \caption{Illustration of the proposed approach based on simulated data. Results in the second and third rows were generated via recursive updates starting with $f_{1:30}$ and ending with $f_{1:100}$. (a) Subset of simulated functions $f_{1:100}$. (b) Marginal posterior means of phase components for the functions in (a) are shown in the same column in black with ground truth in red. (c) Summary of the marginal posterior for the template SRVF using spaghetti plots with transparency (in black) and the ground truth template function (in red). (d) The marginal posterior mean of the template function (in black) and the ground truth template function (in red) in the original function space $\mathcal{F}$.
  (e) Kernel density estimate of the marginal posterior for the error variance $\sigma^2$ in black. The prior distribution of $\sigma^2$ and the ground truth are shown in blue and red, respectively. (f) ESS of weighted samples for a sequence of posterior distributions, $\pi(\cdot\mid f_{1:n}),\ n=31,\ldots,100$, with $n$ on $x$-axis and ESS on $y$-axis.}
    \label{fig:sim1_gt}
\end{figure}

We simulate $100$ functions by deforming a ground truth template $q_\mu$ using randomly sampled phases $\gamma_i,\ i=1,\dots,100$. The template is generated using a linear combination of eight B-spline basis functions and the phases as piecewise linear functions with Dirichlet increments on a uniform partition of size $M_\gamma=5$ with concentration parameter $\kappa=50$. The functional data is then generated via $f_i=Q^{-1}(f_i(0) = 0, q_i)$, with $q_i([t])=(q_\mu,\gamma_i^{-1})([t])+\epsilon_i([t])$ and $\epsilon_i([t])\overset{iid}{\sim} N(0_M,0.03 I_M)$, where $0_M$ is a vector of $M = 100$ zeros. A subset of the $100$ simulated functions is shown in Figure \ref{fig:sim1_gt}(a) with the corresponding subset of ground truth phase components in red in Figure \ref{fig:sim1_gt}(b). We fit the model described in Section \ref{sec:model} with prior hyperparameters $\Sigma_c=20I_{8}$, $\kappa=5$, $\alpha_\sigma=4$, $\beta_\sigma=0.01$, and a uniform partition of size $M_\gamma=5$.

We first obtain an MCMC sample of size $10,000$ from the posterior distribution of the parameters given the first 30 functions, $f_{1:30}$, and use these as equally weighted particle inputs to the subsequent SMC update. The weighted samples are updated recursively using the proposed SMC algorithm adding one function at a time. The resulting summaries of the posterior distribution given the full data $f_{1:100}$ are shown in Figure \ref{fig:sim1_gt}. Panel (b) shows the estimated posterior mean phases in black, noting that they are very similar to the ground truth. The marginal posterior variances obtained using the weighted sample average of the squared $\mathbb{L}^2$ distances of each particle from the marginal posterior mean are very small $(<10^{-4})$. Panel (c) illustrates the posterior uncertainty in the template function in the SRVF space $\mathcal{Q}$ using spaghetti plots (in black) with transparency and the ground truth template function in red. 
The magnitude of uncertainty in the template is small across the domain. The posterior mean template in the original function space $\mathcal{F}$ is shown in panel (d) in black and the ground truth template in red. The eFR distance between the posterior mean SRVF template and the ground truth  is only 0.1080. Panel (e) shows that the true error variance lies within the $99\%$ central credible region of the marginal posterior for $\sigma^2$. Panel (f) shows updated ESS values after resampling the particles when the ESS of the previous particles fell below the threshold of $J/2$ (line 41 of Algorithm 1 in Appendix A in the supplement). The figure suggests that ESS remains relatively large as we add more functions to the data. After augmenting the previous particles with a new phase component and updating the weights using Equation \ref{equ:weight}, the ESS sometimes falls below the threshold, but is still moderately large ($>J/3$).

To compare estimation accuracy, we fit the model to the full data $f_{1:100}$ using both the sequential and MCMC-based batch learning methods. Our implementation of MCMC uses Gibbs and adaptive Metropolis-Hastings updates for a total of $50,000$ iterations, with a burn-in period of $40,000$. Table \ref{tab:sim1_diff} reports the estimated mean squared errors for two posterior summaries of the template basis coefficients and the phase increments, obtained based on 100 runs of the batch and sequential learning algorithms. 
The proposed sequential approach outperforms MCMC-based batch learning in all scenarios. While not presented in this table, similar results were obtained when comparing estimates from intermediate posterior distributions in the sequence to batch learning posterior estimates given the same set of functional data. Furthermore, we see no evidence of degraded accuracy as the number of assimilated functions grows (see Appendix D in the supplement). 
% Figure \ref{fig:sim1_accuracy} illustrates the evolution of the estimated mean squared error given increasing sets of functional data $f_{1:31},\ldots,f_{1:100}$. There appears to be a slight decreasing trend in the errors as the number of observed functions increases.

We further compare the computational efficiency of the two methods. For the proposed sequential approach, we start with posterior samples given $f_{1:30}$, recursively add one function to the data at a time, and update the weighted samples using the proposed algorithm. We do this for the full sequence of posterior distributions until all of the data, $f_{1:100}$, is used. For the MCMC-based batch learning approach, we draw samples from each posterior distribution in the sequence by re-running the full algorithm each time. As before, we use a total of $50,000$ MCMC iterations, with a burn-in period of $40,000$. 
The computation time (in seconds) needed to obtain a Monte Carlo sample of size $10,000$ from each posterior distribution in the sequence given $f_{1:n},\ n=31,\ldots,100$ is decreased by $>80\%$ on average. The computation time to obtain a posterior sample via MCMC-based batch learning given $f_{1:30}, \ldots, f_{1:100}$ is 322,906 seconds. This is substantially longer than the 58,568 seconds for the sequential approach, which includes (i) MCMC-based batch learning to first obtain equally weighted particles from the posterior distribution given $f_{1:30}$, and (ii) SMC-based updates of the posterior given $f_{1:31}, \ldots, f_{1:100}$. This significant gain in computational efficiency is expected due to the sequential nature of the proposed method. A comparison of ESS based on the two approaches is presented in Appendix D in the supplement.

\subsection{Simulated example 2}

\begin{figure}[!t]
\graphicspath{{figures/sim2}}
  \centering
  \begin{tabularx}{13.3cm} { 
  |>{\centering\arraybackslash}X 
   >{\centering\arraybackslash}X 
   >{\centering\arraybackslash}X|
   >{\centering\arraybackslash}X| }
    \hline
    \multicolumn{3}{|c|}{(a1)} & (a2)
    \\
    \begin{minipage}{2.5cm}
    \centering\includegraphics[width=2.5cm]{forig1.pdf}
    \end{minipage}
    &\begin{minipage}{2.5cm}
    \centering\includegraphics[width=2.5cm]{forig2.pdf}
    \end{minipage}
    &\begin{minipage}{2.5cm}
    \centering\includegraphics[width=2.5cm]{forig3.pdf}
    \end{minipage}
    &\begin{minipage}{2.5cm}
    \centering\includegraphics[width=2.5cm]{forig7.pdf}
    \end{minipage}
    \\\hline
    \multicolumn{3}{|c|}{(b1)} & (b2)
    \\
    \begin{minipage}{2.5cm}
    \centering\includegraphics[width=2.5cm]{d1.pdf}
    \end{minipage}
    &\begin{minipage}{2.5cm}
    \centering\includegraphics[width=2.5cm]{d2.pdf}
    \end{minipage}
    &\begin{minipage}{2.5cm}
    \centering\includegraphics[width=2.5cm]{d3.pdf}
    \end{minipage}
    &\begin{minipage}{2.5cm}
    \centering\includegraphics[width=2.5cm]{d7.pdf}
    \end{minipage}
    \\\hline
    \multicolumn{3}{|c|}{(c1)} & (c2)
    \\
    \begin{minipage}{2.5cm}
    \centering\includegraphics[width=2.5cm]{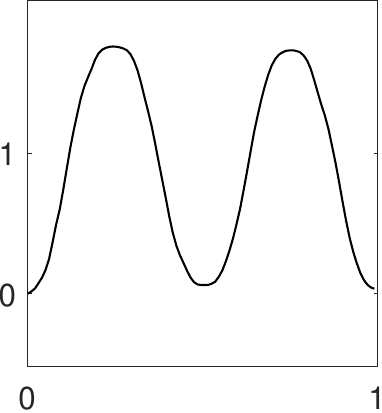}
    \end{minipage}
    &\begin{minipage}{2.5cm}
    \centering\includegraphics[width=2.5cm]{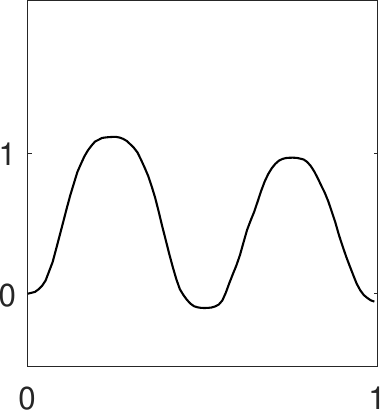}
    \end{minipage}
    &\begin{minipage}{2.5cm}
    \centering\includegraphics[width=2.5cm]{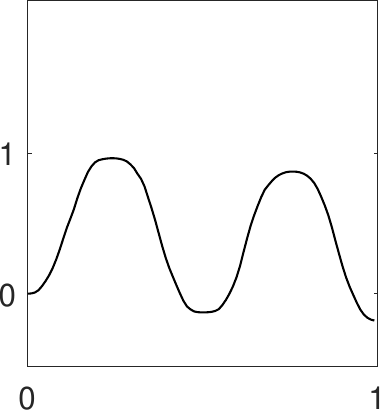}
    \end{minipage}
    &\begin{minipage}{2.5cm}
    \centering\includegraphics[width=2.5cm]{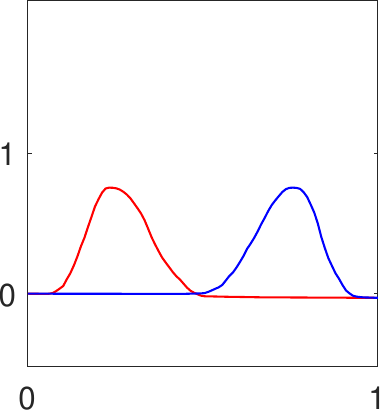}
    \end{minipage}
    \\\hline    %\multicolumn{3}{|c|}{(c) }
    %\\\hline
    %\multicolumn{1}{|c}{}
    \multicolumn{1}{|c|}{(d)}&\multicolumn{1}{c|}{(e)}&\multicolumn{1}{c|}{(f)}&\multicolumn{1}{c|}{(g)}\\
    \multicolumn{1}{|c|}{\begin{minipage}{2.8cm}
    \centering\includegraphics[width=2.8cm]{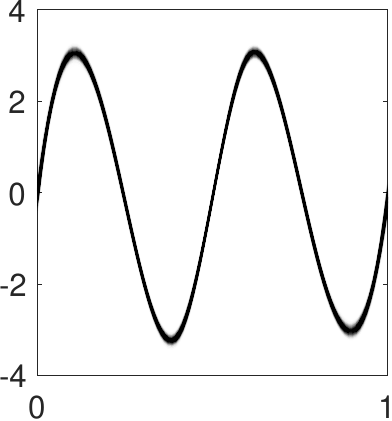}
    \end{minipage}}&\multicolumn{1}{c|}{\begin{minipage}{2.8cm}
    \centering\includegraphics[width=2.8cm]{fmean6.pdf}
    \end{minipage}}&\multicolumn{1}{c|}{\begin{minipage}{2.8cm}
    \centering\includegraphics[width=2.8cm]{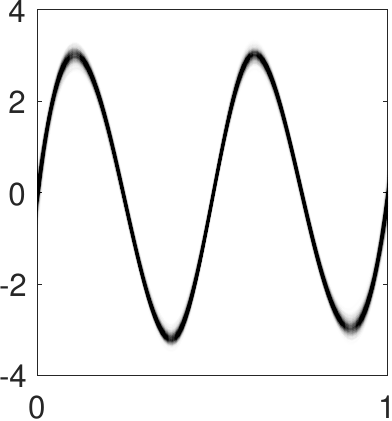}
    \end{minipage}}&\multicolumn{1}{c|}{\begin{minipage}{2.8cm}
    \centering\includegraphics[width=2.8cm]{fmean7.pdf}
    \end{minipage}}
    \\\hline
  \end{tabularx}
    \caption{Illustration of the proposed approach under a bimodal target posterior. (a1) Subset of simulated data $f_{1:6}$ and (a2) simulated data $f_7$ with one peak. (b1) \& (b2) Posterior mean estimates of the corresponding phases. (b2) Weighted mean for each of two modes in the marginal posterior. (c1) \& (c2) Registered functions $f_{1:6}$ and $f_7$, respectively, using posterior mean phases from (b1) \& (b2). (d) \& (f) Marginal posterior over the template SRVF given $f_{1:6}$ and $f_{1:7}$, respectively; transparency reflects magnitude of weights. (e) \& (g) Marginal posterior mean of the template in the original function space $\mathcal{F}$ given $f_{1:6}$ and $f_{1:7}$, respectively.}
    \label{fig:sim2}
\end{figure}

Next, we assess the performance of the proposed sequential approach for registration of functional data when the target posterior density is multimodal. In what follows we reserve the term ``multimodal'' to describe posterior densities rather than the functional observations, some of which will have two peaks. We simulate seven functions: the first six have two peaks and a valley and are simulated in the same fashion as Simulated Example 1, but with an additional random scaling (sampled from a uniform distribution on $[0.7,1.4]$) applied to each of $f_{1:6}$. The seventh function, $f_7$, is simulated such that it only has one peak. We expect the marginal posterior of the template function given $f_{1:7}$ to be unimodal with the mode representing a template that has two peaks, since most of the data has this form. However, since phase is relative with respect to the template, the marginal posterior samples of the phase component for function $f_7$ should cluster into two distinct groups: one that registers the single peak in function $f_7$ to the left peak of the template, and one that registers it to the right peak of the template. 
Thus, we expect the marginal posterior of phase for functions $f_{1:6}$ to be unimodal and for function $f_7$ to be bimodal. 
Figures \ref{fig:sim2}(a1) and (a2) show a subset of the simulated two-peak functions, $f_{1:6}$, and the single-peak function, $f_7$, respectively. All prior hyperparameters in the model are set to the same values as in Simulated Example 1, except for $\alpha_\sigma=40$.

\begin{table}[!t]
    \centering
    \renewcommand{\arraystretch}{1}
    \begin{tabular}{c c c c c}
    \toprule[1pt]\midrule[0.3pt]
         Method & \multicolumn{2}{c}{Template Function ($\textbf{c}$)} & \multicolumn{2}{c}{Phase Components ($\textbf{d}_{1:100}$)}\\
         \midrule
         & Posterior Mean & Posterior Mode & Posterior Mean & Posterior Mode\\\midrule
          SMC & \textbf{0.1712} & \textbf{0.1491} & \textbf{0.0120} & \textbf{0.0079}\\
          \midrule[0.3pt]
          MCMC & 0.2200 & 0.2457 & 0.0121 & 0.0108\\
          \midrule[0.3pt]\bottomrule[1pt]
    \end{tabular}
    \renewcommand{\arraystretch}{1}
    \caption{Estimated mean squared errors of the posterior mean and posterior mode for the template basis coefficients $\textbf{c}$ and phase increments $\mathbf{d}_{1:100}$ based on 100 replications of the batch learning (MCMC) and sequential (SMC) algorithm given $f_{1:100}$. The best performance is shown in bold.}
    \label{tab:sim1_diff}
\end{table}

%% MOVED TO APPENDIX
% \begin{figure}[!t]
%     \centering
%     \begin{tabular}{cc}
%     (a)&(b)\\
%     \includegraphics[width=0.3\linewidth]{figures/sim1/sim1_accuracy_c.pdf}&\includegraphics[width=0.3\linewidth]{figures/sim1/sim1_accuracy_d.pdf}\\
%     \end{tabular}
%     \caption{\textcolor{red}{Estimated mean squared errors for the posterior mode over (a) the template basis coefficients $\textbf{c}$ and (b) phase increments $\mathbf{d}_{1:N}$ given functions $f_{1:N}$, plotted against $N$.}}
%     \label{fig:sim1_accuracy}
% \end{figure}

We initialize the SMC algorithm using MCMC samples from the posterior distribution given $f_{1:3}$. We then assimilate the data $f_{4:7}$, one function at a time, and update the weighted samples sequentially to target the posteriors $\pi(\cdot\mid f_{1:4}),\ldots,\pi(\cdot\mid f_{1:7})$. 
The results of this simulation are presented in Figure \ref{fig:sim2}. As shown in panels (d-e) and (f-g), the template components of the particles indeed appear to have two peaks before and after the addition of $f_7$ to the data. It is evident that posterior uncertainty in (f) is greater than in (d), which is expected due to the very different shape of $f_7$. The marginal posteriors of phase for functions $f_{1:6}$, shown in panel (b1), suggest a unimodal marginal posterior; the marginal posteriors of phase have very small pointwise variances and thus we only visualize the marginal posterior means here. However, the particles of the phase component corresponding to function $f_7$ in panel (b2) clearly fall into two distinct groups as the peak of function $f_7$ can be registered to either of the two peaks in the marginal posterior of the template function, leading to a bimodal marginal posterior density. One mode corresponds to phase functions that fall above the identity with pointwise weighted sample average in red in panel (b2); this corresponds to the registration of the single peak in $f_7$ to the first peak in the template as seen in (c2). The second mode corresponds to phase functions that fall below the identity with pointwise weighted sample average in blue in (b2); this corresponds to the registration of the single peak in $f_7$ to the second peak in the template as seen in (c2). It is well-known that, in practice, MCMC can be quite inefficient in sampling from multimodal posterior distributions. On the other hand, the proposed sequential algorithm is able to sample from both modes in this posterior density relatively easily.

This simulation study also provides empirical evidence that the proposed method is effective at approximating the posterior distribution even when a new observation with a very different shape becomes part of the data. One concern in such a scenario for the proposed SMC algorithm is that the target posterior may change to an extent that would violate some of the assumptions required for the weight update calculations in Section \ref{sec:algo}. In particular, marginal posterior variance for the template function increases from 0.0038 to 0.0089 after the arrival of $f_7$, as seen in panels (d) and (f). At the same time, the distance between the marginal posterior means of the template for the two cases, which are shown in (e) and (g), is small $(< 0.1)$. This shows that adding a function with a different shape does not generally change the marginal posteriors very much, even given a relatively small number of functions prior to the sequential update (we start with only six two-peaked functions in the data). Even after more functions with one peak arrive sequentially, we expect the marginal posterior that is centered at a template function with two peaks to smoothly change toward a marginal posterior centered at a template with one peak due to additional MH-based particle perturbation steps employed in our algorithm. Finally, this example illustrates that initialization of the SMC particles can be based on MCMC samples given a relatively small number of initial functional data. In general, the number of functions required for initialization depends on the expected change in the marginal posterior distribution of the template function as new data arrive. We found through simulation that initializing the particles from an MCMC sample given 10-20 functional observations yields satisfactory results.

\subsection{Real data analysis 1: drought intensity}% in California}
\label{sec:drought}

The proposed method can be effective in analyzing annual functional data related to climate, which is often monitored in real-time.
One annual measurement that exhibits common templates across years, along with phase variation, is drought intensity.
We consider sequential Bayesian registration of trajectories of annual drought intensity near Kaweah River in California. 
Statistical analysis of drought can help understand historical patterns and the variability in timing of drought intensity that can further be accounted for by management and allocation of water to different communities.
Importantly, drought intensity in California has been of great interest to address the state's drought risk management and limited access to water; Californians rely on the West and East Sierra Nevadan water resources due to intense drought in other areas of the state \citep{mann2015}.
Much attention in the literature has been devoted to developing new models for annual drought intensity. However, this task is challenging due to the time-variation and confounding of drought intensity magnitude and relative phase due to meteorologic and anthropogenic changes in the climate \citep{Diffenbaugh2015,son2021}.
In particular, annual drought intensity often exhibits a time lag as well as seasonal variation. Thus, interest not only lies in the magnitude of drought intensity, but also in the times at which drought periods begin and end during a particular year; this feature of drought intensity is captured through phase and can improve our understanding of variation in the drought lead time \citep{Dikshit2021}. While most existing statistical analyses of drought timing variation adopt multivariate data analysis approaches \citep{mohammed2020,ferijal2021}, we apply the proposed method to explore continuous trajectories of historical annual drought intensities and to account for variation in timing of features in the trajectories via registration.

\begin{figure}[!t]
  \centering
  \begin{tabularx}{13.3cm} { 
  |>{\centering\arraybackslash}X 
   >{\centering\arraybackslash}X 
   >{\centering\arraybackslash}X 
   >{\centering\arraybackslash}X
   >{\centering\arraybackslash}X
   >{\centering\arraybackslash}X | }
    \hline
    \multicolumn{6}{|c|}{(a)}
    \\
    \begin{minipage}{2cm}
    \centering\includegraphics[width=2cm]{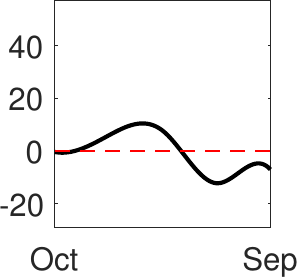}
    \end{minipage}
    &\begin{minipage}{2cm}
    \centering\includegraphics[width=2cm]{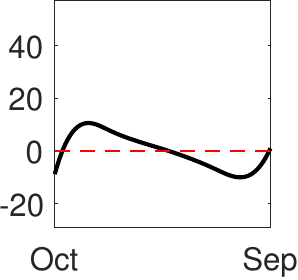}
    \end{minipage}
    &\begin{minipage}{2cm}
    \centering\includegraphics[width=2cm]{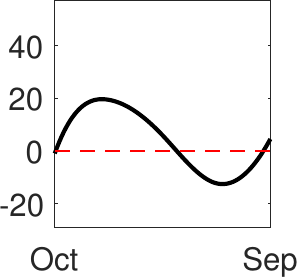}
    \end{minipage}
    &\begin{minipage}{2cm}
    \centering\includegraphics[width=2cm]{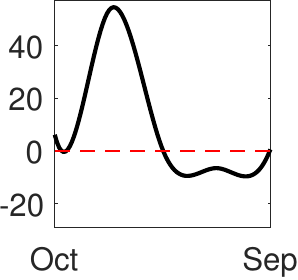}
    \end{minipage}
    &\begin{minipage}{2cm}
    \centering\includegraphics[width=2cm]{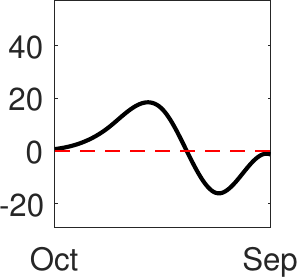}
    \end{minipage}
    &\begin{minipage}{2cm}
    \centering\includegraphics[width=2cm]{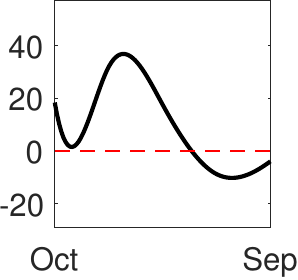}
    \end{minipage}
    \\\hline
    \multicolumn{6}{|c|}{(b)}
    \\
    \begin{minipage}{2cm}
    \centering\includegraphics[width=2cm]{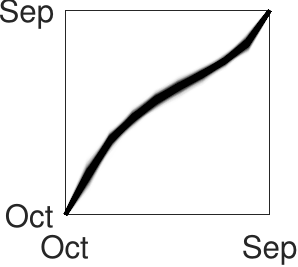}
    \end{minipage}
    &\begin{minipage}{2cm}
    \centering\includegraphics[width=2cm]{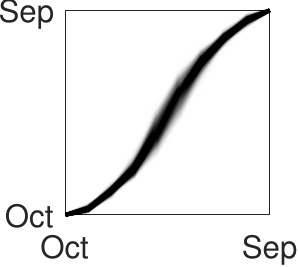}
    \end{minipage}
    &\begin{minipage}{2cm}
    \centering\includegraphics[width=2cm]{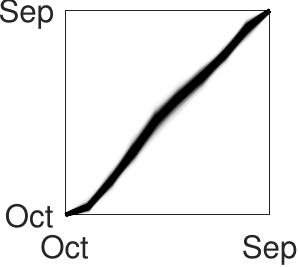}
    \end{minipage}
    &\begin{minipage}{2cm}
    \centering\includegraphics[width=2cm]{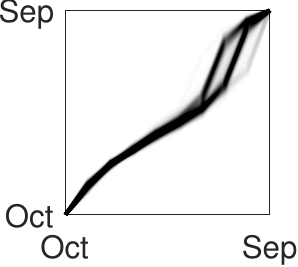}
    \end{minipage}
    &\begin{minipage}{2cm}
    \centering\includegraphics[width=2cm]{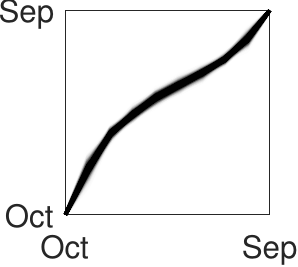}
    \end{minipage}
    &\begin{minipage}{2cm}
    \centering\includegraphics[width=2cm]{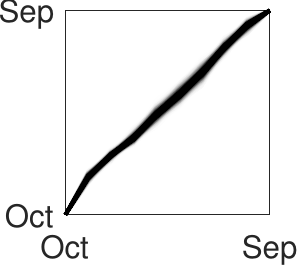}
    \end{minipage}
    \\\hline
    \multicolumn{2}{|c|}{(c)}&\multicolumn{2}{|c|}{(d)}&\multicolumn{2}{|c|}{(e)}\\
    \multicolumn{2}{|c|}{\begin{minipage}{2.5cm}
    \centering\includegraphics[width=2.5cm]{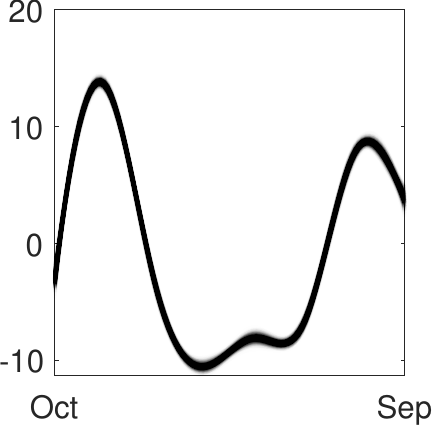}
    \end{minipage}}
    &
    \multicolumn{2}{|c|}{\begin{minipage}{2.5cm}
    \centering\includegraphics[width=2.5cm]{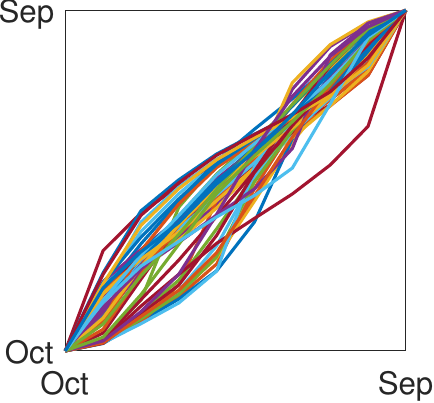}
    \end{minipage}}
    &\multicolumn{2}{|c|}{\begin{minipage}{2.5cm}
    \centering\includegraphics[width=2.5cm]{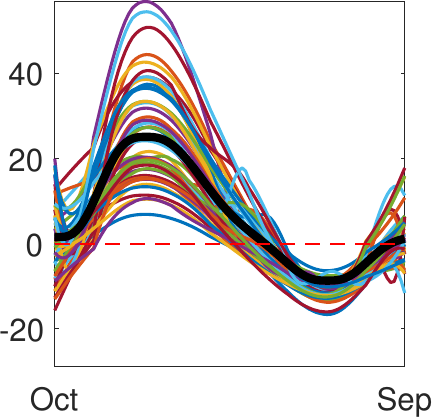}
    \end{minipage}}
    \\\hline
    \multicolumn{2}{|c|}{(f)}&\multicolumn{2}{|c|}{(g)}&\multicolumn{2}{|c|}{(h)}\\
    \multicolumn{2}{|c|}{\begin{minipage}{2.5cm}
    \centering\includegraphics[width=2.5cm]{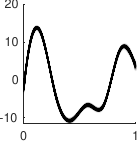}
    \end{minipage}}
    &\multicolumn{2}{|c|}{\begin{minipage}{2.5cm}
    \centering\includegraphics[width=2.5cm]{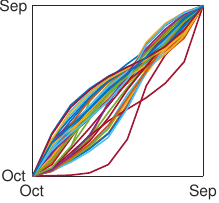}
    \end{minipage}}
    &\multicolumn{2}{|c|}{\begin{minipage}{2.5cm}
    \centering\includegraphics[width=2.5cm]{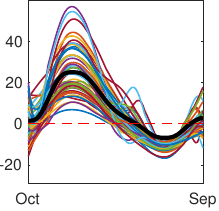}
    \end{minipage}}
    \\\hline
  \end{tabularx}
  \caption{Sequential Bayesian registration of SPEI functions representing drought intensity. (a) Trajectories of annual drought intensity near Kaweah River in California from 2014 to 2019 (hydrological year). Weighted posterior samples of (b) the phase components for the functions in (a), and (c) the template SRVF from the target posterior distribution given SPEI functions from 1970 to 2019. Transparency in (b) and (c) reflects magnitude of the weights. (d) Marginal posterior means of the phase components for SPEI functions from 1970 to 2019. (e) SPEI functions (for all years) registered via the phase components in (d) are shown in colors with the marginal posterior mean of the template in the original function space in black. (f)-(g) Same as (c)-(e), but based on MCMC batch learning. The red dashed line in (a), (e) and (h) is the threshold at which a drought period begins and ends.}
  \label{fig:real1}
\end{figure}

Despite the importance of monitoring the severity of drought, there is no universal definition of drought intensity \citep{mishra2010}. Many types of drought intensity indices have been developed for monitoring purposes, which are derived using a summary of covariates such as precipitation and temperature via a deterministic model. Time lag is a common phenomenon in these drought intensity indices, and some studies have focused on time lag estimation via multivariate time series methods \citep{mishra2011}. 
%data of climate indices through multivariate data analysis techniques without incorporating covariates .
In this study, we extract phase variability from a historical record of a single drought intensity index. In particular, we use the proposed sequential registration approach to explore the variation in the magnitude and phase of annual drought intensity near Kaweah River in California from 1970 to 2019. We use the Standardised Precipitation-Evapotranspiration Index (SPEI), that is obtained via a deterministic equation of latitude, time of year, and precipitation and temperature across time and space \citep{kim2022,spei}. SPEI functional data, which is available at monthly intervals, is pre-processed via spline interpolation and segmentation by hydrological year, starting on October 1 and ending on September 30.  We refer to this data as annual SPEI functions and denote it by $f_{1:50}$. A subset of the data, corresponding to hydrological years 2014 to 2019, is shown in Figure \ref{fig:real1}(a).

Since the structure of SPEI functions is relatively simple, with few peaks and valleys in each functional observation, we model the SRVF template as a linear combination of ten cubic B-spline basis functions with equally spaced knots. We specify the following prior hyperparameters for the template and phase components: $B=10,\ \Sigma_c=20I_B,\ \kappa=5,\ \alpha_\sigma = 4,\ \beta_\sigma=0.01$, and a uniform partition of size $M_\gamma=10$. Given the small number of prominent features on each SPEI, a partition size of $M_\gamma=10$ should be large enough to capture phase variation. We begin by obtaining $10,000$ MCMC samples from the posterior given $f_{1:30}$, after a burn-in period of $40,000$ iterations. The rest of the SPEI functions $f_{31:50}$ are assimilated sequentially, and the initial samples are updated recursively via the proposed SMC algorithm.

Features of the target posterior density, $\pi(\cdot|f_{1:50})$, are visualized in Figure \ref{fig:real1}. Drought is defined as an interval of negative SPEI, with the threshold shown as a red dashed line in panels (a) and (e). In (a), it appears that drought near Kaweah River generally occurs during the summer season, with significant variation in the timing of drought onset from year to year. Panel (c) shows weighted posterior samples of the SRVF template (transparency reflects the magnitude of the corresponding weight), given the annual SPEI functions $f_{1:50}$. 
This marginal posterior has small pointwise uncertainty across the domain. 
Panel (b) shows the weighted samples of phase for SPEI functions $f_{45:50}$, corresponding to hydrological years 2014 to 2019. 
The marginal posterior of the phase for year 2017 (fourth column) appears multimodal. This is due to the SPEI function in 2017 having two valleys during the drought season; either of the two valleys can be aligned to the single valley in the template.

Figure \ref{fig:real1}(d) illustrates annual phase variability of the SPEI functions $f_{1:50}$ via the marginal posterior means of the phase components. There appears to be significant phase variation throughout each hydrological year as most of the posterior marginal phase functions deviate from the identity element. The most phase variability is generally observed around the time when the peak in SPEI functions occurs; overall, there is more phase variation from October to April than from May to September. Such estimates of the relative phases aid in the analysis of variation in drought duration as well as the timing of drought onset across years. Figure \ref{fig:real1}(e) shows the SPEI functions after registration via the marginal posterior means of the phases, which are illustrated using colors. It is evident that amplitude variability in the peak of the SPEI functions is much larger than amplitude variability in the valley; the magnitude of most severe drought (minimum of the SPEI functions) is similar across all years. Panel (e) shows the posterior mean template in the original function space with a thick black line, which effectively captures the prominent features (shape) of the annual SPEI functions. 
We also visualize posterior samples given $f_{1:50}$ obtained via MCMC. Panels (f)-(h) show the marginal posterior of the template SRVF, the marginal posterior means of the phase components, and the registered functions (using the posterior means in (g)) with the marginal posterior mean of the template in the original function space in black, respectively. While the results based on our approach and MCMC-based batch learning are quite similar, there are some differences in the posterior means of the phase components. We hypothesize that this is due to issues with MCMC convergence.

While multivariate analysis methods can be used to analyze discrete drought intensity data, the proposed approach is better suited for inference on the smoothly-varying annual SPEI functions. %Using functional data techniques, we can extract more useful information, e.g., estimation results are not limited to discrete time points.
Often, synchronization of SPEI functions, and extraction of phase variability, is done on the basis of (a few) manually identified landmarks, e.g., peaks and valleys. However, as seen in Figure \ref{fig:real1}(a), systematic selection of landmark points on SPEI functions is difficult as the number of significant features on each function can vary. On the other hand, the sequential registration framework enables full horizontal synchronization of the SPEI functions, without specification of such landmarks and while also providing uncertainty estimates.

\subsection{Real data analysis 2: sea surface salinity}
\label{sec:SSS}

Next, we consider trajectories of annual sea surface salinity (SSS) near Null Island.
Salinity is an important variable in understanding global climate change as it regulates the movement of currents and heat carried within currents based on water density \citep{durack2015}.
As water evaporation and precipitation change over time, the magnitude of SSS fluctuates annually according to seasonal variation and the movement of currents. There are recent studies focusing on using phase variation in SSS to understand ocean dynamics, e.g., on the lag between fluctuations of salinity and precipitation, a major factor that affects ocean salinity \citep{bingham2012}. \cite{bingham2021} compare two sets of SSS observations, which were recorded using different equipment, via estimates of phase variability. Thus, estimation and assessment of phase variation in annual SSS measurements is an important statistical task, and can be subsequently used to understand ocean dynamics and to explore relationships between SSS and other oceanic factors.
In previous studies, \cite{bingham2012,bingham2021} approximated seasonal SSS variability through a one-dimensional measurement of phase corresponding to the time at which maximum SSS magnitude is observed. This, again, is akin to landmark-based registration of SSS functions, with the point of maximum SSS magnitude on each function serving as a single landmark. While intuitive and simple, this approach fails to register prominent features of the SSS functions not captured by the single landmark. Thus, to estimate more flexible and informative phase components of SSS functions, we apply the proposed sequential Bayesian registration approach.

\begin{figure}[!t]
  \centering
  \begin{tabularx}{13.3cm} { | c
  |>{\centering\arraybackslash}X |
   >{\centering\arraybackslash}X |
   >{\centering\arraybackslash}X | }
    \hline
    & (a) & (b) & (c)
    \\\hline
    \rotatebox[origin=c]{90}{El Ni\~{n}o} & \begin{minipage}{2.5cm}
    \centering\includegraphics[width=2.5cm]{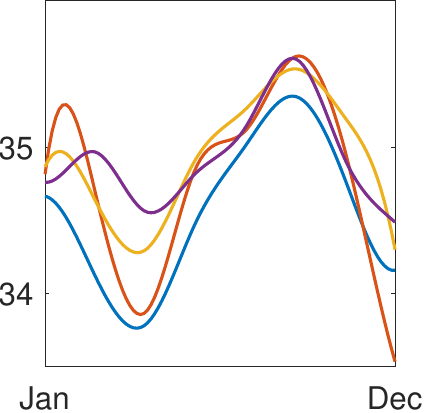}
    \end{minipage}
    &
    \begin{minipage}{2.5cm}
    \centering\includegraphics[width=2.5cm]{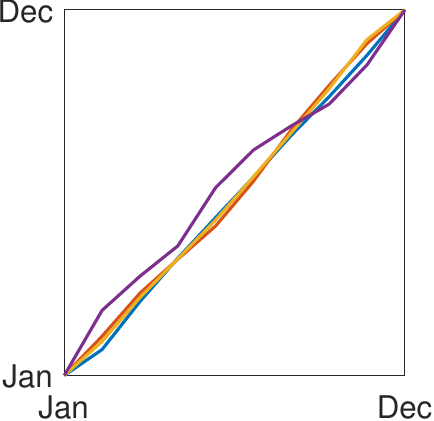}
    \end{minipage}
    &
    \begin{minipage}{2.5cm}
    \centering\includegraphics[width=2.5cm]{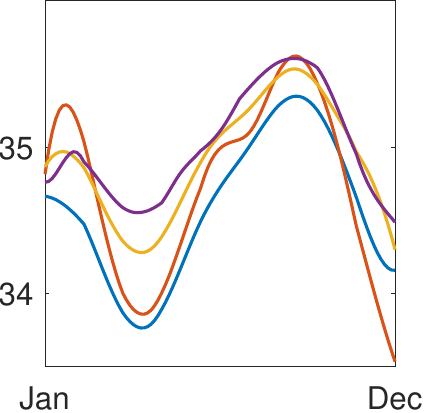}
    \end{minipage}
    \\\hline
    \rotatebox[origin=c]{90}{Neutral} & \begin{minipage}{2.5cm}
    \centering\includegraphics[width=2.5cm]{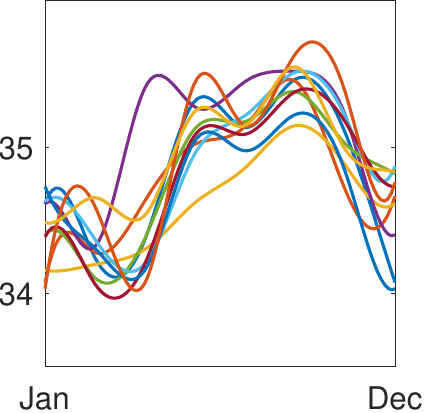}
    \end{minipage}
    &
    \begin{minipage}{2.5cm}
    \centering\includegraphics[width=2.5cm]{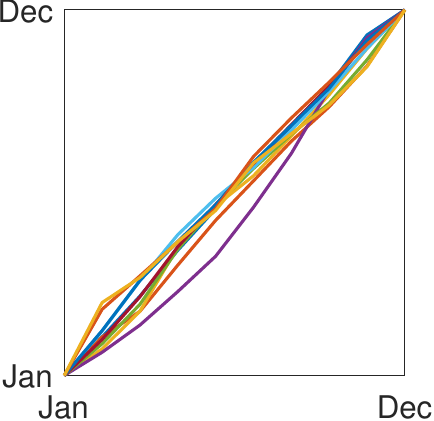}
    \end{minipage}
    &
    \begin{minipage}{2.5cm}
    \centering\includegraphics[width=2.5cm]{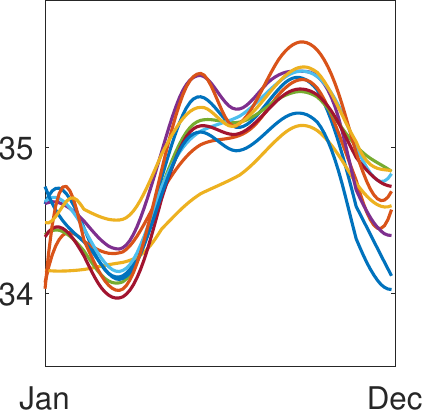}
    \end{minipage}
    \\\hline
    \rotatebox[origin=c]{90}{La Ni\~{n}a} &
    \begin{minipage}{2.5cm}
    \centering\includegraphics[width=2.5cm]{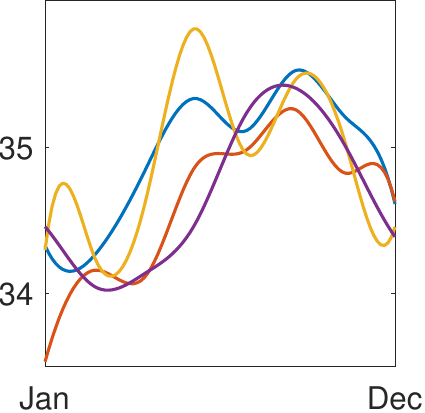}
    \end{minipage}&
    \begin{minipage}{2.5cm}
    \centering\includegraphics[width=2.5cm]{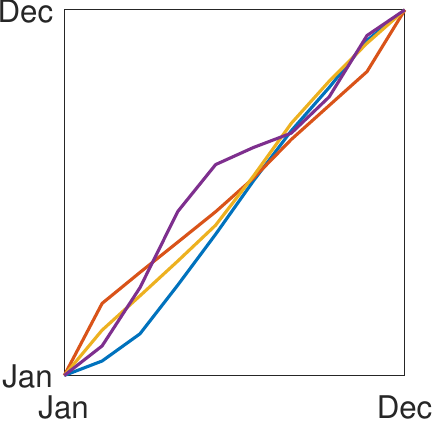}
    \end{minipage}
    &
    \begin{minipage}{2.5cm}
    \centering\includegraphics[width=2.5cm]{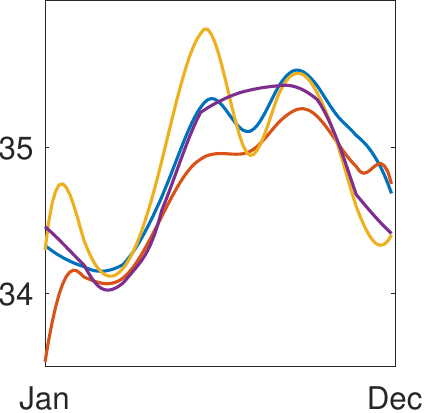}
    \end{minipage}
    \\\hline
  \end{tabularx}
  \caption{Sequential Bayesian registration of SSS functions for three ENSO states. (a) Trajectories of annual SSS near Null Island from 2000 to 2017, plotted by ENSO state. (b) Marginal posterior means of phase components for the SSS functions in (a). (c) SSS functions registered using the phases in (b).}
  \label{fig:real2}
\end{figure}

A climate pattern over the Pacific Ocean that is known to influence global oceanographic variables is El Ni\~{n}o–Southern Oscillation (ENSO). ENSO is a phenomenon that fluctuates among three states: warm state or El Ni\~{n}o, neutral state, and cool state or La Ni\~{n}a; the criteria used to determine which of these three states is occurring in a given year are sea surface temperature, wind speed, surface pressure, and related measurements. This climate pattern has a close relationship with SSS, e.g., SSS tends to decrease (increase) near the equatorial Pacific during El Ni\~{n}o (La Ni\~{n}a) \citep{zhu2014,hackert2020}. In this study, we explore amplitude and phase patterns in SSS functions for the three ENSO states. To do this, we first pre-process monthly EN4 observations of SSS via spline interpolation and segmentation across years 2000 to 2017 \citep{en4,good2013}. We further categorize each year to one of the three ENSO states, and apply the proposed sequential approach to separately register SSS functions for each state. In particular, there are four SSS functions measured during El Ni\~{n}o, $f^E_{1:4}$, ten SSS functions measured during the neutral state, $f^N_{1:10}$, and four SSS functions measured during La Ni\~{n}a, $f^L_{1:4}$; the data is shown in Figure \ref{fig:real2}(a). Note that the SSS functions within each ENSO state tend to share similar amplitude features with clear phase variation. The majority of the El Ni\~{n}o SSS functions have a single distinct peak; the neutral and La Ni\~{n}a SSS functions tend to have more amplitude features.

As in the drought intensity study, we model the SRVF template using a linear combination of ten cubic B-splines basis functions, which provides considerable flexibility for template estimation.
The other prior hyperparameters for the template and phase components are $\Sigma_c = 20I_B,\ \kappa = 5,\ \alpha_\sigma=4,\ \beta_\sigma=0.01$, and a uniform partition of size $M_\gamma=10$ (a partition of this size provides sufficient flexibility for phase estimation). 
For each ENSO state, we initialize the sampler using $10,000$ MCMC samples from the posterior given the first three functions after a burn-in period of $40,000$ iterations. The rest of the SSS functions for each ENSO state are then added to the data sequentially, and the initial samples are updated recursively using the proposed SMC algorithm.

Estimated relative phase and amplitude for the data in Figure \ref{fig:real2}(a), based on posterior densities $\pi(\cdot\mid f^E_{1:4})$, $\pi(\cdot\mid f^N_{1:10})$ and $\pi(\cdot\mid f^L_{1:4})$, are shown in Figure \ref{fig:real2}(b) and (c), respectively. Panels (b) and (c) show the marginal phase posterior means, and registered SSS functions, respectively. Phase variability within each ENSO state is fairly small and registered functions appear to be horizontally synchronized as desired.

\begin{figure}[!t]
  \centering
  \begin{tabularx}{7cm} { |>{\centering\arraybackslash}X |
   >{\centering\arraybackslash}X | }
    \hline
    (a) & (b)
    \\\hline
    \begin{minipage}{2.5cm}
    \centering\includegraphics[width=2.5cm]{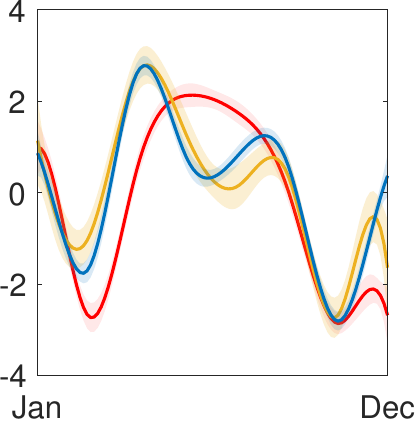}
    \end{minipage}
    &
    \begin{minipage}{2.5cm}
    \centering\includegraphics[width=2.5cm]{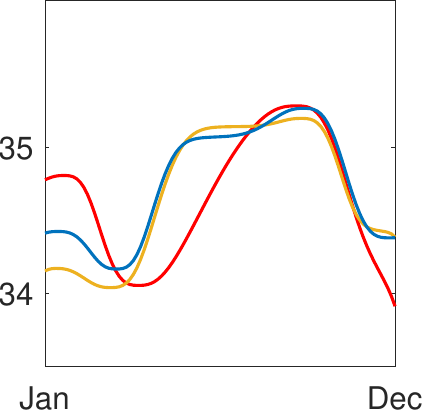}
    \end{minipage}
    \\\hline
  \end{tabularx}
  \caption{Visualization of the marginal posterior for the template for each ENSO state: El Ni\~{n}o in red, neutral in blue, La Ni\~{n}a in yellow. (a) Posterior mean templates in the SRVF space are shown as solid curves. The band around each posterior mean represents the posterior uncertainty: +/- 2 pointwise marginal posterior standard deviations. (b) Marginal posterior mean for the template for each ENSO state in the original function space.}
  \label{fig:real2_template}
\end{figure}

Figure \ref{fig:real2_template} shows the marginal posterior of the template function for each ENSO state: El Ni\~{n}o in red, neutral in blue, and La Ni\~{n}a in yellow. In (a), the marginal posterior mean of the SRVF template is shown as a solid curve with a band corresponding to +/- 2 pointwise marginal posterior standard deviations. The marginal posterior means of the template function in the original function space are illustrated in (b). The marginal posterior means of the template function clearly capture the prominent features of the SSS functions in each of the three ENSO states; the El Ni\~{n}o template has a single valley followed by a fairly sharp peak, while the neutral and La Ni\~{n}a templates have shallower valleys followed by a peak that is spread over a longer time period. In particular, the estimated template functions for the neutral and La Ni\~{n}a states appear to be more similar in shape than the El Ni\~{n}o template. These shape differences can be quantified by computing pairwise $\mathbb{L}^2$ distances between (rescaled to have norm one) SRVFs of the marginal posterior means of the template function for each state. The distances between the El Ni\~{n}o and La Ni\~{n}a/neutral states are both relatively large: 0.7662/0.6867. On the other hand, the distance between the La Ni\~{n}a and neutral states is much smaller: 0.2978. The presented amplitude/phase estimation results for annual SSS functional data, based on the proposed sequential registration framework, can provide insights into the association between amplitude and phase variation in SSS functions across the three ENSO states.

\subsection{Real data analysis 3: segmented PQRST complexes}
\label{sec:ecg}

The electrocardiogram (ECG) is routinely used to assess heart function and diagnose various medical conditions, e.g., myocardial infarction. The data recorded via ECG is a long signal composed of a periodic sequence of a pattern known as the PQRST complex: the P wave corresponds to the first small peak, the QRS wave is composed of a sharp valley followed by a sharp peak followed by another sharp valley, and the T wave represents the last high peak. Prior to statistical analysis, it is beneficial to segment the long ECG signal into its PQRST complexes \citep{kurtek2013}. Our focus here is not on the segmentation problem, but rather on registration of PQRST complexes as the long signal is recorded and segmented sequentially. While there is natural variability among the PQRST complexes along an ECG signal, abnormalities in estimates of the underlying template and variation in timing of PQRST patterns (phase) with respect to that template are beneficial for the aforementioned diagnostic purposes. For example, duration of the QT interval is useful in drug development and approval \citep{zhou2009}.

\begin{figure}[!t]
  \centering
  \begin{tabularx}{13.3cm} { 
  |>{\centering\arraybackslash}X 
   >{\centering\arraybackslash}X 
   >{\centering\arraybackslash}X 
   >{\centering\arraybackslash}X
   >{\centering\arraybackslash}X
   >{\centering\arraybackslash}X | }
    \hline
    \multicolumn{6}{|c|}{(a)}
    \\
    \begin{minipage}{2cm}
    \centering\includegraphics[width=2cm]{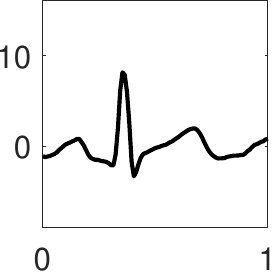}
    \end{minipage}
    &\begin{minipage}{2cm}
    \centering\includegraphics[width=2cm]{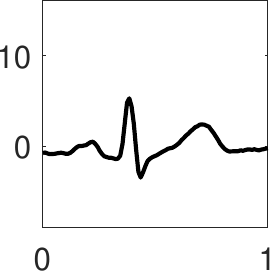}
    \end{minipage}
    &\begin{minipage}{2cm}
    \centering\includegraphics[width=2cm]{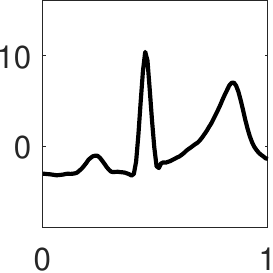}
    \end{minipage}
    &\begin{minipage}{2cm}
    \centering\includegraphics[width=2cm]{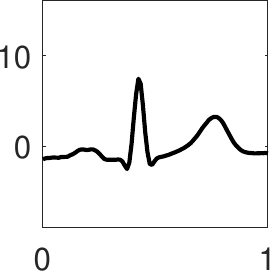}
    \end{minipage}
    &\begin{minipage}{2cm}
    \centering\includegraphics[width=2cm]{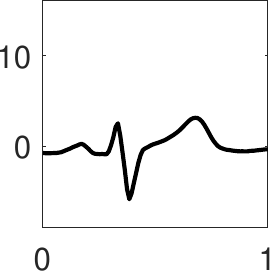}
    \end{minipage}
    &\begin{minipage}{2cm}
    \centering\includegraphics[width=2cm]{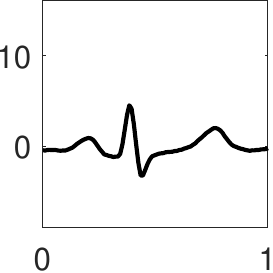}
    \end{minipage}
    \\\hline
    \multicolumn{6}{|c|}{(b)}
    \\
    \begin{minipage}{2cm}
    \centering\includegraphics[width=2cm]{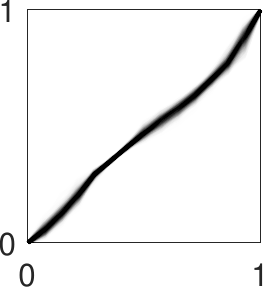}
    \end{minipage}
    &\begin{minipage}{2cm}
    \centering\includegraphics[width=2cm]{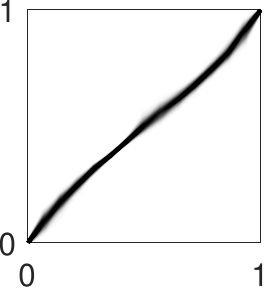}
    \end{minipage}
    &\begin{minipage}{2cm}
    \centering\includegraphics[width=2cm]{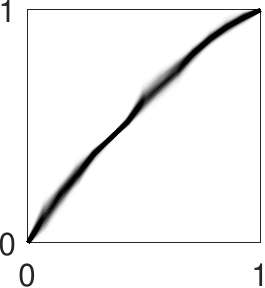}
    \end{minipage}
    &\begin{minipage}{2cm}
    \centering\includegraphics[width=2cm]{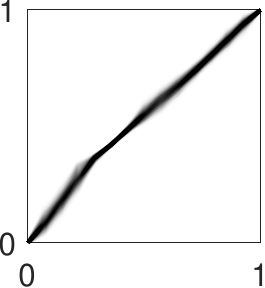}
    \end{minipage}
    &\begin{minipage}{2cm}
    \centering\includegraphics[width=2cm]{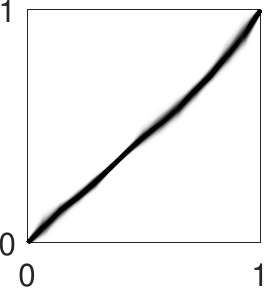}
    \end{minipage}
    &\begin{minipage}{2cm}
    \centering\includegraphics[width=2cm]{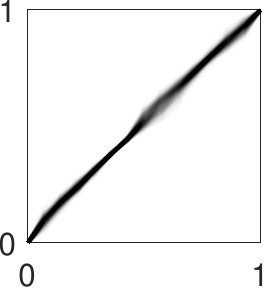}
    \end{minipage}
    \\\hline
    %\multicolumn{3}{|c|}{(c) }
    %\\\hline
    %\multicolumn{1}{|c}{}
    \multicolumn{2}{|c|}{(c)}&\multicolumn{2}{|c|}{(d)}&\multicolumn{2}{|c|}{(e)}\\
    \multicolumn{2}{|c|}{\begin{minipage}{2.5cm}
    \centering\includegraphics[width=2.5cm]{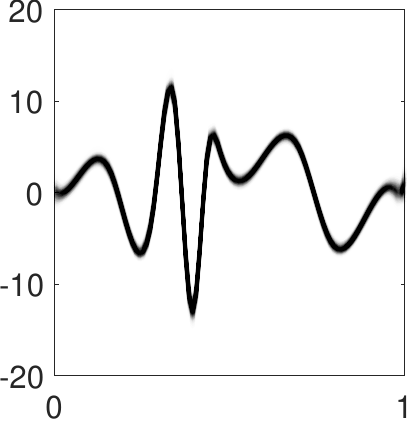}
    \end{minipage}}
    &
    \multicolumn{2}{|c|}{\begin{minipage}{2.5cm}
    \centering\includegraphics[width=2.5cm]{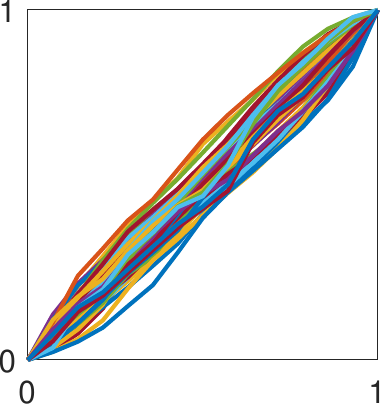}
    \end{minipage}}
    &\multicolumn{2}{|c|}{\begin{minipage}{2.5cm}
    \centering\includegraphics[width=2.5cm]{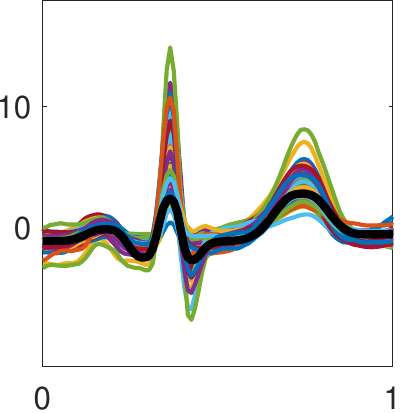}
    \end{minipage}}
    \\\hline
    \multicolumn{6}{|c|}{(f)}\\
    \multicolumn{6}{|c|}{\begin{minipage}{12cm}
    \centering\includegraphics[width=12cm]{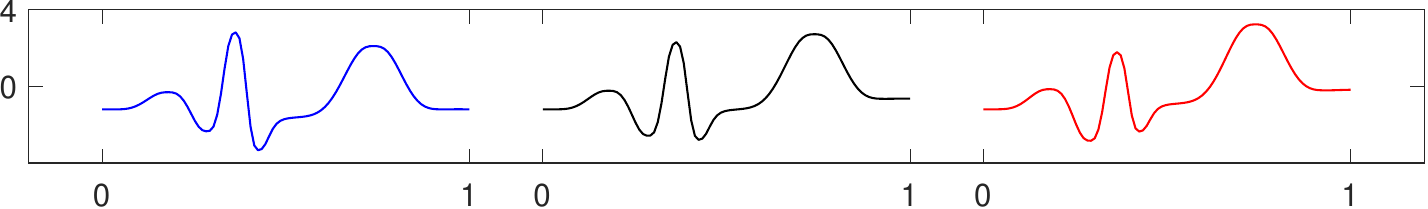}
    \end{minipage}}\\\hline
  \end{tabularx}
  \caption{Sequential Bayesian registration of PQRST complexes extracted from an ECG signal. (a) Subset of 50 PQRST complexes. (b) Weighted posterior samples of relative phases for the functions in (a). (c) Weighted posterior samples of the template SRVF. Transparency in (b) and (c) reflects the magnitude of weights. (d) Posterior mean phases for the 50 functions. (e) Registered PQRST complexes (solid colors) using phase components in (d) and posterior mean template in the original function space (thick black line). (f) Posterior mean template (black), and negative (red) and positive (blue) directions along the first principal component of the template in the original function space.}
  \label{fig:real3}
\end{figure}

We consider 50 PQRST complexes, denoted by $f_{1:50}$, segmented from a long ECG signal \citep{kurtek2013}; six functions from this set are shown in Figure \ref{fig:real3}(a). We initialize the sequential registration algorithm using $10,000$ MCMC samples based on the data $f_{1:30}$ after burn-in. PQRST complexes tend to have ``sharper'' features than the data we encountered in the previous two examples. Thus, we model the template function using a linear combination of $13$ B-spline basis functions ($B = 13$); we set $\Sigma_c=20I_B$ as previously. The concentration parameter in the prior over increments of the phase components is $\kappa=5$ and the size of the uniform partition is $M_\gamma=15$. Finally, the hyperparameters for the error variance are $\alpha_\sigma = 10$ and $\beta_\sigma = 0.01$.

%\label{sec:ecg}

Results based on the last posterior distribution in the sequence, $\pi(\cdot\mid f_{1:50})$, are shown in Figure \ref{fig:real3}.
Panel (b) shows the weighted samples of the phase components for the functions in (a), and panel (c) visualizes the weighted samples of the template SRVF with transparency reflecting the magnitude of weights. The marginal posterior means of the phase components are shown in panel (d). In general, they capture the relative acceleration or delay of the corresponding PQRST complex with respect to the template. Pointwise uncertainty for each phase component during the QRS time interval (sharp valley followed by a sharp peak followed by another sharp valley) tends to be very small compared to other time points along the domain. This is intuitive, as the QRS shape features in a PQRST complex are very prominent resulting in very small uncertainty in their timing relative to the template.
The registered data, obtained by applying the marginal posterior mean phase components in (d) to the corresponding PQRST complexes, is shown in panel (e). The proposed method achieves very good horizontal synchronization of all features of the PQRST complexes.
The marginal posterior mean of the template function in the original function space is illustrated with a thick black curve in panel (e). The estimated template function contains all features of the PQRST complex and is representative of the shape in the data. The uncertainty in the shape of the template is further shown using principal component analysis in panel (f): the black function is the posterior mean template; the blue and red functions show the positive and negative direction along the first principal component with $\alpha=3$, respectively, as described in Section \ref{sec:viz}. Here, we note that most variation occurs in the magnitude of the RST features with very little variation in the magnitude of the PQ features.

\section{Discussion and future work}
\label{sec:summary}

We propose a novel sequential Bayesian registration method for functional data that addresses two common challenges: confounded sources of variation across functions, and availability of an increasing number of functional observations over time. In particular, we propose a Bayesian registration model and an SMC algorithm to update inference when a new function is observed, exploiting an approximate weight updating approach for MH kernels without a closed form. As illustrated in the simulation studies and real data examples, the proposed sequential learning algorithm is computationally more efficient, and more accurate in terms of posterior estimates, compared to the batch MCMC algorithm. It efficiently explores multimodal posterior distributions, which often pose a challenge for many MCMC techniques. Application of this framework to real data studies improves our understanding of the underlying structure of the data, and provides quantitative evidence to study associated research problems. 

An important consideration is that the proposed method leads to approximate posterior inference due to the need to circumvent computation of the MH kernels in the SMC algorithm. The quality of the approximation depends on the assumption that marginals of adjacent posterior distributions are similar. In general, this holds when a new function is assimilated with a relatively large number of functions that are already registered. If the number of previously registered functions is small $(< 5)$, and a function with a very different shape arrives, the difference in the marginal posteriors may lead to non-representative weights. In such cases, we suggest adding annealing steps between adjacent target distributions. With this adjustment, the proposed algorithm is robust to outlying functions that have potentially different shape relative to other functions already in the sample.

Particle degeneracy often arises in SMC sampling for a sequence of distributions with increasing state space dimension, and can lead to unreliable posterior estimates. We partly resolve this issue by initializing new components of each particle such that the updated particles lie in a region of high posterior probability. When phase components are added to the state space as new data arrives, this initialization is performed by drawing from a distribution centered at a phase function obtained by aligning the new functional observation to existing template particles. Furthermore, we monitor ESS as small values of ESS can indicate evidence of particle degeneracy. Degeneracy may be remedied by duplicating existing particles, thereby increasing the number of weighted samples. This was not necessary in our studies, as the ESS was large and no significant decrease in ESS was observed as we added more functions to the data.

The proposed Bayesian registration model may be modified depending on the application of interest. The key assumption that makes the model computationally tractable is that phase components are piecewise linear functions with Dirichlet increments. This reduces the state space dimension relative to similar models in the literature and enables faster posterior inference. Another consideration is that, for the sake of isometry and simple derivations, we build the Bayesian hierarchical model on the SRVF space. This limits our model to functions for which derivatives exist almost everywhere. The proposed method is therefore applicable to various real data studies where interest lies in smooth evolutions of a variable.

In this work, we assume that the functions are independent given the template and phase components. Such an assumption limits forecasting ability beyond the use of the posterior distribution based on historical data for prediction. Thus, a natural extension of the proposed framework is to account for temporal dependence between sequentially observed functions, which is a reasonable assumption for applications such as climate monitoring. This requires a re-formulation of the Bayesian registration model as a functional time series model \citep{bosq2000,kokoszka2012}. Another natural extension is to incorporate time-varying covariates via a functional linear model. We plan to explore these two directions in future work.

The presented Bayesian registration model can be extended to the case of multivariate functional data, i.e., functional data with multiple dependent or independent components. Under the assumption of zero cross-component phase variation within each observation, the main modification to the presented model requires the definition of a prior distribution over the multiple components of the template. However, in the presence of cross-component and cross-observation phase variation, one would have to construct more complicated prior models for the phase component, and ensure that cross-component and cross-observation phase are identifiable. This extension would be especially relevant in the context of ECG signals. While in this work we used segmented PQRST complexes from an ECG signal recorded by a single lead, the standard ECG has 12 leads, each recording the heart's electrical activity based on a different placement of an electrode. Harnessing information recorded across all 12 leads could provide more information about the heart's function for diagnostic purposes.

\backmatter

\bmhead{Supplementary information}

Algorithms and additional results are provided in the supplementary material. 

\bmhead{Acknowledgements}

This project was supported in part by NIH R37-CA214955, NSF CCF-1740761, NSF CCF-1839252, NSF DMS-2015226 and NSF DMS-2413747  (to SK), and NASA-80NSSC18K1322 (to OC).
The authors thank Dr. Frederick Bingham and Susannah Brodnitz for their valuable comments on the sea surface salinity case study and extracting the monthly EN4 data. The authors also thank Dr. Nancy Grulke for her thoughtful comments on the drought intensity case study.

\bibliography{reference}% common bib file

\end{document}

\title[Supplementary Material for “Sequential
Bayesian Registration for Functional Data”]{Supplementary Material for “Sequential
Bayesian Registration for Functional Data”}


\author[1]{\fnm{Yoonji} \sur{Kim}}\email{yoonj2kim@gmail.com}

\author[2]{\fnm{Oksana A.} \sur{Chkrebtii}}\email{oksana@stat.osu.edu}

\author[1]{\fnm{Sebastian A.} \sur{Kurtek}}\email{kurtek.1@stat.osu.edu}

\affil*[1]{\orgdiv{Department of Statistics}, \orgname{The Ohio State University}, \orgaddress{\street{1958 Neil Avenue}, \city{Columbus}, \postcode{43210}, \state{Ohio}, \country{USA}}}

\maketitle

\appendix

\section{Detailed Algorithm}

We present an algorithm that is described in detail in the main manuscript. Algorithm \ref{algo:sequential_registration}, based on Section 4 in the main article, describes the sequential update of the posterior weighted particles as new functions are observed. Here, we assume that an initial set of posterior particles, given functions $f_{1:n_{ini}}$, is available. The algorithm then sequentially updates the posterior weighted particles when new functions, $f_{n_{ini+1}},\ldots,f_{N}$, become part of the data one at a time. Note that due to its length, the full algorithm is split across two pages.

\begin{algorithm}[!h]
\begin{footnotesize}
\caption{Sequential Bayesian Registration Algorithm}
\label{algo:sequential_registration}
\begin{algorithmic}[1]
\IniInputs{(1) SRVFs $q_{1:n_{ini}}$ of $f_{1:n_{ini}}$.\\ 
(2) $J$ weighted samples $\{(\textbf{c}^{(j)},\textbf{d}_{1:n_{ini}}^{(j)},\sigma^{2(j)},w^{(j)}),\ j=1,\ldots,J\}$ from posterior given $f_{1:n_{ini}}$ obtained via, e.g., MCMC.\\
(3) Number of iterations $K$ for Metropolis-Hastings perturbation step.}
\For{$n=n_{ini},\ldots,N-1$}
\Inputs{(1) SRVFs $q_{1:n+1}$ of $f_{1:n+1}$.\\ 
(2) $J$ weighted particles $\{(\textbf{c}^{(j)},\textbf{d}_{1:n}^{(j)},\sigma^{2(j)},w^{(j)}),\ j=1,\ldots,J\}$.}
 % drawn from posterior given $f_{1:n}$.}
\For{$j=1,\ldots,J$}
\State Compute $\hat\gamma_{n+1}^{(j)}\gets\arg\min_{\gamma\in\Gamma}\|\sum_{b=1}^B c_b^{(j)} \phi_b-(q_{n+1}\circ \gamma)\sqrt{\dot{\gamma}}\|^2$ via Dynamic Programming.
\State Replace $\hat\gamma_{n+1}^{(j)}$ with a piecewise linear least squares fit of $\hat\gamma_{n+1}^{(j)}$ with breakpoints at $(s_{1},s_2,\ldots,s_{M_{\gamma}-1},s_{M_{\gamma}})$.
\State Draw $\textbf{d}^p \sim Dir(\frac{\kappa_{ini}}{M_\gamma-1}1_{M_\gamma-1})$, and let $\gamma^{(j)*} = \hat{\gamma}_{n+1}^{(j)} \circ d^{-1}(\textbf{d}^p)$ and $\textbf{d}_{n+1}^{(j)} = d(\gamma^{(j)*})$.
\algstore{alg1}
\end{algorithmic}
\end{footnotesize}
\end{algorithm}

\begin{algorithm}[!h]
\begin{footnotesize}
\begin{algorithmic}[1]
\algrestore{alg1}
\State Update the weights using $\tilde{w}^{(j)}\gets w^{(j)}h(f_{n+1}\mid\textbf{c}^{(j)},\textbf{d}_{n+1}^{(j)},\sigma^{2(j)}) p(\textbf{d}_{n+1}^{(j)})/\tilde{K}(\textbf{d}^{(j)}_{n+1}\mid\textbf{c}^{(j)},f_{n+1})$ where $\tilde{K}(\textbf{d}^{(j)}_{n+1}\mid\textbf{c}^{(j)},f_{n+1}) = \prod_{m=2}^{M_\gamma}((\hat{\gamma}_{n+1}^{(j),-1})'\circ \gamma^{(j)*})(s_m)Dir(\textbf{d}^p;\frac{\kappa_{ini}}{M_\gamma-1}1_{M_\gamma-1})$.
\EndFor
\State Normalize the weights, $[\tilde{w}^{(1)},\ldots,\tilde{w}^{(J)}]^\top \gets \frac{1}{\sum_{j=1}^J\tilde{w}^{(j)}}[\tilde{w}^{(1)},\ldots,\tilde{w}^{(J)}]^\top$.
\If{$(\sum_{j=1}^J(\tilde{w}^{(j)})^2)^{-1}<J/2$}.
\State Resample the particles using multinomial distribution with current weights as parameters, and assign equal weight to each resampled particle.
\EndIf
\State Compute empirical covariance: $\widehat{\Sigma}_c \gets \frac{1}{J-1}\sum_{j=1}^J\tilde{w}^{(j)}\textbf{c}^{(j)}\textbf{c}^{(j)\top}$.
\For{$j=1,\ldots,J$}
    \State $\gamma_{1:n+1}^{(j)}\gets d^{-1}(\textbf{d}_{1:n+1}^{(j)})$.
    \For{$k=1,\ldots,K$}
    \State Sample $\textbf{c}^*\sim N({\textbf{c}}^{(j)},\widehat{\Sigma}_c)$.
    \State Compute $\alpha_c \gets \min\left\{1,\dfrac{h(f_{1:n+1}\mid\textbf{c}^*,\textbf{d}_{1:n+1}^{(j)},\sigma^{2(j)})N(\textbf{c}^{*};\textbf{0}_B,\Sigma_c)}{h(f_{1:n+1}\mid{\textbf{c}}^{(j)},\textbf{d}_{1:n+1}^{(j)},\sigma^{2(j)})N({\textbf{c}}^{(j)};\textbf{0}_B,\Sigma_c)}\right\}$.
    \If{$Unif(0,1)<\alpha_c$}
    \State ${\textbf{c}}^{(j)} \gets \textbf{c}^{*}$.
    \EndIf
    \EndFor
\For{$i=1,\ldots,n+1$}
\For{$k=1,\ldots,K$}
    \State Sample $\textbf{d}^p\sim Dir(\frac{\kappa}{M_\gamma-1}1_{M_\gamma-1})$
    \State $\gamma^p\gets d^{-1}(\textbf{d}^p)$.
    \State $\gamma^*\gets{\gamma}_i^{(j)}\circ\gamma^p$ where ${\gamma}_{i}^{(j)}\gets d^{-1}({\textbf{d}}_{i}^{(j)})$.
    \State $\textbf{d}^*\gets d(\gamma^*)$.
    \State $\alpha_\gamma \gets \min\left\{1,\dfrac{h(f_{i}\mid{\textbf{c}}^{(j)},\textbf{d}^*,\sigma^{2(j)})\pi(\gamma^*)\prod_{m=2}^{M_\gamma}((\gamma^{*,-1})'\circ{\gamma}_{i}^{(j)})(s_m)Dir(d(\gamma^{p,-1}))}{h(f_{i}\mid{\textbf{c}}^{(j)},{\textbf{d}}_{i}^{(j)},\sigma^{2(j)})\pi({\gamma}_{i}^{(j)})\prod_{m=2}^{M_\gamma}(({\gamma}_{i}^{(j),-1})'\circ\gamma^*)(s_m)Dir(d(\gamma^p))}\right\}$.
    %\State Compute $\alpha_\gamma \gets \min\left\{1,\dfrac{h(f_{i}\mid{\textbf{c}}^{(j)},\textbf{d}^*,\sigma^{2(j)})\pi(\gamma^*)\prod_{m=1}^{M_\gamma}(\gamma^{*,-1}\circ{\gamma}_{i}^{(j)}(s_m))Dir(d(\gamma^{p,-1});\frac{\kappa}{M_\gamma}1_{M_\gamma})}{h(f_{i}\mid{\textbf{c}}^{(j)},{\textbf{d}}_{i}^{(j)},\sigma^{2(j)})\pi({\gamma}_{i}^{(j)})\prod_{m=1}^{M_\gamma}({\gamma}_{i}^{(j),-1}\circ\gamma^*(s_m))Dir(d(\gamma^p);\frac{\kappa}{M_\gamma}1_{M_\gamma})}\right\}$.
    \If{$Unif(0,1)<\alpha_\gamma$}
    \State $\textbf{d}_{i}^{(j)} \gets \textbf{d}^*$.
    \EndIf
\EndFor 
\EndFor
\State Center the samples and denote the centered particles by $\tilde{\textbf{c}}^{(j)},\tilde{\textbf{d}}_{1:n+1}^{(j)}$. %(2) Update the weights by the ratio of prior densities prior to and after centering step.
\State Update the weights $\tilde{w}^{(j)}\gets\tilde{w}^{(j)}\dfrac{N(\tilde{\textbf{c}}^{(j)};\textbf{0}_B,\Sigma_c)Dir(\tilde{\textbf{d}}_{1}^{(j)})\cdots Dir(\tilde{\textbf{d}}_{n+1}^{(j)})}{N({\textbf{c}}^{(j)};\textbf{0}_B,\Sigma_c)Dir({\textbf{d}}_{1}^{(j)})\cdots Dir({\textbf{d}}_{n+1}^{(j)})}$, and normalize.
\State Draw $\tilde{\sigma}^{2(j)}\sim IG(\frac{1}{2}(n+1)M+\alpha_\sigma,\frac{1}{2}\sum_{i=1}^{n+1}\sum_{m=1}^{M}(q_i(t_m)-(\tilde{q}_\mu^{(j)},\tilde{\gamma}_i^{(j)-1})(t_m))^2+\beta_\sigma)$.
\State where $\tilde{q}_\mu^{(j)} = \sum_{b=1}^B\tilde{c}_b^{(j)}\phi_b$ and $\tilde{\gamma}_i^{(j)} = d^{-1}(\tilde{\textbf{d}}_i^{(j)}),\ i = 1,\ldots,n+1$.
\EndFor
\State $ESS_n \gets (\sum_{j=1}^J(\tilde{w}^{(j)})^2)^{-1}$.
\Outputs{(1) Posterior draws given $f_{1:n+1}$: $\{(\tilde{\textbf{c}}^{(j)},\tilde{\textbf{d}}_{1:n+1}^{(j)},\tilde{\sigma}^{2(j)},\tilde{w}^{(j)}),\ j=1,\ldots,J\}$.\\
(2) Corresponding effective sample size $ESS_n$.}
\State Update $\{(\textbf{c}^{(j)},\textbf{d}_{1:n+1}^{(j)},\sigma^{2(j)},w^{(j)}),\ j=1,\ldots,J\} \gets \{(\tilde{\textbf{c}}^{(j)},\tilde{\textbf{d}}_{1:n+1}^{(j)},\tilde{\sigma}^{2(j)},\tilde{w}^{(j)}),\ j=1,\ldots,J\}$
\EndFor
\end{algorithmic}
\end{footnotesize}
\end{algorithm}

\newpage

\section{Additional registration results for SSS functions}

\begin{figure}[!t]
  \centering
  \begin{tabularx}{13.3cm} { 
  |>{\centering\arraybackslash}X 
   >{\centering\arraybackslash}X 
   >{\centering\arraybackslash}X 
   >{\centering\arraybackslash}X
   >{\centering\arraybackslash}X
   >{\centering\arraybackslash}X | }
    \hline
    \multicolumn{6}{|c|}{(a)}
    \\
    \begin{minipage}{2cm}
    \centering\includegraphics[width=2cm]{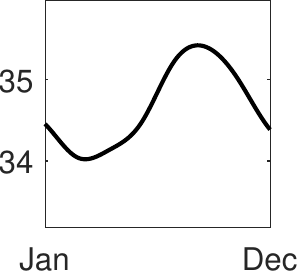}
    \end{minipage}
    &\begin{minipage}{2cm}
    \centering\includegraphics[width=2cm]{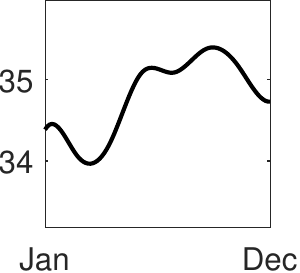}
    \end{minipage}
    &\begin{minipage}{2cm}
    \centering\includegraphics[width=2cm]{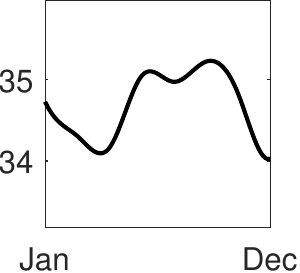}
    \end{minipage}
    &\begin{minipage}{2cm}
    \centering\includegraphics[width=2cm]{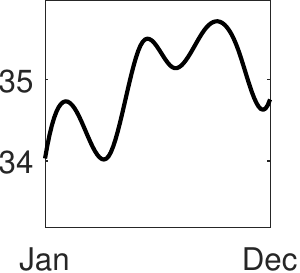}
    \end{minipage}
    &\begin{minipage}{2cm}
    \centering\includegraphics[width=2cm]{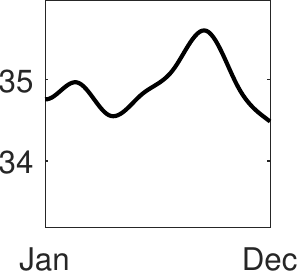}
    \end{minipage}
    &\begin{minipage}{2cm}
    \centering\includegraphics[width=2cm]{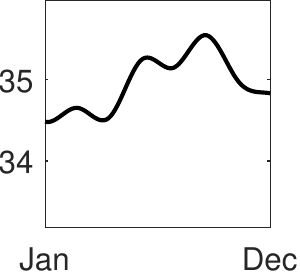}
    \end{minipage}
    \\\hline
    \multicolumn{6}{|c|}{(b)}
    \\
    \begin{minipage}{2cm}
    \centering\includegraphics[width=2cm]{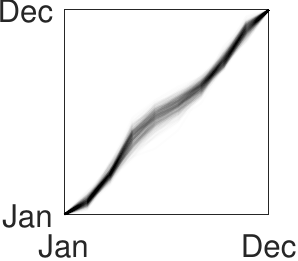}
    \end{minipage}
    &\begin{minipage}{2cm}
    \centering\includegraphics[width=2cm]{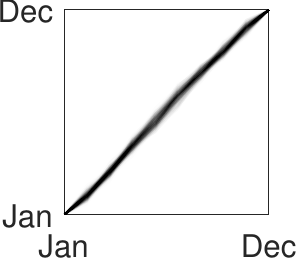}
    \end{minipage}
    &\begin{minipage}{2cm}
    \centering\includegraphics[width=2cm]{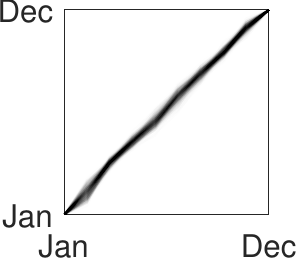}
    \end{minipage}
    &\begin{minipage}{2cm}
    \centering\includegraphics[width=2cm]{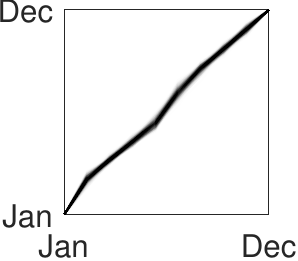}
    \end{minipage}
    &\begin{minipage}{2cm}
    \centering\includegraphics[width=2cm]{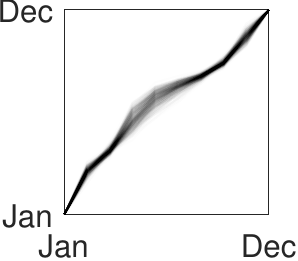}
    \end{minipage}
    &\begin{minipage}{2cm}
    \centering\includegraphics[width=2cm]{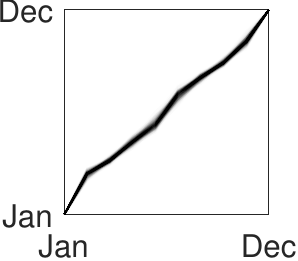}
    \end{minipage}
    \\\hline
    %\multicolumn{3}{|c|}{(c) }
    %\\\hline
    %\multicolumn{1}{|c}{}
    \multicolumn{2}{|c|}{(c)}&\multicolumn{2}{|c|}{(d)}&\multicolumn{2}{|c|}{(e)}\\
    \multicolumn{2}{|c|}{\begin{minipage}{3.5cm}
    \centering\includegraphics[width=3.5cm]{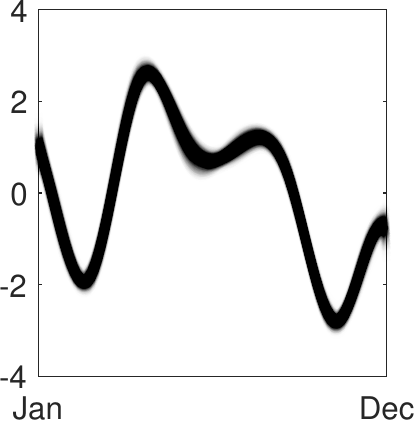}
    \end{minipage}}
    &
    \multicolumn{2}{|c|}{\begin{minipage}{3.5cm}
    \centering\includegraphics[width=3.5cm]{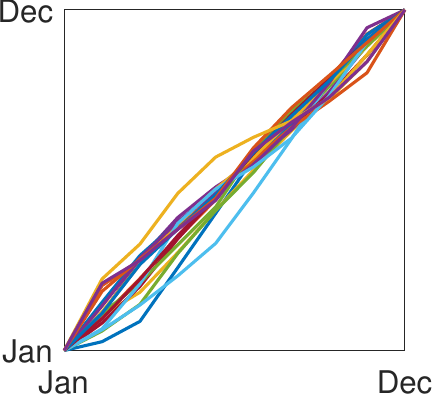}
    \end{minipage}}
    &\multicolumn{2}{|c|}{\begin{minipage}{3.5cm}
    \centering\includegraphics[width=3.5cm]{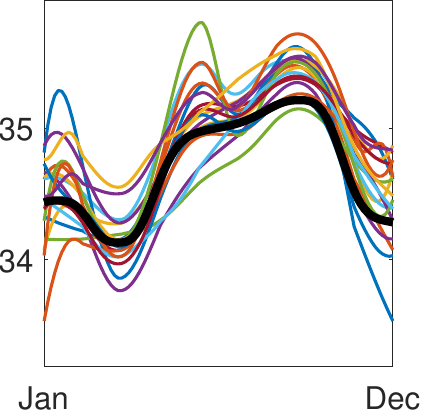}
    \end{minipage}}
    \\\hline
  \end{tabularx}
  \caption{Sequential Bayesian registration of SSS functions. (a) Trajectories of annual SSS near Null Island from 2012 to 2017. (b) Weighted posterior samples of the phase components for the functions displayed in (a). (c) Weighted posterior samples of the template SRVF from the target posterior distribution given SSS functions from 2000 to 2017. Transparency in (b) and (c) reflects the magnitude of the weights. (d) Marginal posterior mean phases for the 18 SSS functions. (e) SSS functions (for all years, solid colors) registered via the phase components in (d) and marginal posterior mean template in the original function space (black).}
  \label{fig:real2}
\end{figure}

%\textcolor{green}{[YK: copy and pasted the previous version of the SSS study from the main text. I think we can delete this section.]}

Here, we present registration results, generated via the proposed sequential approach, for the same annual SSS functional data as in the main manuscript, but without stratification into different ENSO states. The total number of SSS functions is 18. The hyperparameters in the Bayesian model are the same as those used in Section 5.5 in the main manuscript. We initialize the SMC algorithm with $10,000$ MCMC samples from the posterior distribution given $f_{1:10}$ after a burn-in of $40,000$ iterations.
The rest of the SSS functions, $f_{11:18}$, are added to the data sequentially, and the initial weighted samples are updated recursively.
%; thus, the data in this case is $f_{1:18}$, with ea which is displayed in Figure \ref{fig:real2}(a). The SSS functions share a common pattern of local extrema with clear phase variation. In a previous study, \cite{bingham2021} approximated seasonal variability across years through a one-dimensional measurement of phase, corresponding to the time at which maximum SSS magnitude is observed. This is akin to landmark-based registration of the functions, with the point of maximum SSS magnitude on each function serving as a single landmark. This approach is effective at registering the landmarks, but it fails to register other prominent features of the functions. Thus, to estimate more flexible and informative relative phase components of the SSS functional data, we apply the proposed sequential Bayesian registration approach.

%The function observations of sea surface salinity (are expected to) fluctuate smoothly across a year with no more than three peaks. Thus, we assume that the common template is modeled as a linear combination of ten B-spline bases with unknown coefficients.
%Prior hyperparameters for the template and phase components are: $B=10,\ \Sigma_q = 20I_B,\ \kappa = 5,\ \alpha_\sigma=4,\ \beta_\sigma=0.01$, and a uniform partition of size $M_\gamma=10$. 

Features of the target posterior density, $\pi(\cdot\mid f_{1:18})$, are visualized in Figure \ref{fig:real2}.
In panel (b), we show weighted samples of (a subset of) the phase components (transparency of each sample reflects the magnitude of the corresponding weight), given the annual SSS functions $f_{1:18}$, for the subset of SSS functions displayed in panel (a). Since the majority of the annual SSS functions have two distinct peaks, the marginal posterior uncertainty (or band) of a phase component is greater when there is only one distinct peak in the given function (columns 1 and 5). Panel (c) displays the marginal posterior over the template function in the SRVF space $\mathcal{Q}$. The uncertainty in the template appears to be very similar throughout the domain (January to December).
%In particular, the template reflects relatively high SSS from May to November with a maximum around October, while the lowest SSS occurs around March. 
The marginal posterior means of the phase components for all 18 SSS functions are shown in panel (d). 
%It appears that there is more phase variation from January to the first valley. 
%Further, the relative phase components of SSS functions for years 2000, 2005 and 2017 (yellow, dark blue and light blue functions in panel (d)) are much further from the identity than the relative phase components of SSS functions for the other years, which are SSS functions with different numbers of peaks and valleys compared to the majority of others.
Finally, panel (e) shows the registered SSS functions via the posterior mean phase components shown in (d). Based on this result, it is evident that most of the SSS functions have two dominant peaks, the first one around June and the second one around October. The amplitude variability in the second peak is smaller relative to the amplitude variability in the first peak. There is also considerable amplitude variability at the beginning of the year near the valley. The posterior mean template function shown in black clearly captures the prominent features of the given SSS functional data. 
%Similarly, the phase variability is smaller near the second peak. Overall, there is more phase and amplitude variability in the first half of the calendar year and the last prominent peak may be more useful to be used as a landmark for the approach in \cite{bingham2021}.

\begin{figure}[!t]
\graphicspath{{figures/real3_ecg}}
  \centering
  \begin{tabularx}{13.3cm} { 
  | c|>{\centering\arraybackslash}X 
   >{\centering\arraybackslash}X 
   >{\centering\arraybackslash}X 
   >{\centering\arraybackslash}X
   >{\centering\arraybackslash}X| }
    \hline
    &$f_1$ & $f_2$ & $f_3$ & $f_4$ & $f_5$
    \\
    \rotatebox[origin=c]{90}{(a)} &\begin{minipage}{2cm}
    \centering\includegraphics[width=2cm]{forig1.pdf}
    \end{minipage}
    &\begin{minipage}{2cm}
    \centering\includegraphics[width=2cm]{forig2.pdf}
    \end{minipage}
    &\begin{minipage}{2cm}
    \centering\includegraphics[width=2cm]{forig3.pdf}
    \end{minipage}
    &\begin{minipage}{2cm}
    \centering\includegraphics[width=2cm]{forig4.pdf}
    \end{minipage}
    &\begin{minipage}{2cm}
    \centering\includegraphics[width=2cm]{forig5.pdf}
    \end{minipage}
    \\\hline
    &$f_6$ & $f_7$ & $f_8$ & $f_9$ & $f_{10}$
    \\
    \rotatebox[origin=c]{90}{(a)}&\begin{minipage}{2cm}
    \centering\includegraphics[width=2cm]{forig6.pdf}
    \end{minipage}
    &\begin{minipage}{2cm}
    \centering\includegraphics[width=2cm]{forig7.pdf}
    \end{minipage}
    &\begin{minipage}{2cm}
    \centering\includegraphics[width=2cm]{forig8.pdf}
    \end{minipage}
    &\begin{minipage}{2cm}
    \centering\includegraphics[width=2cm]{forig9.pdf}
    \end{minipage}
    &\begin{minipage}{2cm}
    \centering\includegraphics[width=2cm]{forig10.pdf}
    \end{minipage}
    \\
    \rotatebox[origin=c]{90}{(b)}&
    &
    &
    &\begin{minipage}{2cm}
    \centering\includegraphics[width=2cm]{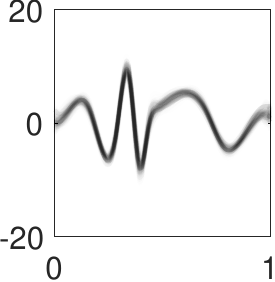}
    \end{minipage}
    &\begin{minipage}{2cm}
    \centering\includegraphics[width=2cm]{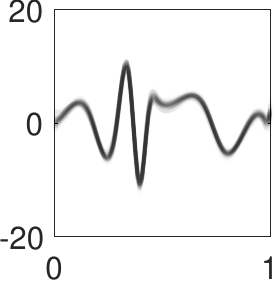}
    \end{minipage}
    \\ 
    \rotatebox[origin=c]{90}{(c)}&
    &
    &
    &\begin{minipage}{2cm}
    \centering\includegraphics[width=2cm]{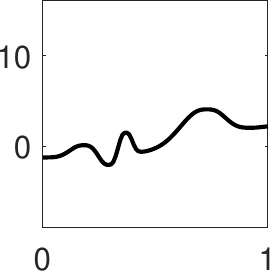}
    \end{minipage}
    &\begin{minipage}{2cm}
    \centering\includegraphics[width=2cm]{fmean10.pdf}
    \end{minipage}
    \\\rotatebox[origin=c]{90}{(d)}&&&&0.3941&0.2381
    \\\hline
    &$f_{11}$ & $f_{12}$ & $f_{13}$ & $f_{14}$ & $f_{15}$
    \\
    \rotatebox[origin=c]{90}{(a)}&\begin{minipage}{2cm}
    \centering\includegraphics[width=2cm]{forig11.pdf}
    \end{minipage}
    &\begin{minipage}{2cm}
    \centering\includegraphics[width=2cm]{forig12.pdf}
    \end{minipage}
    &\begin{minipage}{2cm}
    \centering\includegraphics[width=2cm]{forig13.pdf}
    \end{minipage}
    &\begin{minipage}{2cm}
    \centering\includegraphics[width=2cm]{forig14.pdf}
    \end{minipage}
    &\begin{minipage}{2cm}
    \centering\includegraphics[width=2cm]{forig15.pdf}
    \end{minipage}
    \\
    \rotatebox[origin=c]{90}{(b)}&\begin{minipage}{2cm}
    \centering\includegraphics[width=2cm]{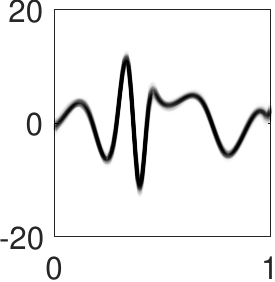}
    \end{minipage}
    &\begin{minipage}{2cm}
    \centering\includegraphics[width=2cm]{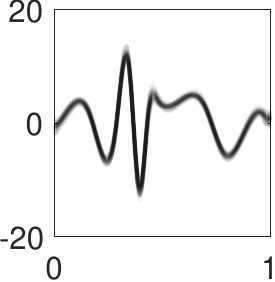}
    \end{minipage}
    &\begin{minipage}{2cm}
    \centering\includegraphics[width=2cm]{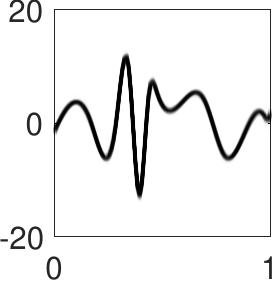}
    \end{minipage}
    &\begin{minipage}{2cm}
    \centering\includegraphics[width=2cm]{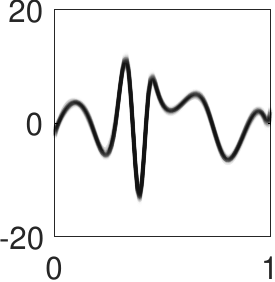}
    \end{minipage}
    &\begin{minipage}{2cm}
    \centering\includegraphics[width=2cm]{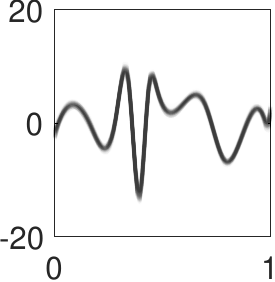}
    \end{minipage}
    \\ 
    \rotatebox[origin=c]{90}{(c)}&\begin{minipage}{2cm}
    \centering\includegraphics[width=2cm]{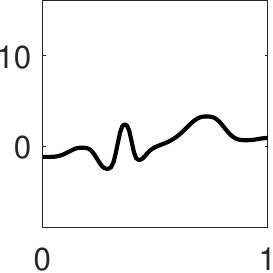}
    \end{minipage}
    &\begin{minipage}{2cm}
    \centering\includegraphics[width=2cm]{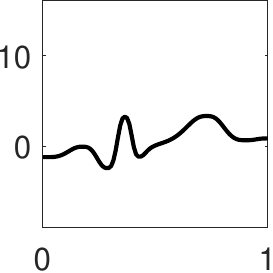}
    \end{minipage}
    &\begin{minipage}{2cm}
    \centering\includegraphics[width=2cm]{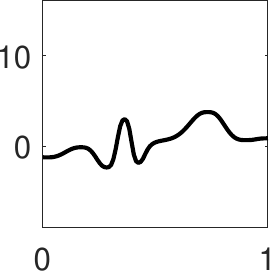}
    \end{minipage}
    &\begin{minipage}{2cm}
    \centering\includegraphics[width=2cm]{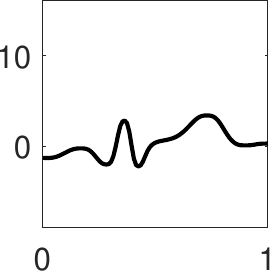}
    \end{minipage}
    &\begin{minipage}{2cm}
    \centering\includegraphics[width=2cm]{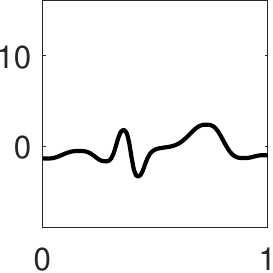}
    \end{minipage}
    \\\rotatebox[origin=c]{90}{(d)}& 0.1832  &  0.2270 &0.0804 &   0.1071   & 0.1235
    \\\hline
  \end{tabularx}
  \caption{(a) 15 PQRST complexes. (b) Marginal posterior samples of the template function in the SRVF space $\mathcal{Q}$ and (c) marginal posterior mean of the template function in the original function space $\mathcal{F}$ given functions 1 through $n$ where $f_n$ is the function above. (d) Marginal posterior variance for the template function.}
  \label{fig:ecg_15functions}
\end{figure}

\section{Additional registration results for segmented PQRST complexes}

We illustrate the evolution of the marginal posterior density of the template function as new data arrives. For this, we use the segmented PQRST complexes that were considered in Section 5.6 in the main article. All modeling settings used in this section are the same as in Section 5.6 in the main article. The first 8 functions in Figure \ref{fig:ecg_15functions}(a) are used to initialize the weighted samples via MCMC-based batch method. Then, functions 9-15 are added sequentially to the data one at a time, and the target posterior distribution over all parameters in the model is updated using the proposed sequential learning approach. Panel (b) shows the evolution of the marginal posterior distribution of the template function in the SRVF space $\mathcal{Q}$. We use spaghetti plots with transparency reflecting magnitude of weights in the weighted samples. Panel (c) shows the evolution of the marginal posterior mean of the template function in the original function space $\mathcal{F}$. Panel (d) reports the overall marginal posterior variance of the template function. 

Based on Figure \ref{fig:ecg_15functions}, we note some interesting changes in the marginal posterior over the template function as new data is added. First, the template appears to undergo significant shape changes when PQRST complex $f_{11}$ is added to the data and then again when PQRST complex $f_{13}$ is added. Overall, marginal posterior variance tends to decrease and PQRST features tend to become more prominent as more data is assimilated. When $f_{11}$ is added, the R peak in the template function becomes sharper since $f_{11}$ has a very sharp R peak. When $f_{13}$ with a very deep S valley is added, the S valley in the template function also becomes deeper. The PQRST complex $f_{15}$ has a relatively low R peak and a high T peak causing the template to change accordingly.

\section{Effective Sample Size Comparisons}

\begin{figure}[!t]
    \centering
    \begin{tabular}{cc}
    (a)&(b)\\
    \includegraphics[width=0.3\linewidth]{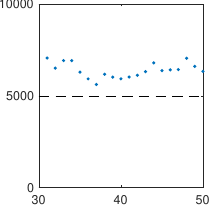}&\includegraphics[width=0.3\linewidth]{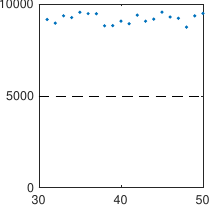}\\
    \end{tabular}
    \caption{ESS of weighted samples for a sequence of posterior distributions, $\pi(\cdot\mid f_{1:n}),\ n=31,\ldots,50$, with $n$ on $x$-axis and ESS on $y$-axis. (a) Real data example 1: drought intensity. (b) Real data example 3: segmented PQRST complexes.}
    \label{fig:real1_ess}
\end{figure}

Next, for all simulated and real data examples, we compare the effective sample size (ESS) for the proposed sequential Bayesian registration algorithm and MCMC-based batch learning. Computation of ESS based on MCMC samples depends on the sample autocorrelation \cite{ess}. In all cases, we use 10,000 posterior (weighted) samples. For all five examples (two simulated and three based on real data), MCMC resulted in ESS between 4,900 and 5,100. The ESSs based on our approach are as follows. For simulated example 1, the evolution of ESS for a sequence of posterior distributions, $\pi(\cdot|f_{1:n})$,\  $n=31,\ldots,100$ is presented in Figure 3(f) in the main manuscript, and is well above 5,000 for most values of $n$. For simulated example 2, the effective sample sizes of weighted samples for a sequence of posterior distributions, given $f_{1:4}, \ldots, f_{1:7}$, are 9,967, 9,960, 9,976, 10,000, respectively. The evolutions of ESS for real data examples 1 and 3 (drought intensity and segmented PQRST complexes) are shown in panels (a) and (b) in Figure \ref{fig:real1_ess}, respectively. For real data example 2 (sea surface salinity), the ESS ranges from 8,246 to 9,845. These results show that ESS resulting from the proposed algorithm compares favorably to ESS based on MCMC posterior samples.

\begin{figure}[!t]
    \centering
    \begin{tabular}{cc}
    (a)&(b)\\
    \includegraphics[width=0.3\linewidth]{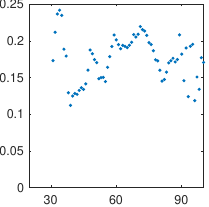}&\includegraphics[width=0.3\linewidth]{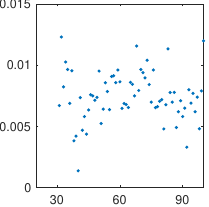}\\
    \end{tabular}
    \caption{Estimated mean squared errors for the posterior mean over (a) the template basis coefficients $\textbf{c}$ and (b) phase increments $\mathbf{d}_{1:n}$ given functions $f_{1:n}$, plotted against the sample size $n$.}
    \label{fig:sim1_accuracy}
\end{figure}

Additionally, Figure \ref{fig:sim1_accuracy} shows estimated mean squared errors of posterior mean template basis coefficients and phase increments plotted against the number of assimilated functions. This shows that estimation accuracy, based on a fixed number of weighted samples, does not decrease with sample size, and may even show a small improvement as the number of sequentially observed functions increases.

\section{Sensitivity Analyses}

For the sensitivity analyses presented in subsequent sections, we use the same data, albeit with sample size 50 rather than 100, and hyperparameter settings as in Simulated Example 1 in Section 5.2 in the main article. While omitted here, in addition to the presented sensitivity analyses, we also explored sensitivity of posterior inference to the choice of $\kappa$ (concentration hyperparamter in the prior for phase components), and the size of the discretization grid for the functional data; in both cases, results showed very little sensitivity under reasonable choices for these parameters.
The marginal posterior variances of the template function are computed using the weighted sample average of the squared eFR distances of each particle from the marginal posterior mean and those of the phase components are obtained using the weighted sample average of the squared $\mathbb{L}^2$ distances of each particle from the marginal posterior mean. 
%number of evaluated points in a function) and showed that the estimation results are not very sensitive to these choices as long as $\kappa$ is small enough $(<5)$ to make the prior on phase components non-informative and $M$ is large enough ($>50$) to represent a function with discretized points.

\subsection{Sensitivity to $\Sigma_c$}
\label{sensitivity_sigmac}

We first assess sensitivity of posterior inference to four different choices of the hyperparameter $\Sigma_c$: (a) $\Sigma_c = 10 \textbf{I}_M,$  (b) $\Sigma_c = 20 \textbf{I}_M,$ (c) $\Sigma_c = 50 \textbf{I}_M,$ and (d) $\Sigma_c$ is a diagonal matrix with diagonal elements $100\left(1,\frac{1}{2^2},\frac{1}{3^2},\ldots,\frac{1}{M^2}\right)$, $M=8$. Again, initialize the sequential algorithm using an MCMC sample of size 10,000 from the posterior distribution given $f_{1:30}$ after burn-in, and recursively update the posterior weighted samples.

\begin{figure}[!t]
\graphicspath{{figure_type2/sensitivity}}
\centering
  \begin{tabularx}{13.3cm} { 
  |>{\centering\arraybackslash}X 
   >{\centering\arraybackslash}X 
   >{\centering\arraybackslash}X 
   >{\centering\arraybackslash}X| }
    \hline
    (a)& (b)& (c)& (d)\\
    \begin{minipage}{3cm}
    \centering\includegraphics[width=3cm]{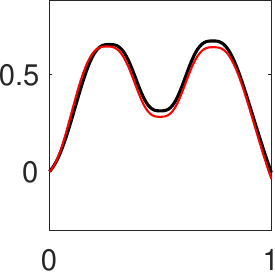}
    \end{minipage}
    &\begin{minipage}{3cm}
    \centering\includegraphics[width=3cm]{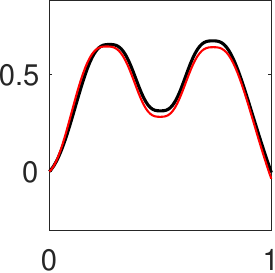}
    \end{minipage}
    &\begin{minipage}{3cm}
    \centering\includegraphics[width=3cm]{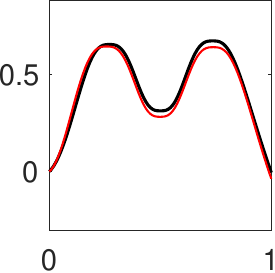}
    \end{minipage}
    &\begin{minipage}{3cm}
    \centering\includegraphics[width=3cm]{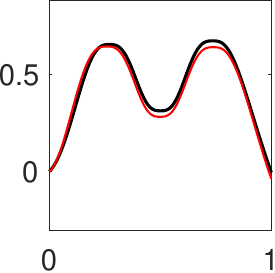}
    \end{minipage}
    \\\hline
    \begin{minipage}{3cm}
    \centering\includegraphics[width=3cm]{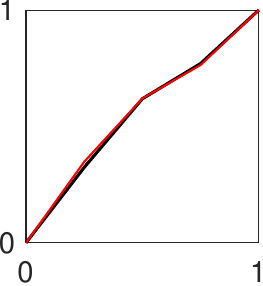}
    \end{minipage}
    &\begin{minipage}{3cm}
    \centering\includegraphics[width=3cm]{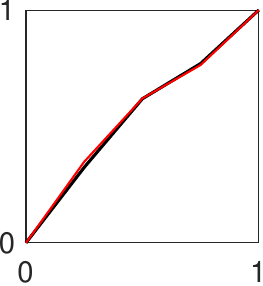}
    \end{minipage}
    &\begin{minipage}{3cm}
    \centering\includegraphics[width=3cm]{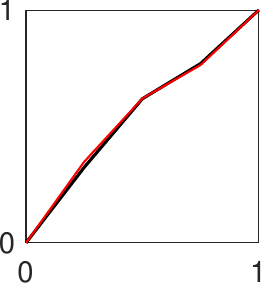}
    \end{minipage}
    &\begin{minipage}{3cm}
    \centering\includegraphics[width=3cm]{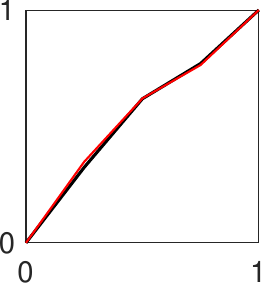}
    \end{minipage}
    \\\hline
    \graphicspath{{figures/sensitivity}}
    \begin{minipage}{3cm}
    \centering\includegraphics[width=3cm]{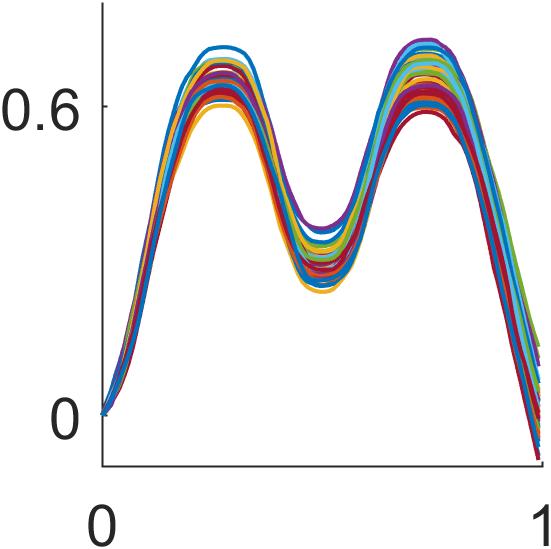}
    \end{minipage}
    &\graphicspath{{figures/sensitivity}}
    \begin{minipage}{3cm}
    \centering\includegraphics[width=3cm]{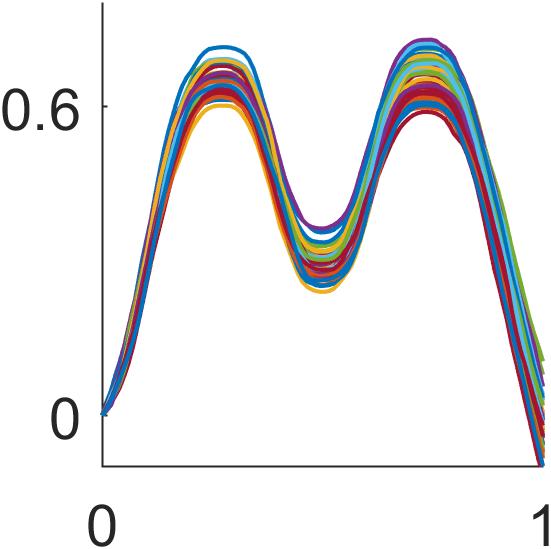}
    \end{minipage}
    &\graphicspath{{figures/sensitivity}}
    \begin{minipage}{3cm}
    \centering\includegraphics[width=3cm]{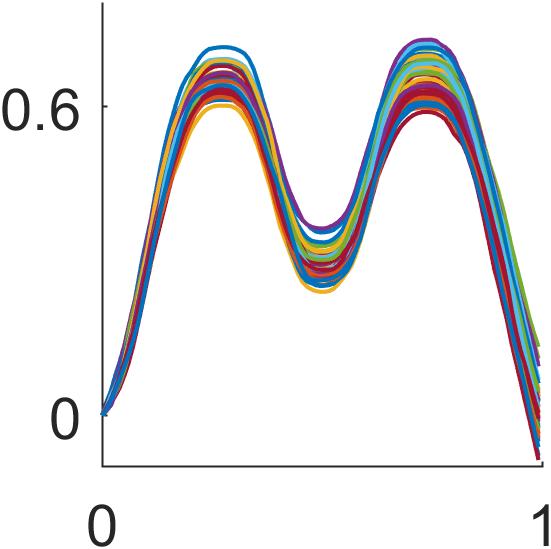}
    \end{minipage}
    &\graphicspath{{figures/sensitivity}}
    \begin{minipage}{3cm}
    \centering\includegraphics[width=3cm]{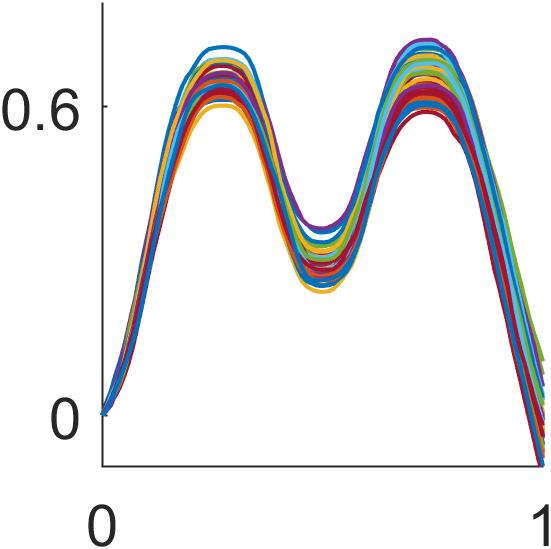}
    \end{minipage}\\
    \hline
  \end{tabularx}
  \caption{Sensitivity of posterior inference to different choices of $\Sigma_c$ based on Simulated Example 1. The first and second rows show the marginal posterior means of the template function and the phase component $\gamma_1$ in black, given $f_{1:50}$, respectively, with the ground truth values shown in red. The third row shows the registered simulated functions using the marginal posterior means of phase components. We consider four different choices of $\Sigma_c$: (a) $\Sigma_c = 10 \textbf{I}_M,$  (b) $\Sigma_c = 20 \textbf{I}_M,$ (c) $\Sigma_c = 50 \textbf{I}_M,$ and (d) $\Sigma_c$ is a diagonal matrix with diagonal elements $100\left(1,\frac{1}{2^2},\frac{1}{3^2},\ldots,\frac{1}{M^2}\right),\ M=8$.}
    \label{fig:Sigma_c}
\end{figure}

Posterior summaries based on the weighted particles of the template function and the relative phase components are illustrated in Figure \ref{fig:Sigma_c}. Panels (a)-(d) correspond to the four different settings for $\Sigma_c$. In the first row, we show the marginal posterior mean of the template function in black and the ground truth template function in red. %``hot'' colors correspond to high pointwise marginal posterior uncertainty and ``cold'' colors correspond to low pointwise marginal posterior uncertainty; the band represents (scaled) pointwise marginal posterior standard deviation around the mean.
%The first row visualizes the marginal posterior of the template function.
%The colored plot is the estimated marginal posterior mean where the color across the domain represents the pointwise marginal posterior standard error and the corresponding colormap is illustrated below.
%Also, we visualize the posterior uncertainty using band plots in grat which is computed by adding/subtracting 15 times the  to the marginal posterior mean.
Visually, the four results are indistinguishable. We can also quantify the difference in shape between the four posterior mean estimates by computing pairwise $\mathbb{L}^2$ distances between the (scaled to have norm one) posterior mean SRVF templates. All pairwise distances are very small ($<0.0015$). Finally, the overall marginal posterior variances for each of the four estimation results are all very similar, between 0.000080 and 0.000083. The second row in Figure \ref{fig:Sigma_c} compares the marginal posterior mean of the first phase component $\gamma_1$; the marginal posterior variances are very small $(<10^{-4})$ in all cases and are not shown. Again, the marginal posterior means are visually very similar. The pairwise $\mathbb{L}^2$ distances between the four estimates are all less than 0.00007.
%The estimates of marginal variance derived using the $\mathbb{L}^2$ distance are all less than $10^{8}$ and have very small differences as well.
The third row in Figure \ref{fig:Sigma_c} illustrates registration of the simulated functional data using the marginal posterior means of phase components. The registration results appear very similar. Thus, we conclude that posterior inference is not sensitive to the choice of $\Sigma_c$.
%as long as the diagonal elements are not too small to make the prior on $\textbf{c}$ informative.

%We can measure the difference among the estimation results by computing the distance between normalized estimates of the marginal posterior mean to account for the scale of the posteriors, i.e., $\left\|\frac{\hat{q}_{1}}{\|\hat{q}_{1}\|}-\frac{\hat{q}_{2}}{\|\hat{q}_{2}\|}\right\| (\leq \sqrt{2})$, where $\hat{q}_{1}$ and $\hat{q}_{2}$ are two estimates of the marginal posterior mean.
%Such distances among four estimation results are all less than 0.0015.

%(We can also visualize the estimates of the pointwise marginal posterior variance as for the template function, but since their values are very small, we need to scale up a lot and would look like the estimation results are quite different from each other, while their differences are very small. We can barely notice the difference when scaling the pointwise marginal standard error by *100.)

%\begin{figure}[!t]
%    \centering
%    \includegraphics[width=3cm]{/orig_func.jpg}
%    \caption{Generated functions in Simulated Example 1.}
%    \label{fig:orig_func}
%\end{figure}

\subsection{Sensitivity to $M_\gamma$}

\begin{figure}[!t]
\graphicspath{{figure_type2/sensitivity}}
\centering
  \begin{tabularx}{13.3cm} { 
  |>{\centering\arraybackslash}X 
   >{\centering\arraybackslash}X 
   >{\centering\arraybackslash}X 
   >{\centering\arraybackslash}X| }
    \hline
    (a)& (b)& (c)& (d)\\
    \begin{minipage}{3cm}
    \centering\includegraphics[width=3cm]{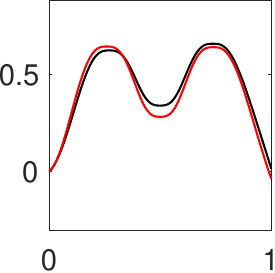}
    \end{minipage}
    &\begin{minipage}{3cm}
    \centering\includegraphics[width=3cm]{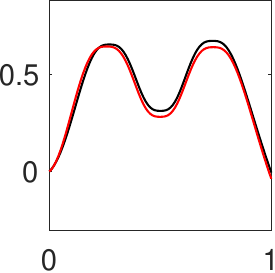}
    \end{minipage}
    &\begin{minipage}{3cm}
    \centering\includegraphics[width=3cm]{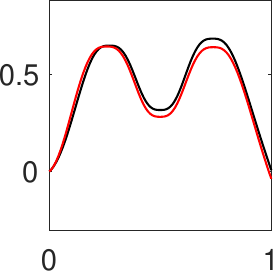}
    \end{minipage}
    &\begin{minipage}{3cm}
    \centering\includegraphics[width=3cm]{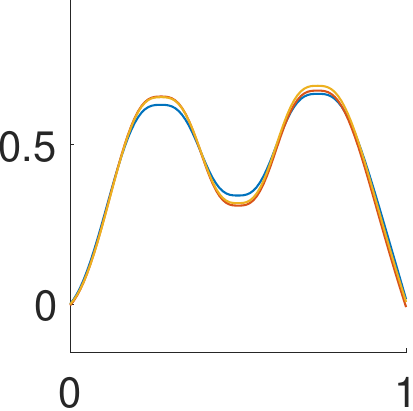}
    \end{minipage}
    \\\hline
    \begin{minipage}{3cm}
    \centering\includegraphics[width=3cm]{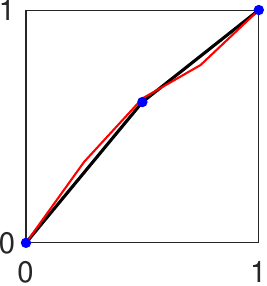}
    \end{minipage}
    &\begin{minipage}{3cm}
    \centering\includegraphics[width=3cm]{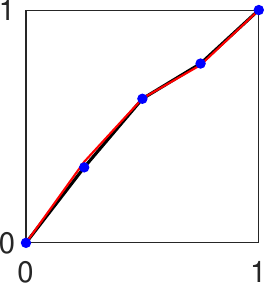}
    \end{minipage}
    &\begin{minipage}{3cm}
    \centering\includegraphics[width=3cm]{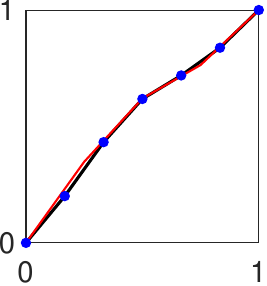}
    \end{minipage}
    &\begin{minipage}{3cm}
    \centering\includegraphics[width=3cm]{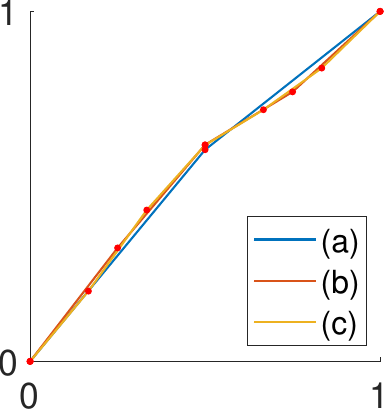}
    \end{minipage}
    \\\hline
    \graphicspath{{figures/sensitivity}}
    \begin{minipage}{3cm}
    \centering\includegraphics[width=3cm]{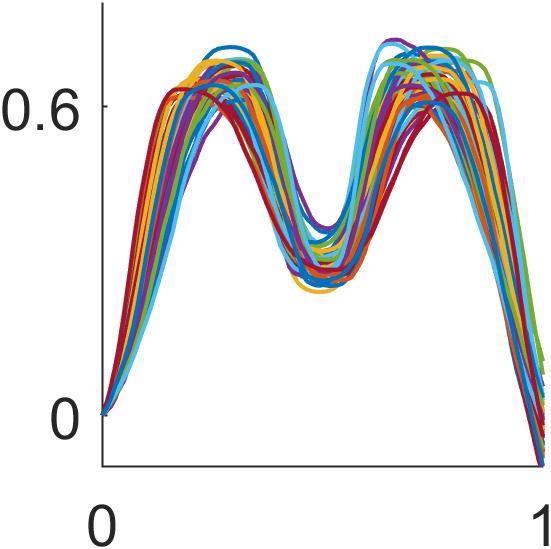}
    \end{minipage}
    &\graphicspath{{figures/sensitivity}}
    \begin{minipage}{3cm}
    \centering\includegraphics[width=3cm]{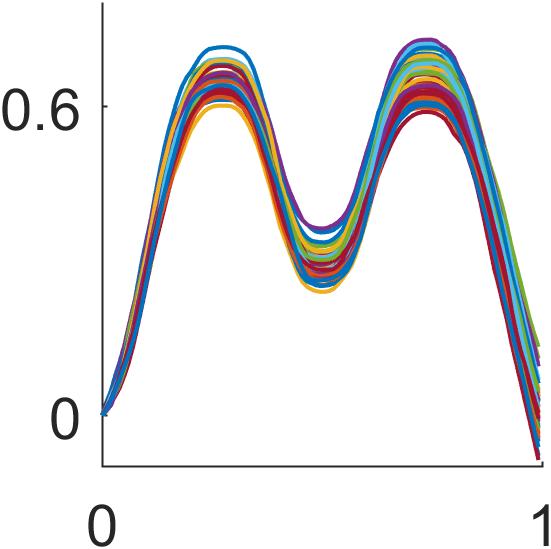}
    \end{minipage}
    &\graphicspath{{figures/sensitivity}}
    \begin{minipage}{3cm}
    \centering\includegraphics[width=3cm]{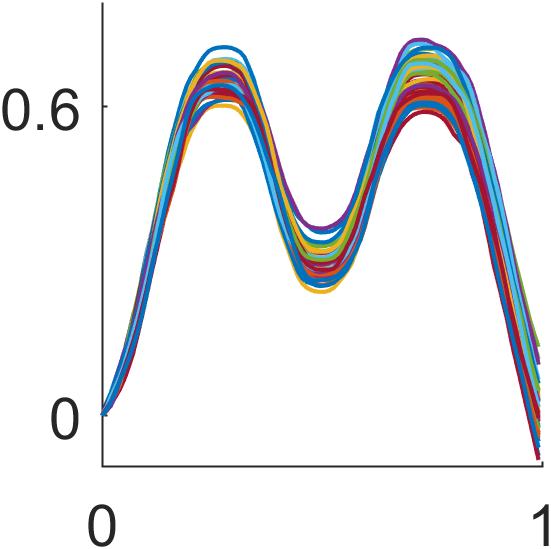}
    \end{minipage}
    &\\
    \hline
  \end{tabularx}
  \caption{Sensitivity of posterior inference to different choices of the partition size, $M_\gamma$, for the phase components based on Simulated Example 1. The first and second rows show the marginal posterior means of the template function and the phase component $\gamma_1$ in black, given $f_{1:50}$, respectively, with the ground truth values shown in red and the blue points indicating the partition. The third row shows the registered simulated functions via the marginal posterior means of phase components. We consider three different choices of $M_\gamma$: (a) $M_\gamma = 3$, (b) $M_\gamma = 5$, and (c) $M_\gamma = 7$. (d) Posterior mean summaries from panels (a)-(c) in blue, red and yellow, respectively, displayed in a single plot.}
  \label{fig:gamma_partition}
\end{figure}

Next, we assess sensitivity of posterior inference to different choices of the partition size $M_\gamma$ for the phase components; we consider (a) $M_\gamma = 3$, (b) $M_\gamma = 5$, and (c) $M_\gamma = 7$. Note that we used $M_\gamma=5$ to generate the simulated data. Marginal posterior summaries of the template and first phase component are displayed in the same manner as in Section \ref{sensitivity_sigmac} in the first two rows of Figure \ref{fig:gamma_partition}. In the first row, we notice that the marginal posterior means over the template function in panels (b) and (c) are very close to each other. %pointwise marginal posterior standard deviation around the template function tends to decrease as the size of the partition increases. 
The estimates of the overall marginal posterior variance for the template function are (a) 0.000341, (b) 0.000082, and (c) 0.000076; the marginal posterior variance is noticeably greater when the partition size is small ($M_\gamma = 3$, i.e., each phase component contains only two linear pieces). The three marginal posterior means of the template function are visualized together in panel (d) in the first row ((a) blue, (b) red, (c) yellow) and all appear very similar. The pairwise distances between their shapes are all very small: distances between (a) and (b), and (a) and (c) are approximately 0.0350, and distance between (b) and (c) is approximately 0.0167. The estimates of the marginal posterior mean of the first phase component, for each choice of partition size (indicated by blue markers), are shown in the second row (a)-(c). In panel (d), we show the four marginal posterior means in the same plot ((a) blue, (b) red, (c) yellow). Again, the results appear very similar for each setting of partition size. The pairwise $\mathbb{L}^2$ distances between the phase components shown in panels (a) and (b), and (a) and (c) are approximately 0.0167, while the distance between the phase components shown in (b) and (c) is 0.0063.
%While they do have differences depending on their partitions, their overall structures are very close to each other.
%The $\mathbb{L}^2$ distance between the estimates of the marginal posterior mean of (b) and (c) is 0.0063 and the distances from (a) to (b) and (c) is around 0.0167.
%Since (a) assumes a smaller partition size than the ground truth, the estimate result is the farthest among three estimation results. 
Finally, the overall marginal posterior variances of the first phase component are (a) 0.000008, (b) 0.000006, and (c) 0.000012, i.e., the smallest marginal posterior variance is achieved when $M_\gamma=5$, which was used to generate the simulated functional data. The third row in Figure \ref{fig:gamma_partition} illustrates registration of the simulated functions using the marginal posterior means of the phase components. The results appear very similar in panels (b) and (c). The horizontal synchronization in panel (a) is not as good due to the partition size $M_\gamma$ being too small. Overall, posterior inference is not very sensitive to the partition size $M_\gamma$ of the phase components, as long as it is not too small. 

%The marginal posterior variances of the phase component obtained using the $\mathbb{L}^2$ distance metric are the smallest when setting the partition size the same as the ground truth value $M_\gamma = 5$, while the differences are very small for all choices of $M_\gamma$.
%For example, 
%As the estimates of the marginal posterior mean of phase components in (b) and (c) are very close to each other, the visualization of the corresponding registration result in the third row in Figure \ref{fig:gamma_partition} are also very similar and shows that the features of given functions are nicely aligned to each other across the domain.
%While (a) assumes a smaller partition size than the ground truth partition size, the registration of given functions achieved by composing the marginal posterior means of phase components show moderate amount of alignment of important features of functions across the domain.
%The presented simulation study show that the target posterior is not very sensitive to the partition size of phase components as long as partition size is not too small. And even when there is only two linear pieces for the phase components, we are able to achieve moderate amount of registration through the marginal posterior means of phase components.

\subsection{Sensitivity to the number and type of basis functions in the prior for the SRVF template}

\begin{figure}[!t]
\graphicspath{{figure_type2/sensitivity}}
  \centering
  \begin{tabularx}{13.3cm} { 
  |>{\centering\arraybackslash}X 
   >{\centering\arraybackslash}X 
   >{\centering\arraybackslash}X| }
    \hline
    (a)& (b)& (c)\\
    \begin{minipage}{3cm}
    \centering\includegraphics[width=3cm]{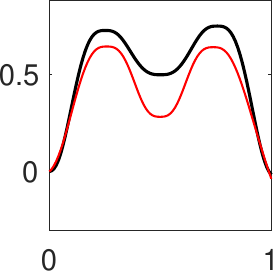}
    \end{minipage}
    &\begin{minipage}{3cm}
    \centering\includegraphics[width=3cm]{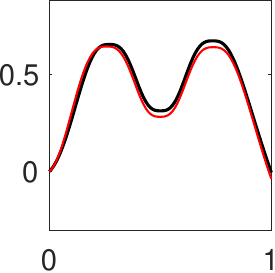}
    \end{minipage}
    &\begin{minipage}{3cm}
    \centering\includegraphics[width=3cm]{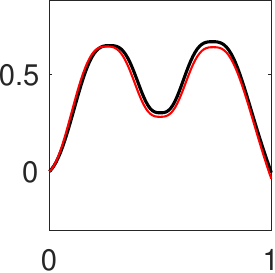}
    \end{minipage}
    \\\hline
    \begin{minipage}{3cm}
    \centering\includegraphics[width=3cm]{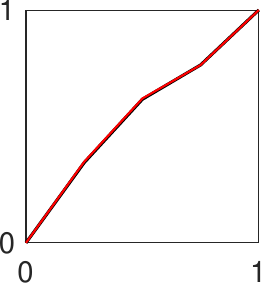}
    \end{minipage}
    &\begin{minipage}{3cm}
    \centering\includegraphics[width=3cm]{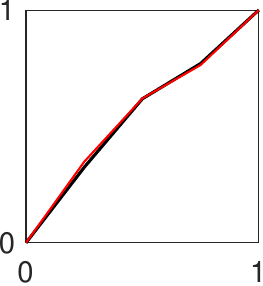}
    \end{minipage}
    &\begin{minipage}{3cm}
    \centering\includegraphics[width=3cm]{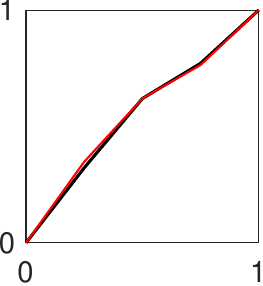}
    \end{minipage}\\\hline
    \begin{minipage}{3cm}\graphicspath{{figures/sensitivity}}
    \centering\includegraphics[width=3cm]{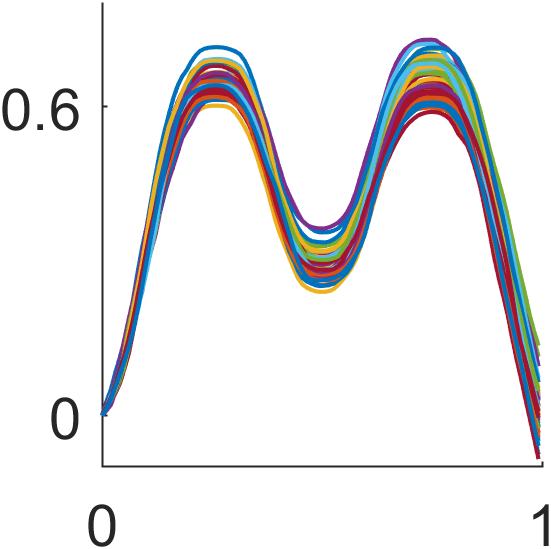}
    \end{minipage}
    &\begin{minipage}{3cm}\graphicspath{{figures/sensitivity}}
    \centering\includegraphics[width=3cm]{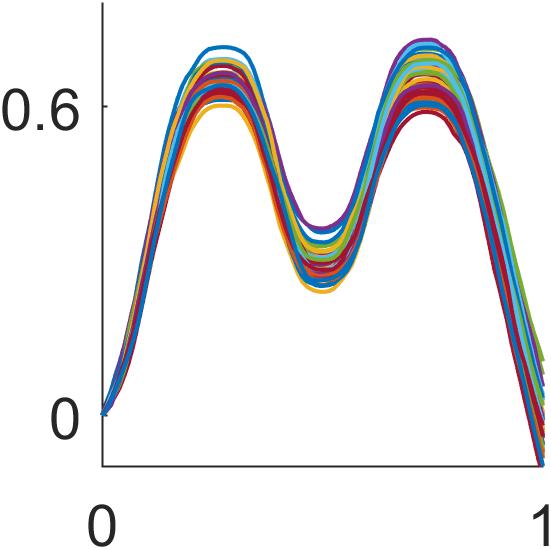}
    \end{minipage}
    &\begin{minipage}{3cm}\graphicspath{{figures/sensitivity}}
    \centering\includegraphics[width=3cm]{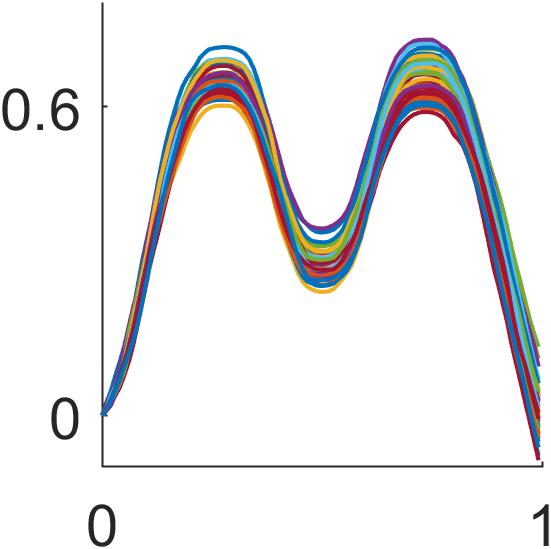}
    \end{minipage}
    \\
    \hline  
    \end{tabularx}
  \caption{Sensitivity of posterior inference to different choices of the number of B-spline basis functions, $B$, used in the SRVF template model based on Simulated Example 1. The first and second rows show the marginal posterior means of the template function and the phase component $\gamma_1$ in black, given $f_{1:50}$, respectively, with ground truth values shown in red. The third row shows the registered simulated functions via the marginal posterior means of phase components. We consider three different choices of $B$: (a) $B = 6$, (b) $B = 8$, and (c) $B = 10$.}
    \label{fig:nbasis}
\end{figure}

We assess sensitivity of posterior inference to different choices of the number and/or type of basis functions used to model the template function. In the main manuscript, we use cubic B-spline basis functions with equally spaced knots, which is a common choice in functional data analysis \citep{telesca2008}. Recall that, to generate functions for Simulated Example 1 in the main manuscript, we used 8 cubic B-splines. We begin by presenting marginal posterior summaries for the template function and the first relative phase component $\gamma_1$ in the first and second rows in Figure \ref{fig:nbasis}; here, panels (a)-(c) correspond to different numbers of B-spline basis functions used in the SRVF template model: (a) 6, (b) 8, and (c) 10. The results are visualized in the same manner as in the two previous sections and report the overall marginal posterior variances of the template function in Table \ref{tab:sensitivity_basis_template_var}. We note that marginal posterior mean template estimates (and associated uncertainty) in panels (b) and (c) are very similar. The posterior mean template estimate in panel (a) has a very similar shape to those in panels (b) and (c), but with larger marginal posterior variance; this is expected since we use only 6 B-spline basis functions to model the template whereas the data was generated using 8 B-spline basis functions. We report the pairwise distances between the (scaled to have norm one) marginal posterior means of the template function in Table \ref{tab:sensitivity_basis_template_mean}. The labels (a)-(c) in both tables correspond to the three panels in Figure \ref{fig:nbasis}. Note that the distance between (b) and (c) is very small. The distances between (a) and (b), and (a) and (c) are larger. %It is also evident that the overall marginal posterior variance of the template function is much larger for (a) than for (b) and (c). 
Examining rows two and three in Figure \ref{fig:nbasis}, it is evident that marginal posterior mean estimates of the first phase component, and the resulting registration results, are not sensitive to the number of B-spline basis functions used to model the SRVF template.

\begin{table}[!t]
    \centering
    \renewcommand{\arraystretch}{1}
    \begin{tabular}{c c c c c c c}
    \toprule[1pt]\midrule[0.3pt]
         (a) & (b) & (c) & (d) & (e) & (f)\\
         \midrule 0.9206  & 0.0807 &  0.0722  & 0.2405  & 0.3069 &  0.4042\\
          \midrule[0.3pt]\bottomrule[1pt]
    \end{tabular}
    \renewcommand{\arraystretch}{1}
    \caption{Estimates of the marginal posterior variance of the template function $(\times 10^3)$ for different choices of type and number of basis functions. (a) B-spline: 6. (b) B-spline: 8. (c) B-spline: 10. (d) Fourier: 6. (e) Fourier: 8. (f) Fourier: 10.}
    \label{tab:sensitivity_basis_template_var}
\end{table}

\begin{figure}[!t]
\graphicspath{{figure_type2/sensitivity}}
  \centering
  \begin{tabularx}{13.3cm} { 
  |>{\centering\arraybackslash}X 
   >{\centering\arraybackslash}X 
   >{\centering\arraybackslash}X| }
    \hline
    (d)& (e)& (f)\\
    \begin{minipage}{3cm}
    \centering\includegraphics[width=3cm]{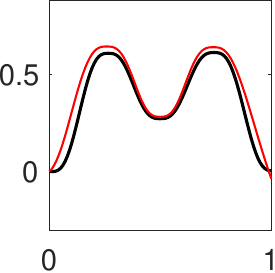}
    \end{minipage}
    &\begin{minipage}{3cm}
    \centering\includegraphics[width=3cm]{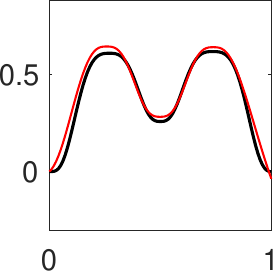}
    \end{minipage}
    &\begin{minipage}{3cm}
    \centering\includegraphics[width=3cm]{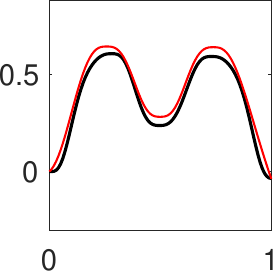}
    \end{minipage}
    \\\hline
    \begin{minipage}{3cm}
    \centering\includegraphics[width=3cm]{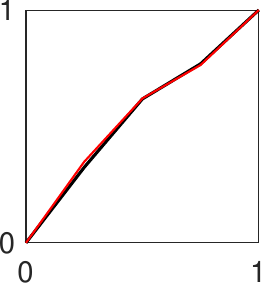}
    \end{minipage}
    &\begin{minipage}{3cm}
    \centering\includegraphics[width=3cm]{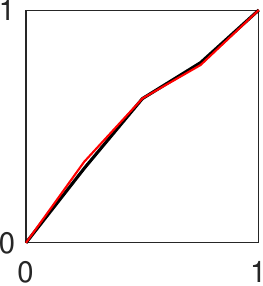}
    \end{minipage}
    &\begin{minipage}{3cm}
    \centering\includegraphics[width=3cm]{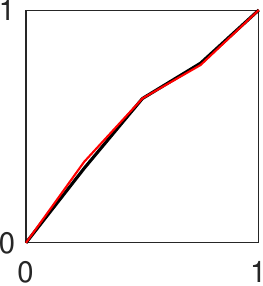}
    \end{minipage}\\\hline
    \begin{minipage}{3cm}\graphicspath{{figures/sensitivity}}
    \centering\includegraphics[width=3cm]{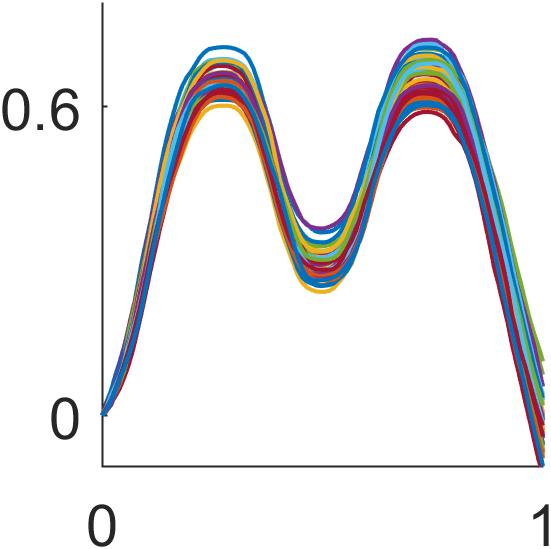}
    \end{minipage}
    &\begin{minipage}{3cm}\graphicspath{{figures/sensitivity}}
    \centering\includegraphics[width=3cm]{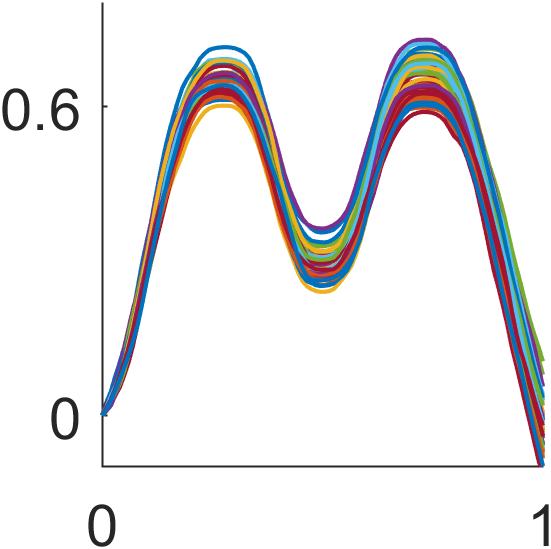}
    \end{minipage}
    &\begin{minipage}{3cm}\graphicspath{{figures/sensitivity}}
    \centering\includegraphics[width=3cm]{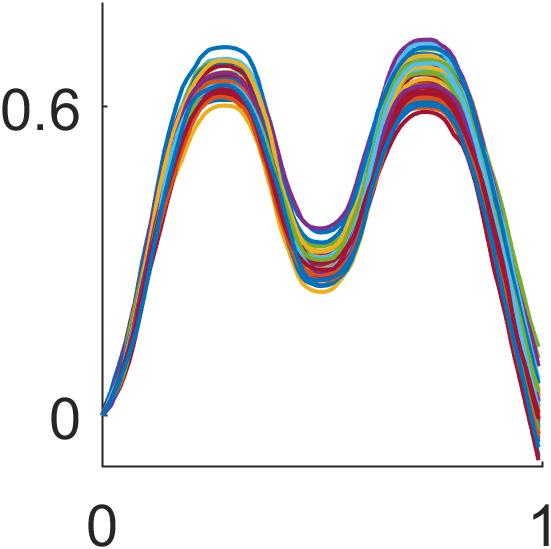}
    \end{minipage}
    \\
    \hline  
    \end{tabularx}
  \caption{Sensitivity of posterior inference to different choices of the number of Fourier basis functions, $B$, used to model the SRVF template based on Simulated Example 1. The first and second rows show the marginal posterior means of the template function and the phase component $\gamma_1$ in black, given $f_{1:50}$, respectively, with the ground truth values shown in red. The third row shows the registered simulated functions via the marginal posterior means of phase components. We consider three different choices of $B$: (d) $B = 6$, (e) $B = 8$, and (f) $B = 10$.}
    \label{fig:fbasis}
\end{figure}

Next, we assess sensitivity of posterior inference to a different choice of basis functions: the Fourier basis. Similarly to the B-spline basis, we considered different numbers of basis functions to model the SRVF template: (d) 6, (e) 8, and (f) 10. Summaries of the posterior are visualized in the same manner as before in Figure \ref{fig:fbasis}. Tables \ref{tab:sensitivity_basis_template_var} and \ref{tab:sensitivity_basis_template_mean} again report  the overall marginal posterior variances as well as pairwise distances between (scaled to have norm one) posterior mean estimates of the SRVF template, respectively. As before, as long as a sufficient number of basis elements is used to model the template, the results do not appear very sensitive to the choice of type of basis, i.e., B-spline vs. Fourier. Furthermore, as seen in rows two and three in Figure \ref{fig:fbasis}, marginal posterior mean estimates of the first phase component, and the resulting registration results, are not sensitive to the choice of the type of basis or the number of basis functions used.

\begin{table}[!t]
    \centering
    \renewcommand{\arraystretch}{1}
    \begin{tabular}{c c c c c c}
    \toprule[1pt]\midrule[0.3pt]
         Distance & (b) & (c) & (d) & (e) & (f)\\
         \midrule
         (a) & 0.1017 & 0.1023 & 0.1506 & 0.1273 & 0.1445\\
         (b) & & 0.0129 & 0.1040 & 0.0688 & 0.0824\\
         (c) & & & 0.1102 & 0.0686 & 0.0826\\
         (d) & & & & 0.0624 & 0.0888\\
         (e) & & & & & 0.0494\\
          \midrule[0.3pt]\bottomrule[1pt]
    \end{tabular}
    \renewcommand{\arraystretch}{1}
    \caption{Pairwise distances between normalized marginal posterior mean estimates of the template function for different choices of type and number of basis functions. (a) B-spline: 6. (b) B-spline: 8. (c) B-spline: 10. (d) Fourier: 6. (e) Fourier: 8. (f) Fourier: 10.}
    \label{tab:sensitivity_basis_template_mean}
\end{table}

%\begin{figure}
%\graphicspath{{figure_type2/sensitivity}}
%    \centering
%    \includegraphics[width=6cm]{/fmean_basis.pdf}
%    \caption{Visualization of the estimates of the marginal posterior mean of the template function given different sets of basis functions.}
%    \label{fig:basis_fmean}
%\end{figure}

\subsection{Sensitivity to $\kappa_{ini}$}

\begin{figure}[!t]
\graphicspath{{figure_type2/sensitivity}}
  \centering
  \begin{tabularx}{13.3cm} { 
  |>{\centering\arraybackslash}X 
   >{\centering\arraybackslash}X 
   >{\centering\arraybackslash}X 
   >{\centering\arraybackslash}X| }
    \hline
    (a)& (b)& (c)& (d)\\
    
    \begin{minipage}{3cm}
    \centering\includegraphics[width=3cm]{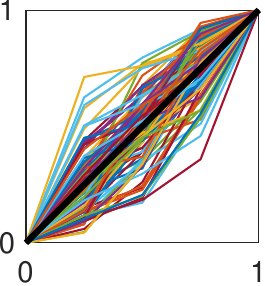}
    \end{minipage}
    &\begin{minipage}{3cm}
    \centering\includegraphics[width=3cm]{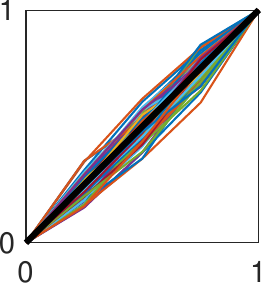}
    \end{minipage}
    &\begin{minipage}{3cm}
    \centering\includegraphics[width=3cm]{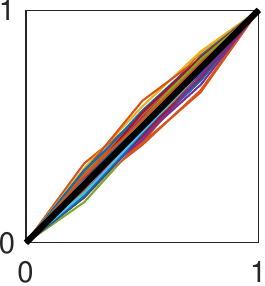}
    \end{minipage}
    &\begin{minipage}{3cm}
    \centering\includegraphics[width=3cm]{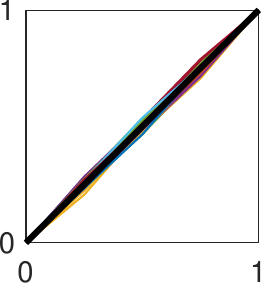}
    \end{minipage}\\
    \hline
    \begin{minipage}{3cm}
    \centering\includegraphics[width=3cm]{/fmean10.pdf}
    \end{minipage}
    &\begin{minipage}{3cm}
    \centering\includegraphics[width=3cm]{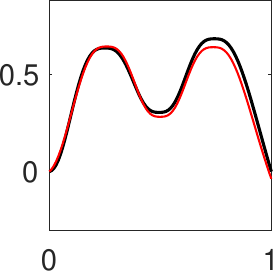}
    \end{minipage}
    &\begin{minipage}{3cm}
    \centering\includegraphics[width=3cm]{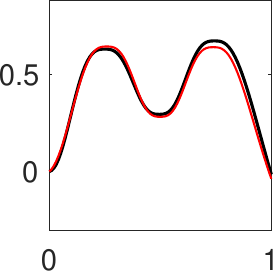}
    \end{minipage}
    &\begin{minipage}{3cm}
    \centering\includegraphics[width=3cm]{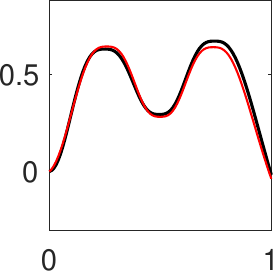}
    \end{minipage}
    \\\hline
    \begin{minipage}{3cm}
    \centering\includegraphics[width=3cm]{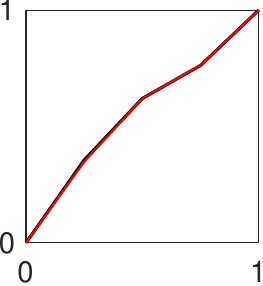}
    \end{minipage}
    &\begin{minipage}{3cm}
    \centering\includegraphics[width=3cm]{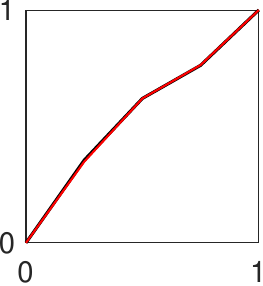}
    \end{minipage}
    &\begin{minipage}{3cm}
    \centering\includegraphics[width=3cm]{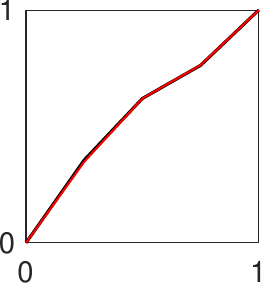}
    \end{minipage}
    &\begin{minipage}{3cm}
    \centering\includegraphics[width=3cm]{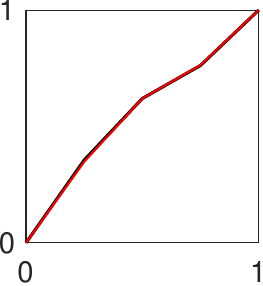}
    \end{minipage}
    \\\hline
   \begin{minipage}{3cm}
    \centering\includegraphics[width=3cm]{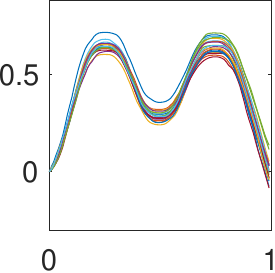}
    \end{minipage}
    &\begin{minipage}{3cm}
    \centering\includegraphics[width=3cm]{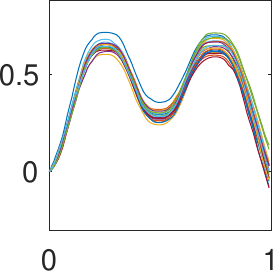}
    \end{minipage}
    &\begin{minipage}{3cm}
    \centering\includegraphics[width=3cm]{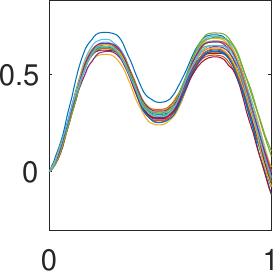}
    \end{minipage}
    &\begin{minipage}{3cm}
    \centering\includegraphics[width=3cm]{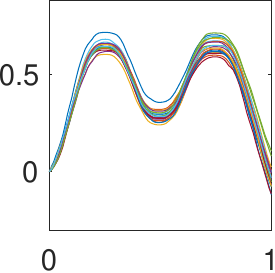}
    \end{minipage}\\
    \hline
  \end{tabularx}
  \caption{Sensitivity of posterior inference to different choices of the concentration parameter for the initialization kernel $\kappa_{ini}$ based on Simulated Example 1. 
  The first row shows 100 random draws of phase components from the Dirichlet distribution centered at the identity warping (black), for different values of $\kappa_{ini}$. The second and third rows show the marginal posterior means of the template function and the phase component $\gamma_1$, given $f_{1:50}$, respectively in black, with the ground truth values in red. The fourth row shows the registered simulated functions via the marginal posterior means of phase components. We consider four different choices of $\kappa_{ini}$: (a) $\kappa_{ini} = 10$, (b) $\kappa_{ini} = 100$, (c) $\kappa_{ini} = 200$, and (d) $\kappa_{ini} = 1000$.}
    \label{fig:kappa}
\end{figure}

We explore sensitivity of posterior inference to the concentration parameter $\kappa_{ini}$ used in the initialization kernel for new phase components as functional data arrive sequentially. We consider four different choices of $\kappa_{ini}$: (a) $\kappa_{ini} = 10$, (b) $\kappa_{ini} = 100$, (c) $\kappa_{ini} = 200$, and  (d) $\kappa_{ini} = 1000$. We illustrate the marginal posterior means over the template function and a phase component $\gamma_1$ in Figure \ref{fig:kappa}.  The first row in Figure \ref{fig:kappa} illustrates random samples from a Dirichlet distribution centered at the identity warping function, for different $\kappa_{ini}$ values. The second, third, and fourth rows are visualized in the same manner as in the previous sections and associated figures; \emph{results computed based on the target posterior look very similar regardless of the value of $\kappa_{ini}$}. The pairwise distances between the marginal posterior means of the template function are all $<0.01$, and those between the marginal posterior means of the first phase component are all $<0.00001$. The differences between marginal posterior variances for the template function and the first phase component are $<0.0001$ and $<0.00001$, respectively. This suggests that posterior inference is not sensitive to different choices of $\kappa_{ini}$.

\bibliography{reference}% common bib file